\documentclass[PAPER, atlasdraft=false, UKenglish, coverpage=false, cernpreprint=true, texmf, orcidlogo]{atlasdoc}
\AtlasTitle{Measurements of Higgs boson production via gluon--gluon fusion and vector-boson fusion using $H\rightarrow WW^\ast \rightarrow \ell\nu\ell\nu$ decays in $pp$ collisions with the ATLAS detector and their effective field theory interpretations}
\AtlasJournal{EPJC}
\AtlasAbstract{Higgs boson production cross-sections via gluon--gluon fusion and vector-boson fusion in proton--proton collisions are measured in the $H\rightarrow WW^\ast \rightarrow \ell\nu\ell\nu$ decay channel. The Large Hadron Collider delivered proton--proton collisions at a centre-of-mass energy of $13\,\textrm{TeV}$ between 2015 and 2018, which were recorded by the ATLAS detector, corresponding to an integrated luminosity of $140\,\textrm{fb}^{-1}$. The total cross-sections for Higgs boson production by gluon--gluon fusion and vector-boson fusion times the $H\rightarrow WW^\ast$ branching ratio are measured to be $12.4^{+1.3}_{-1.2}\,\textrm{pb}$ and $0.79^{+0.18}_{-0.16}\,\textrm{pb}$, respectively, in agreement with the Standard Model predictions. Higgs boson production is further characterised through measurements of Simplified Template Cross-Sections in a total of fifteen kinematic fiducial regions. A new scheme of kinematic fiducial regions has been introduced to enhance the sensitivity to CP-violating effects in Higgs boson interactions. Both schemes are used to constrain CP-even and CP-odd dimension-six operators in the Standard Model effective field theory.}
\usepackage{atlaspackage}
\usepackage{amsmath}
\usepackage{duckuments}
\usepackage{enumitem}
\usepackage{float}
\usepackage{makecell}
\usepackage{multirow}
\usepackage{pdflscape}
\usepackage{siunitx}
\usepackage{slashed}
\usepackage{tabularht}
\usepackage{tabularx}
\newcolumntype{Y}{>{\centering\arraybackslash}X}
\usepackage{atlasbiblatex}

\usepackage{atlasphysics}
\graphicspath{{logos/}{figures/}}

\usepackage{ANA-HIGP-2024-07-PAPER-defs}
\usepackage{eft-defs-top}
\usepackage{eft-diagrams}

\AtlasTitle{%
Measurements of Higgs boson production via gluon--gluon fusion and vector-boson fusion using $H\rightarrow WW^\ast \rightarrow \ell\nu\ell\nu$ decays in $pp$ collisions with the ATLAS detector and their effective field theory interpretations
}

\AtlasAbstract{%
Higgs boson production cross-sections via gluon--gluon fusion and vector-boson fusion in proton--proton collisions are measured in the $H\rightarrow WW^\ast \rightarrow \ell\nu\ell\nu$ decay channel. The Large Hadron Collider delivered proton--proton collisions at a centre-of-mass energy of $13\,\textrm{TeV}$ between 2015 and 2018, which were recorded by the ATLAS detector, corresponding to an integrated luminosity of $140\,\textrm{fb}^{-1}$. The total cross-sections for Higgs boson production by gluon--gluon fusion and vector-boson fusion times the $H\rightarrow WW^\ast$ branching ratio are measured to be $12.4^{+1.3}_{-1.2}\,\textrm{pb}$ and $0.79^{+0.18}_{-0.16}\,\textrm{pb}$, respectively, in agreement with the Standard Model predictions. Higgs boson production is further characterised through measurements of Simplified Template Cross-Sections in a total of fifteen kinematic fiducial regions. A new scheme of kinematic fiducial regions has been introduced to enhance the sensitivity to CP-violating effects in Higgs boson interactions. Both schemes are used to constrain CP-even and CP-odd dimension-six operators in the Standard Model effective field theory.
}

\AtlasRefCode{HIGP-2024-07}

\PreprintIdNumber{CERN-EP-2025-054}

\AtlasJournal{Eur. Phys. J. C}
\AtlasJournalRef{Eur. Phys. J. C 85 (2025) 1403}
\AtlasDOI{10.1140/epjc/s10052-025-14761-2}

\hypersetup{pdftitle={ATLAS document},pdfauthor={The ATLAS Collaboration}}

\begin{document}

\maketitle

\tableofcontents

%
%

%
%
%

\section{Introduction}
\label{sec:intro}

The Higgs boson is a neutral scalar particle produced by the excitation of the Higgs scalar field, whose non-zero vacuum expectation value spontaneously breaks the electroweak (EW) symmetry of the Standard Model (SM), thereby generating masses for the \W and \Z gauge bosons as well as fermions through Yukawa interactions~\cite{Englert:1964et, Higgs:1964ia, Higgs:1964pj, Guralnik:1964eu}. The observation of a new particle consistent with the SM Higgs boson was reported in 2012 by the ATLAS and CMS Collaborations~\cite{HIGG-2012-27, CMS-HIG-12-028}. The Higgs boson exhibits a rich phenomenology that can be probed experimentally. Precision measurements of its characteristics are a powerful test of the SM and can be used to constrain theories of physics beyond the SM (BSM). If BSM physics is present, it can alter the kinematic properties of Higgs boson production and decay. These effects can be probed using the large data sample delivered by the Large Hadron Collider (LHC)~\cite{Evans:2008zzb} at CERN, which allows the measurement of the Higgs boson production cross-sections in different kinematic regions.

This paper describes measurements of Higgs boson production by gluon--gluon fusion (ggF) and vector-boson fusion (VBF) using \hwwlvlv decays in proton--proton (\pp) collisions at a centre-of-mass energy of \qty{13}{\TeV}. The data were recorded by the ATLAS detector~\cite{PERF-2007-01} between the years 2015--2018 (Run~2) of the LHC operation and correspond to an integrated luminosity of \qty{140}{\ifb}. The chosen decay channel takes advantage of the sizeable branching ratio for the \hwwlvlv decay. Final states with both different- and same-flavour charged light leptons (electrons or muons) are considered. Events with \hww decays resulting in final states with two light leptons via intermediate $\tau$-lepton decays are also considered as part of the signal. The measured cross-section in the ggF production mode probes the Higgs boson couplings to heavy quarks or in an effective coupling description to a pair of gluons, while the VBF production mode directly probes its couplings to \W and \Z bosons. Previous measurements of ggF and VBF production cross-sections inclusively and in the Simplified Template Cross-Section (STXS) scheme~\cite{deFlorian:2016spz, Andersen:2016qtm, Berger:2019wnu, Amoroso:2020lgh, LHCHWG-INT-2025-001} have been reported by the ATLAS and CMS Collaborations in the \hwwlvlv decay channel using the full LHC Run~2 data sample~\cite{HIGG-2021-20, CMS-HIG-20-013}. Additionally, the ATLAS Collaboration has performed measurements of other Higgs boson production modes in the \hww decay channel, including production in association with a vector boson (\VH)~\cite{HIGG-2023-09} and with a top quark pair (\ttH)~\cite{HIGG-2018-13}.

Several improvements are made in this paper compared to the previous ATLAS Run~2 analysis~\cite{HIGG-2021-20}: the inclusion of the final state with two same-flavour, oppositely charged leptons, the use of neural networks for all final signal--background discriminants, a more precise estimation and improved rejection of background events with misidentified leptons, a finer granularity of kinematic regions for the Higgs boson production cross-sections measured differentially within the STXS framework, interpretations in the context of the Standard Model effective field theory (SMEFT)~\cite{Brivio:2017vri, ATL-PHYS-PUB-2019-042}, and other experimental improvements. Owing to these improvements, the expected measurement precision reported here improves by up to 36\%. The results presented in this paper supersede those previously reported in Ref.~\cite{HIGG-2021-20}.

A second differential cross-section measurement incorporating a variable sensitive to the combined charge conjugation and parity (CP) inversion symmetries is introduced in the STXS categorisation scheme for the first time to probe CP-violating effects in a Higgs boson production phase space with two jets. The \hww decay channel and full Run~2 data sample were previously used to place constraints on CP-violating anomalous interactions modifying the $HVV$ coupling. ATLAS extracted the constraints via differential measurements of sensitive observables in VBF production~\cite{HIGG-2020-25} and CMS extracted the constraints by using matrix element techniques~\cite{CMS-HIG-22-008}. This analysis enhances the differential measurement of VBF production by incorporating a multidimensional differential approach (see Section~\ref{sec:dphijj_cat}) and by including the same-flavour channel, both of which improve its sensitivity to CP-odd effects.

The outline of this paper is as follows. Section~\ref{sec:overview} provides an overview of the signal characteristics and the analysis strategy. Section~\ref{sec:detector} introduces the ATLAS detector. Section~\ref{sec:samples} describes the data and the simulated event samples. Section~\ref{sec:reconstruction} describes the event reconstruction. Section~\ref{sec:channels} details the various selections used to define the signal and control regions in the analysis. Section~\ref{sec:background} discusses how the backgrounds are estimated. Section~\ref{sec:systematics} discusses the systematic uncertainties, and Section~\ref{sec:statistics} defines the likelihood fit procedure. Finally, the results of the analysis and their interpretations are presented in Sections~\ref{sec:results} and \ref{sec:interpretation}, respectively. The conclusions are summarised in Section~\ref{sec:conclusion}. Additional material on backgrounds with misidentified leptons is shown in Appendix~\ref{app:fakes}, on the STXS and CP-sensitive measurements in Appendix~\ref{app:results}, and on the SMEFT interpretations in Appendix~\ref{app:EFT}.


\section{Analysis overview}
\label{sec:overview}

The \hwwlvlv decay is characterised by two charged leptons and at least two undetected neutrinos in the final state. In the SM, the opening angle between the two charged leptons tends to be small due to the spin-0 nature of the Higgs boson and the chiral structure of the weak force in the decay of the two \W bosons~\cite{Dittmar:1996ss}. This feature of the decay is exploited to separate the Higgs boson signal from the main backgrounds such as continuum production of \WW, where the charged leptons are more likely to have a large opening angle.

In addition to the decay products of the Higgs boson, the final state can be populated by jets either from the quarks participating in the VBF production mode and their final-state radiation or from initial-state radiation from quarks or gluons in both the ggF and VBF production modes. The composition of the background processes changes significantly depending on the number of jets, \Njets, in the final state. Therefore, the analysis is performed separately in three reconstructed `event classes' with \ZeroJet, \OneJet, and \TwoJet. Owing to differences between the background compositions, each event class is also split according to the flavour combination of the final-state lepton pair: different flavour (DF), \emme, and same flavour (SF), \eemm. In the \ZeroJet and \OneJet event classes, only the DF decay channel is considered because the sensitivity of the analysis to SF decay channel is limited by the large \Zeemm background. In the \TwoJet event class, the signal yield is increased relative to the \Zeemm background and both the DF and SF decay channels are considered.

For each event class, selection criteria are applied to enhance the signal purity and result in independent samples of events referred to as signal regions (SRs). A deep neural network (DNN), implemented through Keras~\cite{chollet2015keras} using TensorFlow~\cite{Abadi:2016kic} as the backend, is trained to identify the signal topology targeted in each SR. Additionally, control regions (CRs) that are kinematically close to but statistically independent of each SR, are used to constrain the normalisation of the major backgrounds in each SR. A fit to the signal and background predictions to data in all SRs and CRs is performed. In each SR, the output of the DNN is used as the discriminating variable in the fit, while each CR consists of a single bin. A profile likelihood fit is used to obtain the best-fit values for the set of parameters-of-interest (POIs).

Measurements of the ggF and VBF production cross-sections, multiplied by the \hww branching ratio, are reported. Cross-section measurements are also conducted in the Stage~1.2 STXS category scheme~\cite{deFlorian:2016spz, Andersen:2016qtm, Berger:2019wnu, Amoroso:2020lgh, LHCHWG-INT-2025-001}, which splits the production modes of the simulated Higgs boson into experimentally accessible exclusive kinematic fiducial regions or `categories' that are relevant for theory interpretations. Two STXS template processes are measured. The ggF STXS template process (referred to as \ggH) is defined to be the $\gg\to\H$ process including higher-order quantum chromodynamics (QCD) and EW corrections. The VBF STXS template process (referred to as EW \qqH) includes the VBF production process. The predictions for both \ggH and EW \qqH also include contributions from \VH production with a hadronically decaying vector boson; the latter is only targeted with an analysis reported in Ref.~\cite{HIGG-2023-09}. Simulated events are categorised according to the number of jets (\Njets), the transverse momentum (\pt) of the Higgs boson (\pTH), and, in the case of EW \qqH, the dijet invariant mass of highest-\pt jets (\mjj). For events with at least two jets, this analysis targets events with \mjj above \qty{100}{\GeV} for \ggH and above \qty{350}{\GeV} for EW \qqH; events with lower \mjj are dominated by \VH production with a hadronically decaying vector boson. After merging certain STXS categories to ensure a reasonable sensitivity for all the measured POIs, a total of 15~fiducial cross-sections corresponding to different STXS-bound kinematic categories are measured: seven for \ggH production and eight for EW \qqH production. All STXS categories include a requirement on the Higgs boson rapidity of $|y_H| < 2.5$. These STXS categories, together with the signal regions targeting them, are described in greater detail in Section~\ref{sec:stxs_cat}.

An interpretation of the results is performed in the context of the SMEFT framework. SMEFT provides a nearly model-independent way of probing BSM physics at an energy scale $\Lambda \sim \mathcal{O}(\TeV) \ggmath \nu$, where $\nu$ is the vacuum expectation value of the Higgs field. In this regime, the BSM physics decouples from the SM physics, allowing the SM Lagrangian to be expanded by higher-dimensional operators in powers of $\Lambda^{-n}$, which are built from the SM's fields and retain its symmetries. At each order $n > 1$, the full set of non-redundant operators of mass-dimension $d = 4 + n$ constitute the effective interactions added to the SM Lagrangian. The free parameters of the theory are the Wilson coefficients $c_i$, where $c_i/\Lambda^n$ scales the strength of the corresponding $d = 4 + n$ operator $\mathcal{O}_i$. In the SM, by definition, the value of each Wilson coefficient is equal to zero. In the present interpretation, lepton and baryon number conservation are assumed and so the $d = 5$ operators are excluded; therefore, the \dimSix operators correspond to the lowest higher-dimensional order. Operators of higher dimensions ($d > 6$) are neglected. In this measurement, the Warsaw basis~\cite{Grzadkowski:2010es} in the top-flavour scheme~\cite{Aguilar-Saavedra:2018ksv} is assumed for the \dimSix operators.

In the SMEFT framework, SM scattering amplitudes are modified by the interference with \dimSix operators, leading to corrections to the production cross-sections, partial decay widths, and total decay width of the Higgs boson. The expected STXS cross-sections are parameterised in terms of the Wilson coefficients for sixteen operators and are constrained using the measured STXS cross-sections described above. A second interpretation is performed to set limits on CP violation in the Higgs boson sector. For this interpretation, production cross-sections times the \hww branching ratio are measured relative to their SM prediction (this ratio is referred to as the `signal strength') in an STXS-like scheme that relies on binning in an additional CP-sensitive variable: the azimuthal angular difference between the two leading jets, ordered by rapidity (\dphijjsgn)~\cite{Plehn:2001nj, Klamke:2007cu}. Asymmetry in the measured \dphijjsgn distribution provides sensitivity to CP-odd effects. Selected events in \ggH and EW \qqH STXS categories with two or more jets are further separated according to \dphijjsgn, leading to a total of 20~bins spanning several \pTH and \mjj categories. The measured signal strengths are used to constrain the impacts of two CP-odd effective operators and their CP-even counterparts on Higgs boson production and decay.


\section{The ATLAS detector}
\label{sec:detector}

\newcommand{\AtlasCoordFootnote}{%
ATLAS uses a right-handed coordinate system with its origin at the nominal interaction point (IP) in the centre of the detector and the \(z\)-axis along the beam pipe. The \(x\)-axis points from the IP to the centre of the LHC ring, and the \(y\)-axis points upwards. Polar coordinates \((r,\phi)\) are used in the transverse plane, \(\phi\) being the azimuthal angle around the \(z\)-axis. The pseudorapidity is defined in terms of the polar angle \(\theta\) as \(\eta = -\ln \tan(\theta/2)\) and is equal to the rapidity $ y = \frac{1}{2} \ln \left( \frac{E + p_z}{E - p_z} \right) $ in the relativistic limit. Angular distance is measured in units of \(\Delta R \equiv \sqrt{(\Delta\eta)^{2} + (\Delta\phi)^{2}}\).}

The ATLAS detector~\cite{PERF-2007-01} at the LHC covers nearly the entire solid angle around the collision point.\footnote{\AtlasCoordFootnote} It consists of an inner tracking detector surrounded by a thin superconducting solenoid, electromagnetic and hadronic calorimeters, and a muon spectrometer incorporating three large superconducting air-core toroidal magnets.

The inner-detector system (ID) is immersed in a \qty{2}{\tesla} axial magnetic field and provides charged-particle tracking in the range of \(|\eta| < 2.5\). The high-granularity silicon pixel detector covers the vertex region and typically provides four measurements per track, the first hit generally being in the insertable B-layer (IBL) installed before Run~2~\cite{ATLAS-TDR-19, PIX-2018-001}. It is followed by the SemiConductor Tracker (SCT), which usually provides eight measurements per track. These silicon detectors are complemented by the transition radiation tracker (TRT), which enables radially extended track reconstruction up to \(|\eta| = 2.0\). The TRT also provides electron identification information based on the fraction of hits (typically 30 in total) above a higher energy-deposit threshold corresponding to transition radiation.

The calorimeter system covers the pseudorapidity range \(|\eta| < 4.9\). Within the region \(|\eta|< 3.2\), electromagnetic calorimetry is provided by barrel and endcap high-granularity lead/liquid-argon (LAr) calorimeters, with an additional thin LAr presampler covering \(|\eta| < 1.8\) to correct for energy loss in material upstream of the calorimeters. Hadronic calorimetry is provided by the steel/scintillator-tile calorimeter, segmented into three barrel structures within \(|\eta| < 1.7\), and two copper/LAr hadronic endcap calorimeters. The solid angle coverage is completed with forward copper/LAr and tungsten/LAr calorimeter modules optimised for electromagnetic and hadronic energy measurements, respectively.

The muon spectrometer (MS) comprises separate trigger and high-precision tracking chambers measuring the deflection of muons in a magnetic field generated by the superconducting air-core toroidal magnets. The field integral of the toroids ranges between \num{2.0} and \qty{6.0}{\tesla\metre} across most of the detector. Three layers of precision chambers, each consisting of layers of monitored drift tubes, cover the region \(|\eta| < 2.7\), complemented by cathode-strip chambers in the forward region, where the background is highest. The muon trigger system covers the range \(|\eta| < 2.4\) with resistive-plate chambers in the barrel, and thin-gap chambers in the endcap regions.

The luminosity is measured mainly by the LUCID--2~\cite{LUCID2} detector that records Cherenkov light produced in the quartz windows of photomultipliers located close to the beampipe.

Events are selected by the first-level trigger system implemented in custom hardware, followed by selections made by algorithms implemented in software in the high-level trigger~\cite{TRIG-2016-01}. The first-level trigger accepts events from the \qty{40}{\MHz} bunch crossings at a rate close to \qty{100}{\kHz}, which the high-level trigger further reduces in order to record complete events to disk at about \qty{1.25}{\kHz}.

A software suite~\cite{SOFT-2022-02} is used in data simulation, in the reconstruction and analysis of real and simulated data, in detector operations, and in the trigger and data acquisition systems of the experiment.


\section{Data and simulated event samples}
\label{sec:samples}

The full ATLAS Run~2 data sample is used for this measurement, consisting of \pp collision data produced at $\sqrts = \qty{13}{\TeV}$ and recorded between 2015 and 2018. The data are filtered according to quality criteria~\cite{DAPR-2018-01}, including the removal of events recorded when relevant detector components were not operating correctly. The total integrated luminosity after this cleaning of the data corresponds to \qty{140}{\ifb}~\cite{DAPR-2021-01}. The events were collected using a combination of unprescaled single-lepton and $e\mu$ dilepton triggers~\cite{TRIG-2018-05, TRIG-2018-01} to maximise the total trigger efficiency. The \pt thresholds for the single-electron and single-muon triggers were 24 and \qty{20}{\GeV}, respectively, for the first year of data taking and increased to \qty{26}{\GeV} for both lepton flavours during the remainder of Run~2. The $e\mu$ dilepton trigger had \pt thresholds of 17 and \qty{14}{\GeV} for electrons and muons, respectively.

Monte Carlo (MC) generators were used to model the hard \pp scattering matrix element, parton shower and hadronisation, and underlying event. The generators that were used for modelling signal and background processes are listed in Table~\ref{tab:samples}.

\begin{table}[htbp]
\centering
\caption{Overview of simulation tools used to generate the signal and background processes. Also summarised are the corresponding parton distribution function (PDF) sets and the models used for the underlying event/parton shower (UE/PS) simulation. `Prediction Order' refers to the order of the perturbation theory at which the cross-sections are computed and used to normalise the samples; when higher-orders are specified, QCD is implied if not explicitly stated. Alternative event generators and configurations that are used to estimate systematic uncertainties are shown in parentheses. For a given process, the alternative event generator and configuration used for the matrix element is independent of that used for the UE/PS model.}
\label{tab:samples}
\resizebox{\textwidth}{!}{
\begin{tabular}{l l l l l}
\toprule
Process & Matrix element & PDF set & UE/PS model & Prediction order \\
\midrule
ggF \H & \POWHEGBOX[v2] & \PDFforLHC[15nnlo] & \PYTHIA[8.212] & N$^3$LO QCD + NLO EW \\
& (\MGNLO) & & (\HERWIG[7.13]) & \\
VBF \H & \POWHEGBOX[v2] & \PDFforLHC[15nnlo] & \PYTHIA[8.230] & NNLO QCD + NLO EW \\
& (\MGNLO) & & (\HERWIG[7.13]) & \\
\VH & \POWHEGBOX[v2] & \PDFforLHC[15nnlo] & \PYTHIA[8.212] & NNLO QCD + NLO EW \\
$\gg\to\ZH$ & \POWHEGBOX[v2] & \PDFforLHC[15nnlo] & \PYTHIA[8.212] & NLO + NLL \\
\ttH & \POWHEGBOX[v2] & \NNPDF[3.0nlo] & \PYTHIA[8.230] & NLO QCD + NLO EW \\
\midrule
$\DY\to\eemm$ & \SHERPA[2.2.11] & \NNPDF[3.0nnlo] & \SHERPA[2.2.11] & NNLO \\
$\DY\to\tau\tau$ & \SHERPA[2.2.14] & \NNPDF[3.0nnlo] & \SHERPA[2.2.14] & NNLO \\
\VV/\Vys & \SHERPA[2.2.11/2.2.12] & \NNPDF[3.0nnlo] & \SHERPA[2.2.12] & NLO \\
\Vy & \SHERPA[2.2.8] & \NNPDF[3.0nnlo] & \SHERPA[2.2.8] & NLO \\
$\gg\to\VV$ & \SHERPA[2.2.2] & \NNPDF[3.0nnlo] & \SHERPA[2.2.2] & NLO \\
\VVV & \SHERPA[2.2.2] & \NNPDF[3.0nnlo] & \SHERPA[2.2.2] & NLO \\
EW \Zjj & \HERWIG[7.1.3/7.2.0] & \MMHT[nlo] & \HERWIG[7.1.3/7.2.0] & NLO \\
EW \VVjj & \MADGRAPH & \NNPDF[3.0nlo] & \PYTHIA[8] & LO \\
\ttbar & \POWHEGBOX[v2] & \NNPDF[3.0nlo] & \PYTHIA[8.230] & NNLO + NNLL \\
& (\POWHEGBOX[v2], $\pthard = 1$) & & (\HERWIG[7.04]) & \\
\Wt & \POWHEGBOX[v2] & \NNPDF[3.0nlo] & \PYTHIA[8.230] & NLO + NNLL \\
& (\POWHEGBOX[v2], $\pthard = 1$) & & (\HERWIG[7.04]) & \\
\bottomrule
\end{tabular}
}
\end{table}

The Higgs boson signal samples were simulated with the \HWW decay in the four main production modes: ggF, VBF, \VH, and \ttH. They were generated using a Higgs boson mass of \qty{125}{\GeV} and then normalised to the cross-sections and branching ratios~\cite{deFlorian:2016spz} computed for a mass of \qty{125.09}{\GeV}~\cite{HIGG-2014-14}. The normalisation accounts for the decay branching ratio calculated with \textsc{HDECAY}~\cite{Djouadi:1997yw, Spira:1997dg, Djouadi:2006bz} and \PROPHECY~\cite{Bredenstein:2006rh, Bredenstein:2006nk, Bredenstein:2006ha}. The samples used the \PDFforLHC[15nnlo]~\cite{Butterworth:2015oua} (for ggF, VBF, and \VH) or \NNPDF[3.0nlo]~\cite{Ball:2014uwa} (for \ttH) parton distribution function (PDF) sets. They were interfaced to \PYTHIA[8.212]~\cite{Sjostrand:2014zea} (for ggF and \VH) or \PYTHIA[8.230] (for VBF and \ttH) for parton showering and hadronisation, with parameters set according to the AZNLO~\cite{STDM-2012-23} (for ggF, VBF, and \VH) or the A14~\cite{ATL-PHYS-PUB-2014-021} (for \ttH) set of tuned parameters (tune). The decays of bottom and charm hadrons were simulated using \EVTGEN[1.6.0]~\cite{Lange:2001uf}.

The ggF events were simulated at next-to-next-to-leading-order (NNLO) accuracy in QCD using the \POWHEGBOX[v2]~NNLOPS program~\cite{Nason:2004rx, Frixione:2007vw, Alioli:2010xd, Hamilton:2013fea, Hamilton:2015nsa}. The simulation achieved NNLO accuracy for arbitrary inclusive $\gg\to\H$ observables by reweighting the Higgs boson rapidity spectrum in \textsc{Hj}-\MINLO~\cite{Campbell:2012am, Hamilton:2012np, Hamilton:2012rf} to that of HNNLO~\cite{Catani:2007vq}. The prediction was normalised to the next-to-next-to-next-to-leading-order (N$^3$LO) cross-section in QCD plus EW corrections at next-to-leading-order (NLO)~\cite{Actis:2008ug, Actis:2008ts, Harlander:2009bw, Harlander:2009mq, Pak:2009dg, Harlander:2009my, Anastasiou:2015ema, Anastasiou:2016cez, deFlorian:2016spz, Bonetti:2018ukf, Dulat:2018rbf}.

VBF production was simulated using \POWHEGBOX[v2]~\cite{Nason:2009ai}. The \POWHEG prediction is accurate to NLO in QCD corrections and tuned to match calculations with effects due to finite heavy-quark masses and soft-gluon resummations up to next-to-next-to-leading-logarithm (NNLL). The prediction is normalised to an approximate-NNLO QCD cross-section with NLO EW corrections~\cite{Ciccolini:2007jr, Ciccolini:2007ec, Bolzoni:2010xr}.

The uncertainties due to the parton shower and hadronisation model for the ggF and VBF Higgs boson signal samples are evaluated using the events in the nominal sample generated with \POWHEG but interfaced to \HERWIG[7.13]~\cite{Bahr:2008pv, Bellm:2015jjp} instead of \PYTHIA[8]. To estimate the uncertainties related to the matching of matrix elements and parton showers for ggF and VBF production, events were simulated using \MGNLO~\cite{Alwall:2014hca} at NLO in QCD with the \NNPDF[3.0nlo] PDF set and interfaced to \PYTHIA[8] (for ggF) or \HERWIG[7] (for VBF). For events interfaced to \HERWIG[7], the H7UE tune~\cite{Bellm:2015jjp} and the \MMHT[lo]~\cite{Harland-Lang:2014zoa} PDF set were used for the underlying event; for those interfaced to \PYTHIA[8], the A14 tune and the \NNPDF[2.3lo]~\cite{Ball:2012cx} PDF were used.

Higgs boson production in association with a vector boson was simulated using \POWHEGBOX[v2]~\MINLO~\cite{Alioli:2008gx, Luisoni:2013cuh}. The \POWHEG prediction is accurate to NLO for \VH plus one jet production, and each sample was normalised to a cross-section calculated at NNLO in QCD and at NLO in EW theory~\cite{Ciccolini:2003jy, Brein:2003wg, Brein:2011vx, Denner:2014cla, Brein:2012ne}. The subdominant $\gg\to\ZH$ process was simulated at leading order (LO) in QCD with \POWHEGBOX[v2] and normalised to a cross-section calculated at NLO in QCD with next-to-leading logarithmic (NLL) corrections~\cite{Altenkamp:2012sx, Harlander:2014wda}.

The production of \ttH events was simulated using the \POWHEGBOX[v2]~\cite{Frixione:2007nw, Hartanto:2015uka} generator at NLO and normalised to a cross-section calculated at NLO in QCD and NLO in EW accuracy~\cite{deFlorian:2016spz}.

The main background processes include single boson, diboson, triboson, and top quark production.

Production of \DY in association with zero or more jets (\Zjets) was using the \SHERPA[2.2.11]~\cite{Bothmann:2019yzt} generator for simulation, in the case of \eemm final states, or the \SHERPA[2.2.14] generator, in the case of $\tau\tau$ final states. The simulation is accurate to NLO in QCD for up to two jets and LO in QCD for up to five jets, calculated in the five-flavour scheme using \textsc{Comix}~\cite{Gleisberg:2008fv}. Approximate NLO EW corrections were applied to the samples following the additive approach described in Ref.~\cite{Kallweit:2014xda}. The samples were normalised to cross-sections computed at NNLO in QCD~\cite{Anastasiou:2003ds}.

Fully leptonic final states consisting of \VV/\Vys (including \WW, \WZ, and \ZZ) and \Vy were simulated using \SHERPA[2.2.12] and \SHERPA[2.2.8], respectively, while semileptonic final states consisting of \VV/\Vys were simulated using \SHERPA[2.2.11]. The simulation is accurate to NLO QCD for up to one jet and at LO for up to three jets. Similarly to the \Zjets samples, approximate NLO EW corrections were applied to the \VV samples following the additive approach. The loop-induced $\gg\to\VV$ process was simulated by \SHERPA[2.2.2] at LO in QCD for up to one jet and normalised to a cross-section computed at NLO in QCD~\cite{Caola:2015rqy, Cascioli:2013gfa, Grazzini:2018owa}. On-shell \VVV processes were simulated by \SHERPA[2.2.2] at NLO in QCD inclusively and at LO for up two additional jets.

For all \SHERPA samples, the matrix element calculations were matched and merged with the \SHERPA parton shower based on Catani--Seymour dipole factorisation~\cite{Schumann:2007mg, Gleisberg:2008fv} using the \MEPSatNLO prescription~\cite{Catani:2001cc, Hoeche:2009rj, Hoeche:2011fd, Hoeche:2012yf}. Virtual QCD NLO corrections were provided by the \OPENLOOPS library~\cite{Cascioli:2011va, Denner:2016kdg, Buccioni:2017yxi, Buccioni:2019sur}. The \NNPDF[3.0nnlo]~\cite{Ball:2014uwa} set of PDFs was used, along with the dedicated set of tuned parton shower parameters developed by the \SHERPA authors.

EW \Z production in association with two jets (EW \Zjj) was simulated using \HERWIG[7.1.3] and \HERWIG[7.2.0] for $\Z\to{ee}/\tau\tau$ and $\Z\to\mu\mu$, respectively. The simulation is accurate to NLO in QCD for up to two additional parton emissions and used the \MMHT[nlo]~\cite{Harland-Lang:2014zoa} set of PDFs.

EW diboson production in association with two jets (EW \VVjj) was simulated using \MADGRAPH~\cite{Alwall:2014hca} with LO matrix elements and used the \NNPDF[3.0nlo] set of PDFs. The events were interfaced to \PYTHIA[8] for the parton shower.

The production of \ttbar events was simulated using \POWHEGBOX[v2] at NLO in QCD with the \NNPDF[3.0nlo] PDF set and the \hdamp parameter\footnote{The \hdamp parameter is a resummation damping factor and one of the parameters that controls the matching of \POWHEG matrix elements to the parton shower and thus effectively regulates the high-\pt radiation against which the \ttbar system recoils.} set to $1.5\times\mtop$~\cite{ATL-PHYS-PUB-2016-020}. To correct for a known mismodelling of the leading lepton \pt due to missing higher-order corrections, an NNLO reweighting was applied to the sample. The events were normalised to a cross-section computed at NNLO in QCD with NNLL corrections~\cite{Beneke:2011mq, Cacciari:2011hy, Czakon:2011xx, Baernreuther:2012ws, Czakon:2012zr, Czakon:2012pz, Czakon:2013goa}.

The production of top quarks in association with \W bosons (mainly \Wt) was modelled using the \POWHEGBOX[v2]~\cite{Re:2010bp} generator at NLO in QCD. The generation used the five-flavour scheme and the \NNPDF[3.0nlo] set of PDFs. The diagram removal scheme~\cite{Frixione:2008yi} was used to remove interference and overlap with \ttbar production. The events were normalised to a cross-section computed at NLO in QCD with NNLL soft-gluon corrections~\cite{Kidonakis:2010ux, Kidonakis:2013zqa}.

For all top quark processes, the events were interfaced to \PYTHIA[8.230] to model the parton shower, hadronisation, and underlying event, with parameters set according to the A14 tune and using the \NNPDF[2.3lo] set of PDFs. The decays of bottom and charm hadrons were simulated using \EVTGEN[1.6.0].

In a manner analogous to that used for signal processes, the uncertainties due to the parton shower and hadronisation model for top quark processes are evaluated using the events in the nominal sample simulated with \POWHEG but interfaced to the alternative showering program \HERWIG[7.04] instead of \PYTHIA[8]. The simulation of the underlying event used the H7UE tune and the \MMHT[lo]~\cite{Harland-Lang:2014zoa} set of PDFs. To estimate the uncertainty related to the matching of the matrix element and parton shower for \ttbar and \Wt production, \POWHEGBOX[v2] events were showered using \PYTHIA[8] with the \pthard parameter set to 1 (nominally 0)~\cite{ATL-PHYS-PUB-2023-029}, which regulates the radiation phase space for \PYTHIA[8]~\cite{Bierlich:2022pfr}. They are accurate to NLO in QCD corrections and utilised the \NNPDF[3.0nlo] PDF set. The simulation of the underlying event used the A14 tune and the \NNPDF[2.3lo] PDF set.

The \Wjets and multijet backgrounds are estimated from data. Simulated \Wjets and \Zjets event samples are used to validate the estimate and to determine its corresponding uncertainties. These MC samples are generated using \SHERPA[2.2.11] or using \POWHEG interfaced to \PYTHIA[8]. The data-driven estimation is described in greater detail in Section~\ref{sec:background}.

All samples were processed through the \GEANT-based~\cite{Agostinelli:2002hh} ATLAS detector simulation~\cite{SOFT-2010-01} and the standard ATLAS reconstruction software~\cite{SOFT-2022-02}. The effect of pile-up, defined as additional \pp interactions in the same and neighbouring bunch crossings, was modelled by overlaying the hard-scattering event with simulated inelastic \pp events generated with \PYTHIA[8.186]~\cite{Sjostrand:2007gs} using the \NNPDF[2.3lo] PDF set and the A3~\cite{ATL-PHYS-PUB-2016-017} tune. The simulated event samples were weighted to reproduce the distribution of the average number of interactions per bunch crossing observed in the data, referred to as `pile-up reweighting'.

In this paper, the `particle-level' event corresponds to the state of a given simulated event following generation, parton showering, and hadronisation but before detector simulation. In the particle-level event, the four-momentum of each charged lepton was corrected (`dressed') with the four-momenta of all photons in a $\DR < 0.1$ cone around the lepton direction~\cite{ATL-PHYS-PUB-2015-013}.


\section{Event reconstruction}
\label{sec:reconstruction}

Primary vertices in the event are reconstructed from tracks in the ID with $\pt > \qty{500}{\MeV}$~\cite{ATL-PHYS-PUB-2015-026}. Events are required to have at least one primary vertex with at least two associated tracks. The hard-scatter vertex is selected as the vertex with the highest $\sum\pt^2$, where the sum is calculated using all tracks associated with that particular vertex.

Electron candidates are reconstructed by matching energy clusters in the electromagnetic calorimeter to well-reconstructed tracks that are extrapolated to the calorimeter~\cite{EGAM-2021-01}. All candidate electron tracks are fitted using a Gaussian sum filter to account for bremsstrahlung energy losses. Electron candidates are required to satisfy $\pt > \qty{15}{\GeV}$ and $|\eta| < 2.47$, excluding the transition region $1.37 < |\eta| < 1.52$ between the barrel and end caps of the LAr calorimeter.

Muon candidates are reconstructed from a global fit to matching tracks from the ID and the MS~\cite{MUON-2018-03}. They are required to satisfy $\pt > \qty{15}{\GeV}$ and $|\eta| < 2.5$.

To reject particles misidentified as prompt leptons\footnote{A prompt lepton is an electron or muon resulting from a direct decay of a $\W\inclvirt$ boson, a $\Z\inclvirt$ boson, or a massive $\gamma\virt$ boson. A non-prompt lepton originates from a photon conversion or hadron decay. Electrons and muons resulting from fully leptonic $\tau$-lepton decays are considered as prompt leptons.}, several identification requirements as well as isolation and impact parameter criteria~\cite{EGAM-2021-01, MUON-2018-03} are applied. For electrons, a likelihood-based identification method~\cite{EGAM-2018-01} is employed, which takes into account a number of discriminating variables such as the shape of the electromagnetic shower, track properties, the transition radiation response, and the quality of the cluster-to-track matching. Electron candidates in the \ZeroJet and \OneJet event classes must satisfy the `Tight' likelihood working point, which has prompt and non-prompt lepton efficiencies of $\sim$79\% and $\sim$0.46\%, respectively, for electrons with $20 < \pt < \qty{50}{\GeV}$, while electron candidates in the \TwoJet event class must satisfy the `Medium' likelihood working point, which has prompt and non-prompt lepton efficiencies of $\sim$87\% and $\sim$0.75\%, respectively, for electrons with $20 < \pt < \qty{50}{\GeV}$. The main reason for this separate treatment is to reduce further the contribution from misidentified leptons (see Section~\ref{sec:fakes}) in the \ZeroJet and \OneJet event classes, while in the \TwoJet event class, the better signal efficiency of the `Medium' likelihood working point is of higher importance to maintain sufficient data statistics. For muons, a cut-based identification method~\cite{MUON-2018-03} is employed, using the `Tight' working point with an efficiency of $\sim$95\% to maximise the sample purity. The impact parameter requirements are $|z_0\sin\theta| < \qty{0.5}{mm}$ and $|d_0|/\sigma_{d_0} < 5\ (3)$ for electrons (muons).\footnote{The transverse impact parameter, $d_0$, is defined by the point of closest approach of the track to the beamline in the $r$--$\phi$ plane with an estimated uncertainty $\sigma_{d_0}$, while the longitudinal impact parameter, $z_0$, is given by the longitudinal distance to the hard-scatter vertex from this same point.} Leptons are required to be isolated from other activity (e.g., tracks in the ID or energy deposits in the calorimeters) in the event. A multivariate method based on a boosted decision tree classifier (BDT) is employed for this purpose. This is an improved version of that described in Ref.~\cite{HIGG-2017-02} and leads to a substantial improvement --- more than a factor of 2 --- in the rejection of misidentified leptons compared to previously available methods in ATLAS while maintaining high efficiency for selecting prompt leptons. Lepton candidates must satisfy the `Tight' working point of this improved lepton isolation BDT.

At least one of the offline reconstructed leptons must be matched to an online object that triggered the recording of the event. In the case where an $e\mu$ trigger is solely responsible for the event being recorded, each lepton must correspond to one of the trigger objects. This trigger matching scheme also requires the \pt of the lepton to be at least \qty{1}{\GeV} above the trigger-level threshold, to ensure that the lepton lies in the trigger efficiency plateau.

Jets are reconstructed using the anti-$k_t$ algorithm with a radius parameter of $R = 0.4$ and particle-flow objects as input~\cite{Cacciari:2008gp, Fastjet, PERF-2015-09}. The four-momentum of the jets is corrected for the response of the non-compensating calorimeter, signal losses due to noise threshold effects, energy loss in passive material, and contamination from pile-up~\cite{JETM-2018-05}. For jets entering the analysis, a kinematic selection of $\pt > \qty{20}{\GeV}$ and $|\eta| < 4.5$ is applied. For the purpose of jet counting, only jets with $\pt > \qty{30}{\GeV}$ are considered. A selection on the jet vertex tagger, a multivariate discriminant, is applied to jets with $20 < \pt < \qty{60}{\GeV}$ and $|\eta| < 2.4$. This reduces contamination from pile-up~\cite{PERF-2014-03} by utilising calorimeter and tracking information to separate hard-scatter jets from pile-up jets~\cite{PERF-2014-03}. Furthermore, to suppress pile-up jets in the forward region, $|\eta| > 2.5$, jet shapes and topological correlations in pile-up interactions are exploited~\cite{ATL-PHYS-PUB-2019-026}.

Jets with $\pt > \qty{20}{\GeV}$ and $|\eta| < 2.5$ containing $b$-hadrons (\bjets) are identified using a neural network discriminant, DL1r, based on a number of lower-level taggers that utilise relevant quantities such as the associated track impact parameters and information from secondary vertices. The working point that is adopted has an average 85\% \bjet tagging efficiency and rejection factors of $\sim$3 and $\sim$50 for jets containing charm- and light-flavour hadrons, respectively, as estimated from simulated \ttbar events~\cite{FTAG-2019-07, FTAG-2018-01}.

Two leptons or a lepton and a jet may be close in $\eta$--$\phi$ space. The following procedure is adopted in the case of overlapping objects in order to avoid double-counting the same detector signal in two different reconstructed objects. If two electrons share an ID track, the electron with the lower transverse projection of the energy measured in the calorimeter is removed, reducing electron-track mis-assignment. If a muon shares an ID track with an electron, the electron is removed. For electrons and jets, the jet is removed if $\DR(\mathrm{jet}, e) < 0.2$ and the jet is not tagged as a \bjet, to remove electrons mis-reconstructed as jets. For any surviving jets, the electron is removed if $\DR(\mathrm{jet}, e) < 0.4$, which seeks to remove electrons from hadron decays that are already taken into account by the jet reconstruction algorithm. For muons and jets, the jet is removed if $\DR(\mathrm{jet}, \mu) < 0.2$, the jet has less than three associated tracks with $\pt > \qty{500}{\MeV}$, and the jet is not tagged as a \bjet; this seeks to remove mis-reconstructed jets that originate from muons with high energy bremsstrahlung photons. For any surviving jets with $\pt > \qty{20}{\GeV}$, the muon is removed if $\DR(\mathrm{jet}, \mu) < 0.4$, with the goal of removing muons from in-flight hadron decays.

The missing transverse momentum \vmet, with magnitude \met, is calculated as the negative vector sum of the \vpt of all the selected leptons and jets, together with the reconstructed tracks that are not associated with these objects but are consistent with originating from the primary vertex (i.e., the `soft term')~\cite{JETM-2020-03}. An object-based \met significance~\cite{JETM-2020-03}, \metsig, is used to reject events where the \met arises from the mis-reconstruction of the physics objects entering the calculation. The \met significance quantifies the likelihood that the reconstructed quantity corresponds to genuine missing transverse momentum.


\section{Event selection and classification}
\label{sec:channels}

Events are required to satisfy data quality and trigger criteria and contain exactly two opposite-sign leptons, identified using the selections described in the previous section. Both the DF and SF decay channels are considered. Events with additional leptons with $\pt > \qty{10}{\GeV}$ and satisfying `Medium' identification and `Loose' isolation criteria are vetoed to suppress backgrounds with additional leptons, such as \WZ production, and to guarantee statistical independence with a measurement of the \VH production mode~\cite{HIGG-2023-09}. Additionally, the higher-\pt (`leading') lepton is required to satisfy $\pt > \qty{22}{\GeV}$ and the lower-\pt (`subleading') lepton is required to satisfy $\pt > \qty{15}{\GeV}$. As the SF channel utilises only single-lepton triggers, as discussed in Section~\ref{sec:samples}, this effectively raises the leading lepton requirement to $\pt > \qty{27}{\GeV}$ for the data recorded in the years 2016--2018. For data recorded in 2015, the SF dimuon channel remains unaffected by the trigger \pt threshold, while for the SF dielectron channel, the leading electron requirement is effectively $\pt > \qty{25}{\GeV}$. Dilepton backgrounds from low-mass \Zjets production and meson decays are removed by requiring a dilepton invariant mass $\mll > 10$ and \qty{12}{\GeV} for the DF and SF channels, respectively. For the SF channel, additional selections on the \met significance, $\metsig > 4$, and the dilepton \pt, $\ptll > \qty{40}{\GeV}$, are used to further reject \Zjets production. The above criteria constitute the event preselection.

\begin{figure}[htbp]
\centering
\includegraphics[width=\textwidth]{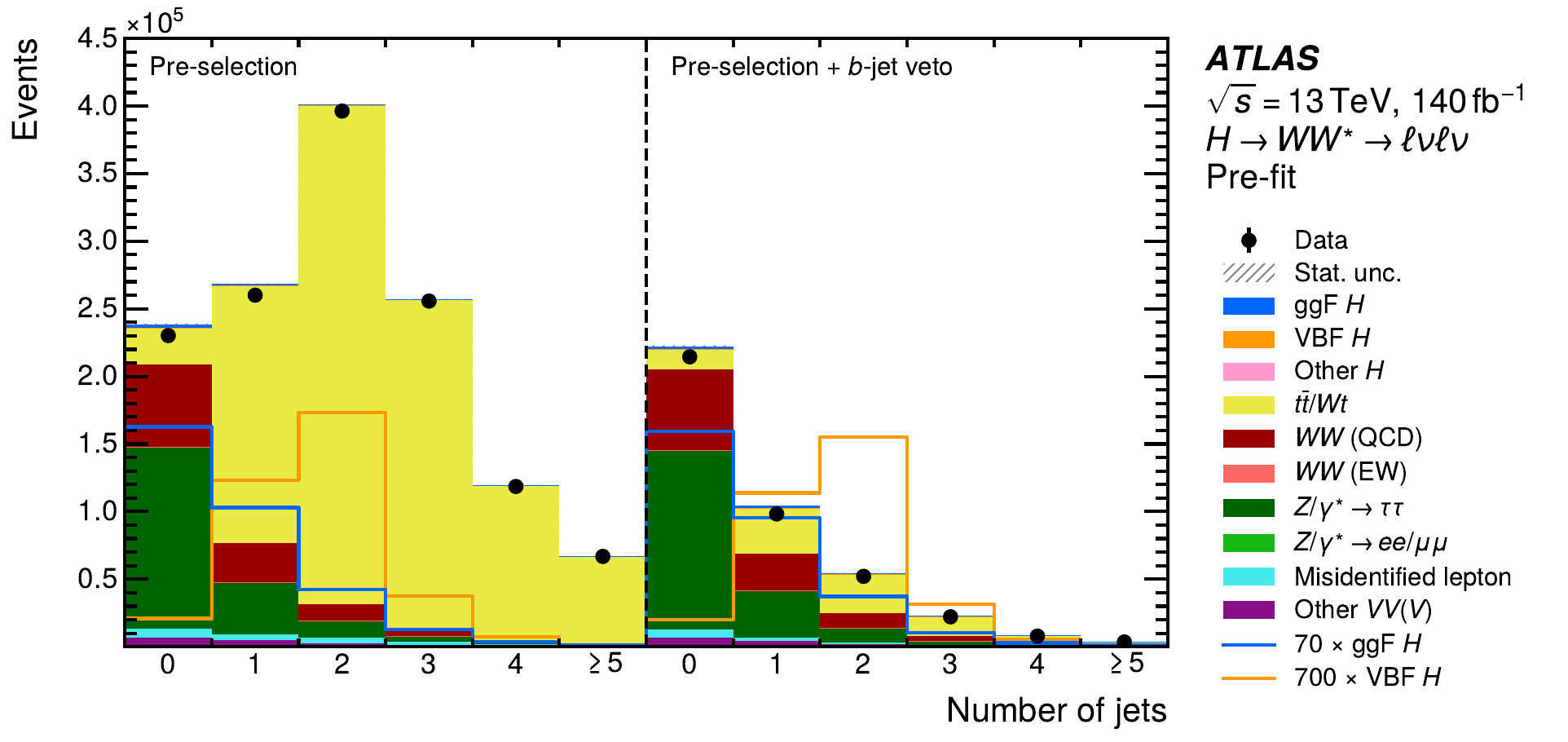}
\caption{Multiplicity distributions for jets with $\pt > \qty{30}{\GeV}$ and $|\eta| < 4.5$ in the DF channel, (left) after applying the preselection criteria and (right) after additionally vetoing events containing at least one \bjet using the 85\% efficiency working point for jets with $\pt > \qty{20}{\GeV}$ and $|\eta| < 2.5$. Shown are pre-fit distributions (see Section~\ref{sec:results} for its definition), assuming SM Higgs boson production. The very thin hatched band represents the small statistical uncertainty of the samples relative to their yields. The expected ggF (VBF) signal contribution is additionally overlaid and scaled up by a factor of 70 (700) for visibility.}
\label{fig:nJets}
\end{figure}

Following this preselection, events are divided into \ZeroJet, \OneJet, and \TwoJet event classes. This is mainly due to the very different background composition which varies as a function of the number of jets, as can be seen exemplified in the DF channel after preselection in Figure~\ref{fig:nJets} on the left. Only the DF channel is considered in the \ZeroJet and \OneJet event classes, while both the DF and SF channels are considered in the \TwoJet event class. To reject background from top quark production, events containing \bjets with $\pt > \qty{20}{\GeV}$ are vetoed in all event classes. As jet counting uses a jet $\pt$ threshold of $\qty{30}{\GeV}$, this implies specifically for the \ZeroJet event class that events are additionally rejected if at least one \bjet with $20 < \pt < \qty{30}{\GeV}$ is found. The background composition after applying the preselection criteria as a function of the number of jets can be seen in Figure~\ref{fig:nJets} on the right. The remaining selections used to define the SRs are described separately for the \ZeroJet, \OneJet, and \TwoJet event classes in Sections~\ref{sec:0j_cat}, \ref{sec:1j_cat}, and \ref{sec:2j_cat}, respectively, and are summarised in Table~\ref{tab:SR_selections}; the listed selections are motivated and defined in those sections. Sections~\ref{sec:stxs_cat} and \ref{sec:dphijj_cat} describe the STXS and STXS-like \dphijjsgn (referred to as `\STXSCP') classification schemes, that are used to further subdivide each jet multiplicity event class. Finally, Section~\ref{sec:control_regions} describes the CR selections used in each event class.

The leading and subleading leptons are denoted by $\ell_0$ and $\ell_1$, respectively. Similarly, the (\pt-ordered) leading and subleading jets are denoted by $j_0$ and $j_1$, respectively. Dilepton and dijet variables are always constructed using these leptons and jets.

\begin{table}[htbp]
\centering
\caption{Event selection criteria used to define the SRs, grouped according to their purpose. The SF channel is only defined in the \TwoJet phase space, and selections that are specific to the SF channel are denoted in parentheses. The definitions of the variables can be found in the main text. The mass of the leptonically decaying $\tau$-lepton pair, \mtt, is defined in Section~\ref{sec:1j_cat}, while central-jet-veto (CJV) and outside-lepton-veto (OLV) are defined in Section~\ref{sec:2j_cat}. `\VH orthogonality' corresponds to either $|\mjj - \qty{85}{\GeV}| > \qty{15}{\GeV}$, or $|\dyjj| > 1.2$, or both.}
\label{tab:SR_selections}
\resizebox{\textwidth}{!}{
\begin{tabularx}{1.30\textwidth}{c || Y | Y | Y | Y}
\toprule
& \multicolumn{4}{c}{Event class} \\
\multirow{2}{*}{Purpose} & \multirow{2}{*}{\ZeroJet} & \multirow{2}{*}{\OneJet} & \TwoJet & \TwoJet \\
& & & ggF-enriched & VBF-enriched \\
\midrule
\multirow{5}{*}{Preselection} & \multicolumn{4}{c}{Two isolated, different-flavour (same-flavour) leptons with opposite charge} \\
& \multicolumn{4}{c}{$\ptlead > \qty{22}{\GeV}$ and $\ptsublead > \qty{15}{\GeV}$} \\
& \multicolumn{4}{c}{$\mll > \qty{10}{\GeV}$ (\qty{12}{\GeV})} \\
& \multicolumn{2}{c|}{} & \multicolumn{2}{c}{($\metsig > 4$ and $\ptll > \qty{40}{\GeV}$)} \\
& \multicolumn{2}{c|}{Different-flavour leptons} & \multicolumn{2}{c}{Different- or same-flavour leptons} \\
\midrule
\multirow{5}{*}{Background suppression} & \multicolumn{4}{c}{$\nbjets\,[\pt > \qty{20}{\GeV}] = 0$} \\
& $\met > \qty{20}{\GeV}$ & \multicolumn{3}{c}{$\mtt < \mZ - \qty{25}{\GeV}$} \\
& \multicolumn{2}{c|}{$\dphill < \qty{2.0}{rad}$} & $\dphill < \qty{1.8}{rad}$ & -- \\
& \multicolumn{2}{c|}{$\mll < \qty{55}{\GeV}$} & $\mll < \qty{55}{\GeV}$ & $\mll < \qty{70}{\GeV}$ \\
& $\drll > 0.6$ & & & \\
\midrule
\multirow{2}{*}{Signal topology} & \multirow{2}{*}{$\dphillmiss > \qty{1.57}{rad}$} & \multirow{2}{*}{--} & \VH orthogonality & $\mjj > \qty{350}{\GeV}$ \\
& & & Fail CJV or Fail OLV & Pass CJV and Pass OLV \\
\bottomrule
\end{tabularx}
}
\end{table}

\subsection{Event class with zero jets}
\label{sec:0j_cat}

A requirement of $\met > \qty{20}{\GeV}$ ensures that the direction of \vmet in the transverse plane is well defined and suppresses backgrounds with low \met. Continuum \WW production and resonant Higgs boson production can be separated by exploiting the spin-0 property of the Higgs boson, which, when combined with the chiral structure of the \W boson decay, leads to a small opening angle, \dphill, between the charged leptons. A requirement on the dilepton invariant mass of $\mll < \qty{55}{\GeV}$, which combines the small lepton opening angle with the kinematics of a low-mass Higgs boson ($\mH \approx \qty{125}{\GeV}$), significantly reduces the \WW and top quark backgrounds as well as several other small background processes. A requirement of $\dphill < \qty{2.0}{rad}$ reduces the remaining \Zjets process while retaining most of the signal. A selection on the angular separation of the leading leptons, $\drll > 0.6$, is used to remove a region of phase space for which misidentified leptons are poorly modelled; doing so has a negligible impact on the expected sensitivity. Events with a significant mis-measurement of the missing transverse momentum are suppressed by requiring \vmet to point away from the dilepton transverse momentum, $\dphillmiss > \qty{1.57}{rad}$, where \dphillmiss is defined as the azimuthal angular difference between the dilepton system and the missing transverse momentum. This requirement also reduces contributions from \Ztt decays, where \vmet generally points in the same transverse direction as the dilepton system. The combination of these selection criteria defines the inclusive \ZeroJet SR, where the ggF production mode is the dominant signal with an expected ggF signal purity of about 11\%. The purity is defined as the number of expected signal events divided by the number of expected events from all sources.

The \ZeroJet event class targets exclusively ggF production, as VBF production without reconstructed jets in the final state cannot be experimentally distinguished from ggF production; a DNN discriminant is therefore trained to distinguish between ggF events and background events. It is trained in the inclusive \ZeroJet SR with simulated ggF signal events and simulated events of all main background processes, with the exception of events containing misidentified leptons that are modelled by a data-driven approach (see Section~\ref{sec:fakes}). As input, the DNN receives the \pt, $\eta$, and $\phi$ of the two leading leptons as well as the magnitude and azimuthal angle of the missing transverse momentum. Rotational and mirror symmetries\footnote{Rotational symmetry is imposed by setting the leading lepton $\phi$ to 0 and rotating all other leptons, jets, and \vmet accordingly. When the leading lepton $\eta$ is negative, the sign of $\eta$ of all objects is inverted, exploiting this mirror symmetry. When the subleading lepton $\phi$ is negative, its $\phi$ value is inverted to be positive, as well as mirroring the $\phi$ value of all other jets and \vmet accordingly, exploiting this last symmetry.} of the detector are incorporated directly into the DNN training, leading to seven degrees of freedom as input. By doing so, the DNN does not need to learn these symmetries and may exploit the training data more efficiently. The DNN training is performed on a part of the sample, the training sample, while for optimising the DNN's architecture and hyperparameters, an independent validation sample is used. The hyperparameters of the DNN are optimised using random searches and the \textsc{Hyperband} algorithm~\cite{Li:2016kel}, resulting in approximately 27,000 trainable parameters. DNN architectures with fewer free parameters than the default one perform systematically worse. For each task, four independent DNNs are trained and later on used. To maximise the available sample size for training and thus improving the performance for each DNN, the simulated event samples are divided in four equal-size folds. Four DNNs are trained, where each DNN uses two folds as training samples, one fold as a validation sample, and one fold as a testing sample. The four DNNs each use a different fold for validation and testing that are non-overlapping. This technique is referred to as `four-fold cross-validation'. The subsequent analysis uses the inference on the testing sample that is never used for training or validation.

\subsection{Event class with one jet}
\label{sec:1j_cat}

Compared to the \ZeroJet event class, the suppression of the \Ztt background in the \OneJet event class is more challenging. Using the direction and magnitude of the measured missing transverse momentum and projecting it along the directions defined by the two reconstructed charged leptons, the mass of the leptonically decaying $\tau$-lepton pair, \mtt, can be reconstructed using the collinear approximation~\cite{Plehn:1999xi}. A requirement of $\mtt < \mZ - \qty{25}{\GeV}$ significantly reduces the remaining \Zjets contribution.\footnote{For this selection, \mZ is set to \qty{91.1876}{\GeV}~\cite{ParticleDataGroup:2024cfk}.} The same \dphill and \mll selections as described in Section~\ref{sec:0j_cat} are also applied in the \OneJet event class. The combination of these selection criteria defines the inclusive \OneJet SR, where the ggF production mode is the dominant signal with an expected purity of about 8.3\%, while the expected VBF purity is only about 1.0\%.

Both ggF and VBF signal events can be reconstructed with exactly one jet in addition to the $\lnu\lnu$ system. This can occur, for instance, from a QCD initial-state radiation resulting in an extra jet for ggF or when one of the two resulting jets is outside the kinematic acceptance for VBF. The \OneJet event class therefore seeks to independently measure the ggF and VBF 1-jet contributions. The DNN discriminant in the \OneJet event class is therefore trained to separately distinguish between ggF events, VBF events, and background events, where all such events are included in the training with the exception of the data-driven misidentified lepton background (see Section~\ref{sec:fakes}). The events from the three input classes (ggF, VBF, and background) are not normalised to the same integral, as this would overemphasise the importance of the VBF production process. Instead, the sum of weights for the ggF and VBF events, each first individually normalised to their expected production cross-section, is normalised to the same sum of weights as the total background. Otherwise, the same procedure for the optimisation and training of the DNN as in Section~\ref{sec:0j_cat} is employed, resulting in approximately 105,000 trainable parameters. The DNN receives the same inputs as the DNN used in the \ZeroJet event class, supplemented by the \pt, $\eta$, and $\phi$ of the single jet. The same symmetry transformations are applied to those inputs, leading to ten degrees of freedom.

\subsection{Event class with two or more jets}
\label{sec:2j_cat}

The \TwoJet event class targets both ggF and VBF production modes and is the only event class that includes events from both the DF and SF channels. In the DF channel, top quark production is the dominant background, while in the SF channel, both top quark and \Zjets production are large and important backgrounds. In addition to the preselection criteria summarised in Table~\ref{tab:SR_selections}, this event class imposes the same $\mtt < \mZ - \qty{25}{\GeV}$ selection for reducing the \Zjets contribution as described in Section~\ref{sec:1j_cat}.

The VBF process exhibits a $t$-channel exchange of heavy weak bosons and an absence of colour flow between the two incoming partons at leading order. This gives it a distinctive final-state topology, with two forward jets separated by a large rapidity gap without hadronic activity. This topology difference relative to the ggF production mode is exploited to render the VBF and ggF SRs orthogonal to one another. VBF SR events are required to satisfy a central-jet-veto (CJV) that rejects events with jets with $\pt > \qty{30}{\GeV}$ that lie in the pseudorapidity range between the two tagging jets, i.e.\ the leading and subleading jets. Events must also satisfy an outside-lepton-veto (OLV), which requires that the two charged leptons lie within the rapidity interval defined by the two tagging jets. The dilepton invariant mass \mll is required to be less than \qty{70}{\GeV} to ensure orthogonality with the ATLAS measurement of off-shell Higgs boson production in the \WW decay channel~\cite{HIGP-2024-05}. This selection on \mll removes about 1.5\% (0.7\%) of the VBF signal events and about 57\% (59\%) of the top quark background events in the DF (SF) channel. Finally, the dijet invariant mass, \mjj, is required to be above \qty{350}{\GeV} to align with the STXS categorisation, which also ensures orthogonality with a measurement of the \VH production mode~\cite{HIGG-2023-09}. After applying these VBF SR selection criteria, resulting in the inclusive VBF \TwoJet SR (and also referred to as the `\TwoJet VBF-enriched event class'), the VBF production mode is the largest signal component with an expected purity of about 7.8\% (3.2\%) in the DF (SF) channel. The purity of the ggF production mode is 4.0\% (1.5\%) in the DF (SF) channel.

The SRs targeting the ggF production mode also exploit the characteristic spin correlation of the signal by requiring $\dphill < \qty{1.8}{rad}$ and $\mll < \qty{55}{\GeV}$. The orthogonality to the VBF SRs in the \TwoJet event class is ensured by requiring events to fail to satisfy the CJV, or the OLV, or both. The orthogonality with the \VH production mode is ensured by requiring $|\mjj - \qty{85}{\GeV}| > \qty{15}{\GeV}$, or $|\dyjj| > 1.2$, or both. After applying these ggF signal selection criteria, resulting in the inclusive ggF \TwoJet SR  (and also referred to as the `\TwoJet ggF-enriched event class'), the ggF production mode is the dominant signal with an expected purity of about 4.1\% (1.4\%) in the DF (SF) channel. The purity of the VBF production mode is 0.6\% (0.2\%) in the DF (SF) channel.

Following these selections, the DNNs are trained separately for the DF and SF channels to account for their very different background compositions; the \Zjets contribution in particular is much larger in the SF channel. The training of these two DNNs is performed using events satisfying the \TwoJet preselection plus the veto on events with \bjets (see Table~\ref{tab:SR_selections}). All simulated signal and background events are included in the training, except for the data-driven misidentified lepton background (see Section~\ref{sec:fakes}). The DNNs are structured to have three output nodes: one for ggF signal events, one for VBF signal events, and one background output node. As input, the DNNs receive 20~variables, which can be grouped into those that target the VBF topology, those that target the \hwwlvlv decay, and those which suppress top quark production. The VBF final-state topology results in a large dijet invariant mass, \mjj, and a large dijet rapidity separation, $|\dyjj|$. Furthermore, the Higgs boson decay products tend to be in between these two tagging jets in rapidity space, with little additional hadronic activity. These latter attributes are exploited via lepton-jet invariant masses, $m_{\ell_xj_y}$ ($x, y = 0, 1$), angular separations, $\Delta{R}_{\ell_xj_y}$ ($x, y = 0, 1$), and the lepton $\eta$-centralities, $\centlep{x} = |2\eta_{\ell_x} - \sum\eta_j| / \Delta\eta_{jj}$ ($x = 0, 1$), of the leading and subleading leptons~\cite{Barger:1994zq}. Also included is the dilepton transverse mass:
\begin{equation*}
\mt = \sqrt{(E_{\ell\ell} + \met)^2 - |\vptll + \vmet|^2} \,,
\label{eqn:mT}
\end{equation*}
where $E_{\ell\ell}$ is the energy of the dilepton system in the laboratory frame. These variables are summarised in Table~\ref{tab:2j_DNN_inputs}.

\begin{table}[htbp]
\centering
\caption{Input variables to the DNNs used in the \TwoJet event class, grouped according to their purpose.}
\label{tab:2j_DNN_inputs}
\resizebox{\textwidth}{!}{
\begin{tabular}{l | l}
\toprule
Purpose & Variable(s) \\
\midrule
\multirow{6}{*}{VBF topology} & $\eta$-centralities of the leading and subleading leptons, \centlep{0} and \centlep{1} \\
& Transverse momenta of the leading, subleading, and (if present; otherwise \qty{0}{\GeV}) third jets \\
& Dijet invariant mass, \mjj \\
& Dijet rapidity separation, $|\dyjj|$ \\
& Lepton-jet invariant mass combinations,  $m_{\ell_0j_0}$, $m_{\ell_0j_1}$, $m_{\ell_1j_0}$, and $m_{\ell_1j_1}$\\
& Lepton-jet angular separation combinations, $\Delta{R}_{\ell_0j_0}$, $\Delta{R}_{\ell_0j_1}$, $\Delta{R}_{\ell_1j_0}$, and $\Delta{R}_{\ell_1j_1}$ \\
\midrule
\multirow{3}{*}{\hwwlvlv decay} & Dilepton invariant mass, \mll \\
& Dilepton aziumuthal separation, \dphill \\
& Transverse mass, \mt \\
\midrule
\multirow{2}{*}{Top quark suppression} & Magnitude of the vectorial sum of the \pt of all leptons, jets, and \met, \pTtot \\
& Object-based \met significance, \metsig \\
\bottomrule
\end{tabular}
}
\end{table}

\subsection{Event classification for the STXS measurement}
\label{sec:stxs_cat}

\begin{figure}[htbp]
\centering
\includegraphics[width=\textwidth]{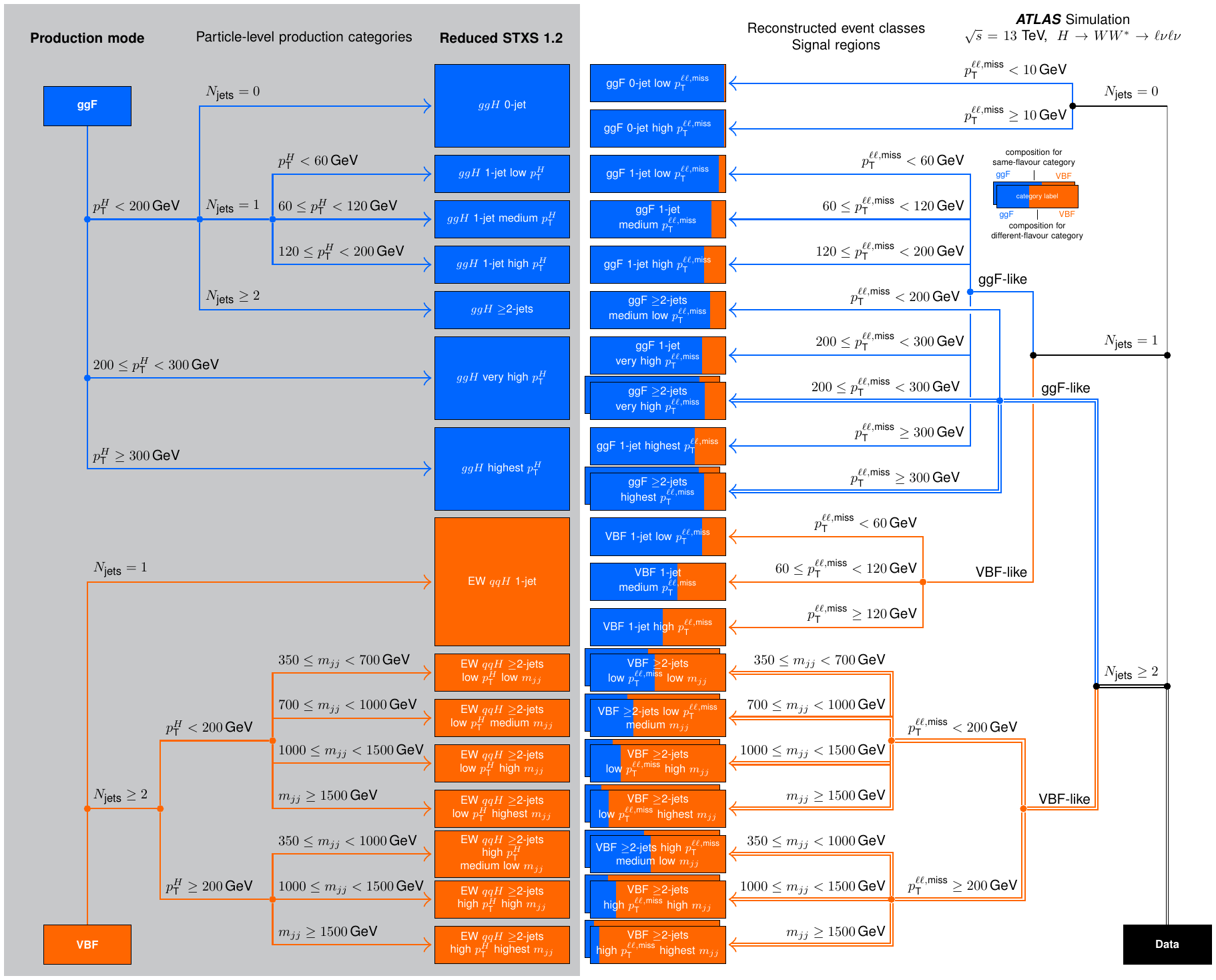}
\caption{Two sets (`Production mode' and `Reduced STXS~1.2') of exclusive phase-space regions (`production categories') defined at particle-level for (left panel, shaded) the measurement of the Higgs boson production cross-sections and (right panel, blank) the corresponding reconstructed SRs. The description of the production categories and reconstructed SRs is given in Section~\ref{sec:channels}. All Reduced STXS~1.2 production categories include a requirement on the Higgs boson rapidity of $|y_H| < 2.5$. Single lines indicate SRs that exist in only the DF channel; double lines indicate SRs that exist in both the DF and SF channels. The colours of each SR box indicate the SM expected relative fractions of ggF (blue, left part of each SR box) and VBF (orange, right part of each SR box). In cases where a SF SR is present, its box is shown beneath the corresponding DF SR box.}
\label{fig:STXS_cat}
\end{figure}

\begin{figure}[htbp]
\centering
\includegraphics[width=\textwidth]{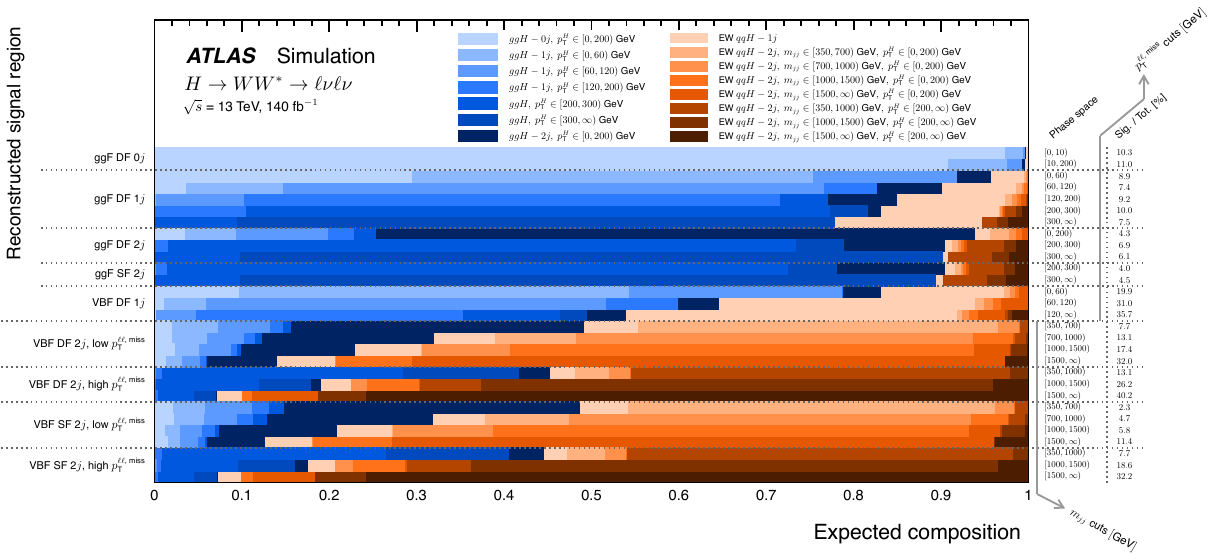}
\caption{Relative SM signal composition in terms of the measured STXS production categories for each reconstructed SR. The phase space selections based on \pTllmiss and \mjj summarised on the right are described in Section~\ref{sec:stxs_cat}. The signal purity (`Sig. / Tot.') is defined as the sum of the yields for the measured STXS production categories, i.e.\ all those shown in the legend, divided by the sum of the yields for all physics processes. For each SR, the \ggH contributions start on the left, going from lighter to darker shades, followed by the EW \qqH, again going from lighter to darker shades.}
\label{fig:STXS_comp}
\end{figure}

In the STXS measurement, the reconstruction-level phase space is designed to reflect the particle-level phase space per production mode to which the measured results are unfolded. The categories defined in the Stage~1.2 STXS categorisation scheme are merged or `reduced' into a coarser set of categories that can be measured with sufficient precision. This coarser set of categories is referred to as the Reduced Stage~1.2 categorisation scheme and consists of 15~\ggH and EW \qqH categories that are differential in particle-level \Njets, \pTH, and \mjj. Selections on the reconstructed \pTH and \mjj are used to subdivide the inclusive \Njets SRs, described in Sections~\ref{sec:0j_cat}, \ref{sec:1j_cat}, and \ref{sec:2j_cat} and summarised in Table~\ref{tab:SR_selections}, into regions that target one of the categories in the reduced scheme. As a proxy for \pTH, the norm of the vector sum of the \pt of the dilepton and missing transverse momentum system, \pTllmiss, is used. Both the reduced STXS scheme and the reconstructed SRs are illustrated in Figure~\ref{fig:STXS_cat}, while Figure~\ref{fig:STXS_comp} shows the relative contributions of each category in the reduced scheme in a given reconstructed SR.

The inclusive \ZeroJet SR is further split into two regions with a boundary of $\pTllmiss = \qty{10}{\GeV}$, resulting in the ggF 0-jet low \pTllmiss SR and the ggF 0-jet high \pTllmiss SR. The DNN described in Section~\ref{sec:0j_cat} is used as the final discriminating variable, applied separately in these two SRs. However, this analysis is not able to reliably measure separately the cross-section in the \ZeroJet STXS categories with $\pTH < \qty{10}{\GeV}$ and $\geq \qty{10}{\GeV}$, which are better probed by other decay channels such as $\H\to\gamma\gamma$~\cite{HIGG-2021-23}, due to the rather large migration of $\gtrapprox 30\%$ between particle-level and reconstruction-level signal events of these two categories. This is mainly due to the rather poor \met resolution. Nonetheless, this reconstruction-level split is still useful when combining the results of this analysis with other analyses, as done previously, e.g.\ in Ref.~\cite{HIGG-2021-23}. Thus, both SRs jointly target the \ggH 0-jet STXS category.

The inclusive \OneJet SR predominantly targets the ggF production mode while also containing sensitivity to the VBF production mode, as discussed in Section~\ref{sec:1j_cat}. This inclusive \OneJet SR is subdivided into eight SRs, based on the reconstructed transverse momentum of the Higgs boson candidate, \pTllmiss. Three of the SRs are constructed to target the VBF production mode. These SRs cover $\pTllmiss < \qty{60}{\GeV}$, $60 \leq \pTllmiss < \qty{120}{\GeV}$, and $\pTllmiss \geq \qty{120}{\GeV}$ and are respectively called the VBF 1-jet low \pTllmiss SR, VBF 1-jet medium \pTllmiss SR, and VBF 1-jet high \pTllmiss SR. All three SRs jointly target the EW \qqH 1-jet STXS category. Splitting the VBF \OneJet SR into these three SRs based on \pTllmiss facilitates a consistent treatment of the background, which is described in Section~\ref{sec:control_regions}, with the corresponding ggF SRs described in the next paragraph. Events are classified into these VBF-enriched regions using the VBF output node of the DNN described in Section~\ref{sec:1j_cat}. For each \pTllmiss region, the most VBF-like events --- defined as the 30\% of VBF signal events with the highest purity --- are assigned to the VBF-enriched SR, while the remaining, more ggF-like events are directed to the ggF-enriched SR.\footnote{Here, `VBF-like events' refers to all events that the 1-jet DNN VBF output node classifies as being more likely VBF events, while `ggF-like events' refers to events that are classified by the same DNN output to be more likely ggF events.} A looser cut on the VBF score for the classification would increase the yields in the VBF SRs and improve the sensitivity to VBF at a cost of sensitivity to ggF. In practice, the improvement for VBF would be small due to the small VBF purity of the data that would move into the VBF SRs; the chosen classification provides good performance for both ggF and VBF.

The ggF-enriched \OneJet SRs are split kinematically into five SRs. Three of them cover $\pTllmiss < \qty{60}{\GeV}$, $60 \leq \pTllmiss < \qty{120}{\GeV}$, and $120 \leq \pTllmiss < \qty{200}{\GeV}$ and are respectively called the ggF 1-jet low \pTllmiss SR, ggF 1-jet medium \pTllmiss SR, and ggF 1-jet high \pTllmiss SR. They target the \ggH 1-jet STXS categories with the corresponding \pTH splits. The remaining two ggF-like \OneJet SRs are defined via $200 \leq \pTllmiss < \qty{300}{\GeV}$, called the ggF 1-jet very high \pTllmiss SR, and via $\pTllmiss \geq \qty{300}{\GeV}$, called the ggF 1-jet highest \pTllmiss SR. These target the jet-inclusive \ggH very high \pTH and \ggH highest \pTH STXS categories, respectively, which are also targeted by ggF $\geq$2-jets SRs discussed in the following paragraph.

The inclusive ggF \TwoJet SR targets the ggF production mode. It is further split into $\pTllmiss < \qty{200}{\GeV}$, $200 \leq \pTllmiss < \qty{300}{\GeV}$,  and $\pTllmiss \geq \qty{300}{\GeV}$ resulting in the ggF $\geq$2-jets medium low \pTllmiss SR, ggF $\geq$2-jets very high \pTllmiss SR, and ggF $\geq$2-jets highest \pTllmiss SR, respectively. The latter two SRs exist in both DF and SF channels, while only the DF channel is considered for the first, as the SF channel would have negligible sensitivity in this region due to a very low signal-to-background ratio. These SRs target the \ggH $\geq$2-jets, \ggH very high \pTH, and \ggH highest \pTH STXS categories, respectively.

The inclusive VBF \TwoJet SR targets the VBF production mode. It is split into seven SRs for both the DF and SF channels using identical kinematic selections and resulting in a total of 14~VBF $\geq$2-jets SRs. Events with $\pTllmiss < \qty{200}{\GeV}$ are further divided into:
\begin{itemize}
\item $350 \leq \mjj < \qty{700}{\GeV}$, resulting in the VBF $\geq$2-jets low \pTllmiss low \mjj SR, targeting the EW \qqH $\geq$2-jets low \pTH low \mjj STXS category;
\item $700 \leq \mjj < \qty{1000}{\GeV}$, resulting in the VBF $\geq$2-jets low \pTllmiss medium \mjj SR, targeting the EW \qqH $\geq$2-jets low \pTH medium \mjj STXS category;
\item $1000 \leq \mjj < \qty{1500}{\GeV}$, resulting in the VBF $\geq$2-jets low \pTllmiss high \mjj SR, targeting the EW \qqH $\geq$2-jets low \pTH high \mjj STXS category; and
\item $\mjj \geq \qty{1500}{\GeV}$, resulting in the VBF $\geq$2-jets low \pTllmiss highest \mjj SR, targeting the EW \qqH $\geq$2-jets low \pTH highest \mjj STXS category.
\end{itemize}
Events with $\pTllmiss \geq \qty{200}{\GeV}$ are further divided into:
\begin{itemize}
\item $350 \leq \mjj < \qty{1000}{\GeV}$, resulting in the VBF $\geq$2-jets high \pTllmiss medium low \mjj SR, targeting the EW \qqH $\geq$2-jets high \pTH medium low \mjj STXS category;
\item $1000 \leq \mjj < \qty{1500}{\GeV}$, resulting in the VBF $\geq$2-jets high \pTllmiss high \mjj SR, targeting the EW \qqH $\geq$2-jets high \pTH high \mjj STXS category; and
\item $\mjj \geq \qty{1500}{\GeV}$, resulting in the VBF $\geq$2-jets high \pTllmiss highest \mjj SR, targeting the EW \qqH $\geq$2-jets high \pTH highest \mjj STXS category.
\end{itemize}

\subsection{Event classification for the CP-sensitive measurement}
\label{sec:dphijj_cat}

\begin{figure}[htbp]
\centering
\includegraphics[width=\textwidth]{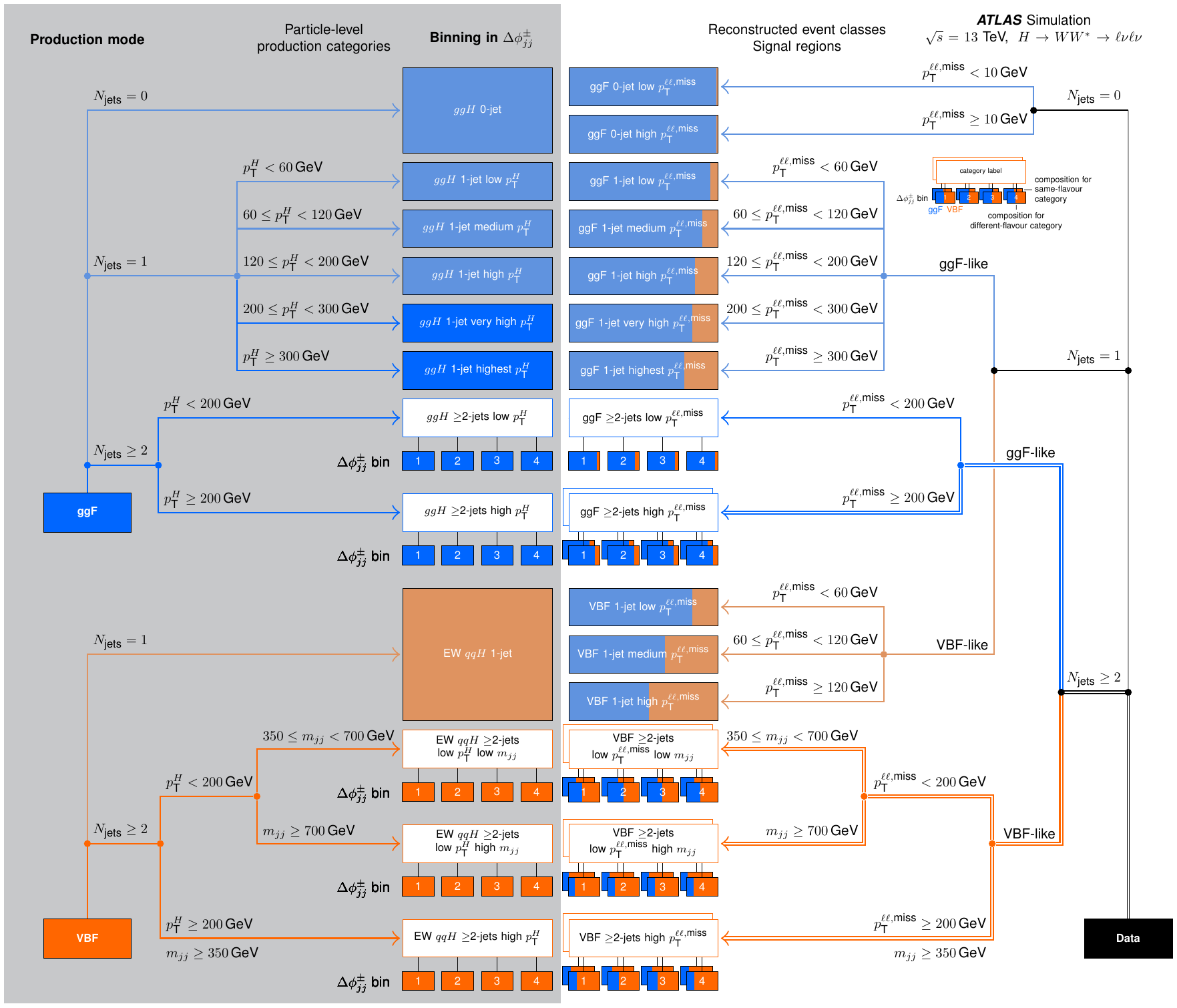}
\caption{\STXSCP categorisation scheme, modified from the Stage~1.2 STXS categorisation scheme shown in Figure~\ref{fig:STXS_cat}, showing (left panel, shaded) the particle-level production categories and (right panel, blank) the corresponding reconstructed SRs. The description of the production categories and reconstructed SRs is given in Section~\ref{sec:channels}. All production categories in the \STXSCP categorisation scheme include a requirement on the Higgs boson rapidity of $|y_H| < 2.5$. The \ggH 0/1-jet $200 \leq \pTH < \qty{300}{\GeV}$ and $\pTH \geq \qty{300}{\GeV}$ categories branch off of the ggF 1-jet line to avoid complicating the figure. Single lines indicate SRs that exist in only the DF channel; double lines indicate SRs that exist in both the DF and SF channels. The colours of each SR box indicate the SM expected relative fractions of ggF (blue, left part of each SR box) and VBF (orange, right part of each SR box). In cases where a SF SR is present, its box is shown beneath the corresponding DF SR box. The pale blue and orange boxes indicate production categories and SRs that are unchanged from the Stage~1.2 STXS categorisation scheme.}
\label{fig:STXS_dphijj}
\end{figure}

The STXS measurement described in the previous section is of high importance and can easily be combined with similar measurements using other Higgs boson decay channels. However, the Stage~1.2 STXS categorisation scheme has not been designed to be sensitive to the CP structure of potential effective Higgs boson couplings. To directly target the CP coupling structure, this analysis introduces an alternative \STXSCP categorisation scheme, which is shown diagrammatically in Figure~\ref{fig:STXS_dphijj}. This alternative categorisation affects the STXS categories targeting the production of a Higgs boson in association with two or more jets; the STXS categories with zero or one jets are unchanged relative to STXS Stage~1.2. The CP-sensitive variable used for the categorisation is the signed azimuthal angular difference between the two leading jets, \DPhijjsgn, defined as the azimuthal angular difference between the forward (positive $\eta$) and backward jet, mapped onto the range $[-\pi, +\pi)\,\si{rad}$~\cite{Plehn:2001nj, Klamke:2007cu}. Differential measurements in this variable were used previously to measure the CP nature of the Higgs boson couplings, via reconstruction-level information using a partial Run~2 data sample~\cite{HIGG-2017-13} and via profile likelihood unfolding using the Run~2 data sample in the ggF and VBF phase space~\cite{HIGG-2019-13, HIGG-2020-25, HIGG-2022-07}.

The particle-level categorisation used for this CP-sensitive measurement (or `\STXSCP measurement') is as follows. The EW \qqH $\geq$2-jets phase space is split at $\pTH = \qty{200}{\GeV}$, as the shape of the SM in the \DPhijjsgn distribution changes between these two phase space regions, shown in Ref.~\cite{Amoroso:2020lgh}. A split at $\mjj = \qty{700}{\GeV}$ in the low \pTH regions is introduced to match the background normalisation scheme for the STXS measurement (see Section~\ref{sec:control_regions}), which ensures that background processes are accurately measured. Each of the resulting three EW \qqH $\geq$2-jets categories is split into four new CP-sensitive \STXSCP bins using \DPhijjsgn: $[-\pi, -\pi/2)$, $[-\pi/2, 0)$, $[0, +\pi/2)$, and $[+\pi/2, +\pi)\,\si{rad}$. This binning is chosen for consistency with other measurements probing similar effects using \DPhijjsgn, e.g.\ Refs.~\cite{HIGG-2020-25, CMS-HIG-22-008, HIGG-2019-13, HIGG-2022-07}.

In the STXS Stage~1.2 categorisation scheme for \ggH, the $\pTH \geq \qty{200}{\GeV}$ category is inclusive in the number of jets. The \STXSCP categorisation scheme modifies this approach by additionally splitting the $\pTH \geq \qty{200}{\GeV}$ category into $\Njets < 2$ and \TwoJet subcategories.\footnote{No \ggH 0-jet event considered by this measurement is expected to have $\pTH > \qty{200}{\GeV}$ at particle-level. Therefore, the treatment of 0-jet events is in practice equivalent in Figures~\ref{fig:STXS_cat} and \ref{fig:STXS_dphijj}.} Following this, the \ggH $\geq$2-jets phase space is split at $\pTH = \qty{200}{\GeV}$ to match the background normalisation scheme of the STXS measurement. The two \pTH regions are split again into the same four \DPhijjsgn bins as for EW \qqH. Splitting the \ggH phase space in a similar way to the EW \qqH phase space improves the ability to measure contamination from the former in SRs targeting the latter. However, the anomalous couplings modifying EW \qqH production are not significantly affected by those modifying \ggH production, shown in Section~\ref{sec:cp_interp}.

To improve sensitivity by reducing the correlations between POIs, reconstruction-level SRs are defined to closely resemble the \STXSCP categorisation to minimise migrations. Therefore, the final SRs are constructed, after splitting in \pTH and \mjj, by splitting into four equidistant \DPhijjsgn bins, matching the particle-level \STXSCP categories and \DPhijjsgn bins. Furthermore, the \ggH high \pTH categories with $\Njets < 2$ and \TwoJet have corresponding SRs that are split in the same way using the reconstructed number of jets. As in the STXS measurement, the same DF and SF channels are used. These newly constructed SRs are shown alongside their matching \STXSCP categories in Figure~\ref{fig:STXS_dphijj}.

\subsection{Control regions}
\label{sec:control_regions}

For each event class introduced in the previous sections, corresponding CRs are defined that are enriched in the dominant background processes, including \qqWW (labelled as \WW), top quark, \Ztt, and \Zeemm production. These are used to normalise the rates for these processes from data via normalisation factors. The use of CRs additionally leads to a partial cancellation of systematic uncertainties that affect SRs and their corresponding CRs in similar ways for the normalised background processes, improving the measurement precision.

The CRs are defined using inverted SR selections such that all CRs and SRs are non-overlapping, i.e.\ statistically independent. Each CR applies selections on \pTllmiss and \mjj. Typically the same selections as used to define an SR are applied, but in some cases coarser selections corresponding to the \pTllmiss and \mjj range of multiple SRs are chosen. This is done when the normalisation factors are observed to be consistent within statistical uncertainties between CRs defined using finer splitting in \pTllmiss and \mjj, mitigating the potential for statistical fluctuations in simulation and data. Any SRs occupying a subset of the phase space defined by the \pTllmiss or \mjj selections of a CR are coherently normalised via that same CR using a common normalisation factor. The \WW, top quark, and \Ztt CRs defined in the DF channel of the \TwoJet event class are used to normalise these background processes simultaneously in the DF and SF channels, whereas the normalisation factors obtained using the \Zeemm CRs are only applied to the SF channel. In the statistical analysis, each CR consists of only a single bin; however, all SR DNNs are validated by comparing their predicted shape and normalisation with those in data in their respective CRs, and no significant mismodelling is observed. The selections used to define the CRs are summarised in Table~\ref{tab:CR_selections}.

For some of the SRs in the \TwoJet event class, the normalisation of a particular background process is obtained from a fit using exclusively the SR and not an accompanying CR. This is done where it is difficult to define a CR sufficiently close to the SR in phase space and where the SR has sufficient statistics to estimate the normalisation of that background. In particular, the background-rich regions of that SR are what primarily determine the normalisation.

The \ZeroJet, \OneJet, and \TwoJet ggF-enriched event classes utilise \WW CRs that are made statistically independent of their respective SRs using inverted selections on the dilepton invariant mass, \mll, following the DF channel preselection. In the \TwoJet ggF-enriched event class, the \mTTwo variable~\cite{Lester:1999tx} is additionally used to separate diboson and top quark production and is defined as:
\begin{equation*}
\mTTwo = \sqrt{\min_{\substack{x, y = 0, 1 (x \neq y) \\ z = 0, 1}}\left(\min_{\ptslash^1 + \ptslash^2 = \ptslash}\left[\max\left\{\mt^2(\pt^{\ell_x+j_z}, \ptslash^1), \mt^2(\pt^{\ell_y}, \ptslash^2)\right\}\right]\right)} \,,
\label{eqn:mT2}
\end{equation*}
where the minimisation is over all possible two-component transverse-momentum vectors, $\ptslash^i$ ($i = 1, 2$), such that their sum gives the observed missing transverse momentum, \ptslash, and where $\pt^{\ell_x+j_z}$ is the combined transverse momentum of a charged lepton and a jet and $\pt^{\ell_y}$ is the transverse momentum of the other charged lepton. All combinations of the leading and subleading leptons and jets are considered in the calculation. The \mTTwo variable provides a lower bound on the mass of a parent particle that decays into visible and invisible products for the case where there are two such parent particles in the event; this is the case for $\qqWW\to\lnu\lnu$. Additionally, the inclusion of the transverse momentum of a single jet in the visible decay products improves the ability of this variable to separate \qqWW and top quark production.

All event classes utilise top quark CRs that are made statistically independent of their corresponding SRs by inverting the \bjet veto into a \bjet requirement. The \Njets requirement is retained, in accordance with the DF channel preselection. In the \ZeroJet event class, only sub-threshold jets, $20 < \pt < \qty{30}{\GeV}$, are considered for the \bjet requirement, which ensures that the phase space of the CR remains close to that of the SR.

The \OneJet and the \TwoJet ggF- and VBF-enriched event classes utilise \Ztt CRs that are made statistically independent of their respective SRs using inverted selections on the mass of the $\tau$-lepton pair, \mtt, following the DF channel preselection. No \Ztt CR is defined for the \ZeroJet SRs, as this background constitutes only about 1\% of the total background in these SRs.

Finally, the \TwoJet ggF- and VBF-enriched event classes utilise \Zeemm CRs that are made statistically independent of their respective SRs by requiring the dilepton invariant mass, \mll, to be within \qty{15}{\GeV} of the \Z pole following the SF channel preselection.

In the \ZeroJet event class, the \WW and top quark CRs are each split into two regions defined by $\pTllmiss < \qty{10}{\GeV}$ and $\pTllmiss \geq \qty{10}{\GeV}$. In the \OneJet event class, the \WW, top quark, and \Ztt CRs are each split into three regions defined by $\pTllmiss < \qty{60}{\GeV}$, $60 \leq \pTllmiss < \qty{120}{\GeV}$, and $\pTllmiss \geq \qty{120}{\GeV}$.

In the \TwoJet ggF-enriched event class, the \WW and \Ztt CRs are each split into two regions defined by $\pTllmiss < \qty{200}{\GeV}$ and $\pTllmiss \geq \qty{200}{\GeV}$. The top quark CR requires $\pTllmiss \geq \qty{200}{\GeV}$, as including an additional CR for the $\pTllmiss < \qty{200}{\GeV}$ phase space does not improve the expected measurement precision. Instead, the normalisation of the top quark background in this region is obtained from a fit to data in the corresponding SR. The \Zeemm CR also requires $\pTllmiss \geq \qty{200}{\GeV}$, as the $\pTllmiss < \qty{200}{\GeV}$ phase space is not considered in the SF channel due to negligible expected sensitivity.

In the \TwoJet VBF-enriched event class, the top quark and \Ztt CRs are each split into three regions. They are first divided into regions of $\pTllmiss < \qty{200}{\GeV}$ and $\pTllmiss \geq \qty{200}{\GeV}$, and then the former is subdivided into regions with $350 \leq \mjj < \qty{700}{\GeV}$ and $\mjj \geq \qty{700}{\GeV}$. The \Zeemm CR requires $\pTllmiss \geq \qty{200}{\GeV}$, as including additional CRs does not improve the expected measurement precision. Instead, in the $\pTllmiss < \qty{200}{\GeV}$ phase space, the normalisations of the \Zeemm background in the regions defined by $350 \leq \mjj < \qty{700}{\GeV}$, $700 \leq \mjj < \qty{1000}{\GeV}$, and $\mjj \geq \qty{1000}{\GeV}$ are each obtained from a fit to data in the corresponding SRs. The DNNs in the SRs provide excellent shape separation between the signal and the \Zeemm background, which allows the normalisation of the \Zeemm background to be estimated and the signal to be measured simultaneously.

\begin{figure}[htbp]
\centering
\includegraphics[width=\textwidth]{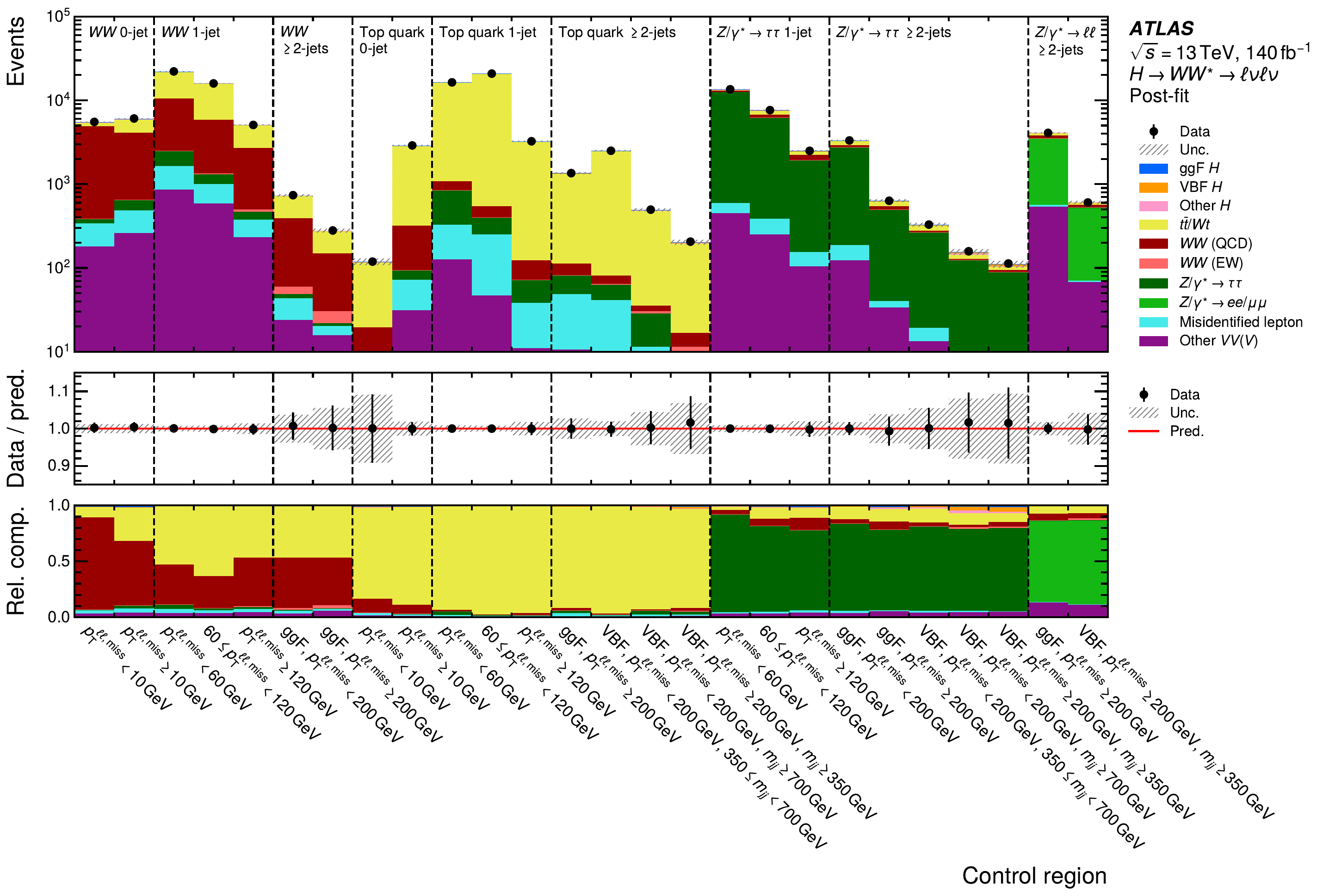}
\caption{Yield distributions in the CRs of the STXS measurement after a full fit to data (i.e., `Post-fit'). The vertical dashed lines delineate CRs targeting different backgrounds, where `$\DY\to\ell\ell$' corresponds to `\Zeemm'. The middle panel shows the ratio of the data to the sum of the fitted signal and background, and the bottom panel shows the relative composition among the fitted signal and background processes. In the top and middle panels, the hatched band shows the total uncertainty, including both statistical and systematics sources.}
\label{fig:STXS_CRs}
\end{figure}

The yield and composition of each CR is shown in Figure~\ref{fig:STXS_CRs}, following the full fit to data described in Section~\ref{sec:results_STXS}. Additionally, the corresponding values of the background normalisation factors obtained from this fit are consistent with unity and shown in Figure~\ref{fig:normalization_factors} of Appendix~\ref{app:results}.

\begin{table}[htbp]
\centering
\caption{Event selection criteria used to define the CRs. The definitions of the variables can be found in the main text. The \WW, top quark, and \Ztt CRs follow the DF channel preselection in Table~\ref{tab:SR_selections}; the \Zeemm CRs follow from the SF channel preselection in Table~\ref{tab:SR_selections}. Selections in bold indicate those that ensure the CRs are statistically independent of their respective SRs. `\VH orthogonality' corresponds to either $|\mjj - \qty{85}{\GeV}| > \qty{15}{\GeV}$, or $|\dyjj| > 1.2$, or both.}
\label{tab:CR_selections}
\resizebox{\textwidth}{!}{
\begin{tabularx}{1.55\textwidth}{c || Y | Y | Y | Y}
\toprule
& \multicolumn{4}{c}{Event class} \\
Control region & \ZeroJet & \OneJet & \TwoJet ggF-enriched & \TwoJet VBF-enriched \\
\midrule
\multirow{7}{*}{\WW} & \multicolumn{3}{c|}{DF channel preselection} & \multirow{7}{*}{N/A} \\
& \multicolumn{3}{c|}{$\nbjets\,(\pt > \qty{20}{\GeV}) = 0$} & \\
& $\boldsymbol{55 < \mll < \qty{100}{\GeV}}$ & $\boldsymbol{\mll > \qty{80}{\GeV}}$ & $\boldsymbol{\mll > \qty{80}{\GeV}}$ \\
& $\dphill < \qty{2.0}{rad}$ & $\mt > \qty{110}{\GeV}$ & $\mTTwo > \qty{165}{\GeV}$ \\
& $\met > \qty{20}{\GeV}$ & & $\mtt < \mZ - \qty{25}{\GeV}$ \\
& $\dphillmiss > \qty{1.57}{rad}$ & & \VH orthogonality \\
& $\drll > 0.6$ & & Fail CJV or Fail OLV \\
\midrule
\multirow{8}{*}{Top quark} & \multicolumn{4}{c}{DF channel preselection} \\
& \multirow{2}{*}{$\boldsymbol{N_{b\textrm{\textbf{-jets}}}\,(20 < p_\textrm{\textbf{T}} < \qty{30}{\GeV}) = 1}$} & $\boldsymbol{N_{b\textrm{\textbf{-jets}}}\,(p_\textrm{\textbf{T}} > \qty{30}{\GeV}) = 1}$ & \multicolumn{2}{c}{\multirow{2}{*}{$\boldsymbol{N_{b\textrm{\textbf{-jets}}}\,(p_\textrm{\textbf{T}} > \qty{20}{\GeV}) = 1}$}} \\
& & $\nbjets\,(20 < \pt < \qty{30}{\GeV}) = 0$ \\
& $\mll < \qty{100}{\GeV}$ & $\mll < \qty{100}{\GeV}$ & $\mll < \qty{55}{\GeV}$ & $\mll < \qty{70}{\GeV}$ \\
& $\met > \qty{20}{\GeV}$ & $\mtt < \mZ - \qty{25}{\GeV}$ & $\mtt < \mZ - \qty{25}{\GeV}$ & $\mtt < \mZ - \qty{25}{\GeV}$ \\
& $\dphill < \qty{2.0}{rad}$ & & $\dphill < \qty{1.8}{rad}$ & -- \\
& $\dphillmiss > \qty{1.57}{rad}$ & & \VH orthogonality & $\mjj \geq \qty{350}{\GeV}$ \\
& $\drll > 0.6$ & & Fail CJV or Fail OLV & Pass CJV and Pass OLV \\
\midrule
\multirow{6}{*}{\Ztt} & \multirow{6}{*}{N/A} & \multicolumn{3}{c}{DF channel preselection} \\
& & \multicolumn{3}{c}{$\nbjets\,(\pt > \qty{20}{\GeV}) = 0$} \\
& & \multicolumn{2}{c|}{$\boldsymbol{\mtt > \mZ - \qty{25}{\GeV}}$} & $\boldsymbol{|\mtt - \mZ| < \qty{25}{\GeV}}$ \\
& & $\mll< \qty{80}{\GeV}$ & $\mll< \qty{55}{\GeV}$ & $\mll< \qty{70}{\GeV}$ \\
& & & \VH orthogonality & $\mjj \geq \qty{350}{\GeV}$ \\
& & & Fail CJV or Fail OLV & Pass CJV and Pass OLV \\
\midrule
\multirow{6}{*}{\Zeemm} & \multicolumn{2}{c|}{\multirow{6}{*}{N/A}} & \multicolumn{2}{c}{SF channel preselection} \\
& \multicolumn{2}{c|}{\phantom{--}} & \multicolumn{2}{c}{$\nbjets\,(\pt > \qty{20}{\GeV}) = 0$} \\
& \multicolumn{2}{c|}{\phantom{--}} & \multicolumn{2}{c}{$\boldsymbol{|\mll - \mZ| < \qty{15}{\GeV}}$} \\
& \multicolumn{2}{c|}{\phantom{--}} & -- & $\dphill < \qty{1.8}{rad}$ \\
& \multicolumn{2}{c|}{\phantom{--}} & \VH orthogonality & $\mjj \geq \qty{350}{\GeV}$ \\
& \multicolumn{2}{c|}{\phantom{--}} & Fail CJV or Fail OLV & Pass CJV and Pass OLV \\
\bottomrule
\end{tabularx}
}
\end{table}

\paragraph{Control region modifications for the CP-sensitive measurement}

The CR selections in the \STXSCP measurement are exactly the same as in the STXS measurement, including the regions for which particular backgrounds are normalised in SRs without dedicated CRs. However, there is one exception: unlike the STXS measurement, the \TwoJet VBF-enriched event class in the \STXSCP measurement does not use a split at $\mjj = \qty{1000}{\GeV}$ for $\pTH < \qty{200}{\GeV}$, and so the \Zeemm background in the region $\mjj \geq \qty{700}{\GeV}$ is normalised to data in the corresponding SR via a single normalisation factor.


\section{Background estimation}
\label{sec:background}

Background processes are modelled in three ways: using MC predictions normalised to data, pure data-driven predictions, or pure MC predictions.

The \WW, top quark, and \Zjets processes are modelled using MC samples and normalised to data. The corresponding rates are fitted using the CRs and SRs defined in Section~\ref{sec:channels}. The differential distributions or `shapes' of relevant observables for those same backgrounds are extracted from simulation.

The backgrounds that are estimated by using pure data-driven predictions correspond to processes with misidentified leptons. The estimation of these backgrounds is described in detail in Section~\ref{sec:fakes}.

All other physics processes are estimated by using pure MC predictions, including Higgs boson production not considered as signal (e.g., \VH, \ttH, and \Htautau), \WZ/\ZZ/\Vy diboson production, and triboson production.

\subsection{Backgrounds with misidentified leptons}
\label{sec:fakes}

The backgrounds originating from either one or two misidentified leptons are primarily due to \Wjets and multijet processes, respectively. They are estimated by using a data-driven technique~\cite{Lehmann:2023zlz} where estimation regions (ERs) are defined by applying all nominal selections, except that one of the two lepton candidates is required to fail the full identification criteria outlined in Section~\ref{sec:reconstruction} while still satisfying a looser set of identification requirements. These lepton candidates are referred to as anti-identified leptons. The expected misidentified lepton background yields in the SRs and CRs are extrapolated from the observed number of events in the corresponding ERs with anti-identified leptons, after subtracting the expected contribution from processes with two prompt leptons using simulated samples. The method appropriately accounts for all processes with misidentified leptons, as long as they are present in the sample with one anti-identified lepton. The small contribution from multijet processes with two misidentified leptons is accounted for in the extrapolation by applying a correction term evaluated in an ER where both lepton candidates are anti-identified.

The extrapolation factor that is used to extrapolate the expected misidentified lepton background yields in the ERs to the SRs is determined using a sample of \Zjets-enriched events, where a three-lepton selection is applied to target events with a leptonically decaying \Z boson plus a misidentified-lepton candidate recoiling against the \Z boson. The extrapolation factor is defined as the ratio between the number of events in which the misidentified-lepton candidate satisfies the nominal identification criteria and the number of events in which it satisfies the anti-identified criteria. The main sources of misidentified leptons in the \Zjets-enriched sample used to evaluate the extrapolation factor can be different from those in the SR, where misidentified leptons mostly come from \Wjets events. A correction factor is introduced to account for this difference in sample composition between \W and \Z events (`flavour correction'). This is determined from the ratio of the extrapolation factors measured in samples of \Wjets and \Zjets events using MC simulation. For electrons, the flavour correction increases from $\sim$0.9 for $15 < \pt < \qty{20}{\GeV}$ to $\sim$1.6 for $\pt > \qty{35}{GeV}$; for muons, it increases from $\sim$0.8 for $15 < \pt < \qty{20}{\GeV}$ to $\sim$1.5 for $\pt > \qty{35}{GeV}$. The product of the extrapolation and correction factors is referred to as the corrected extrapolation factor.

\begin{figure}[htbp]
\centering
\subfloat[]{
\includegraphics[width=0.48\textwidth]{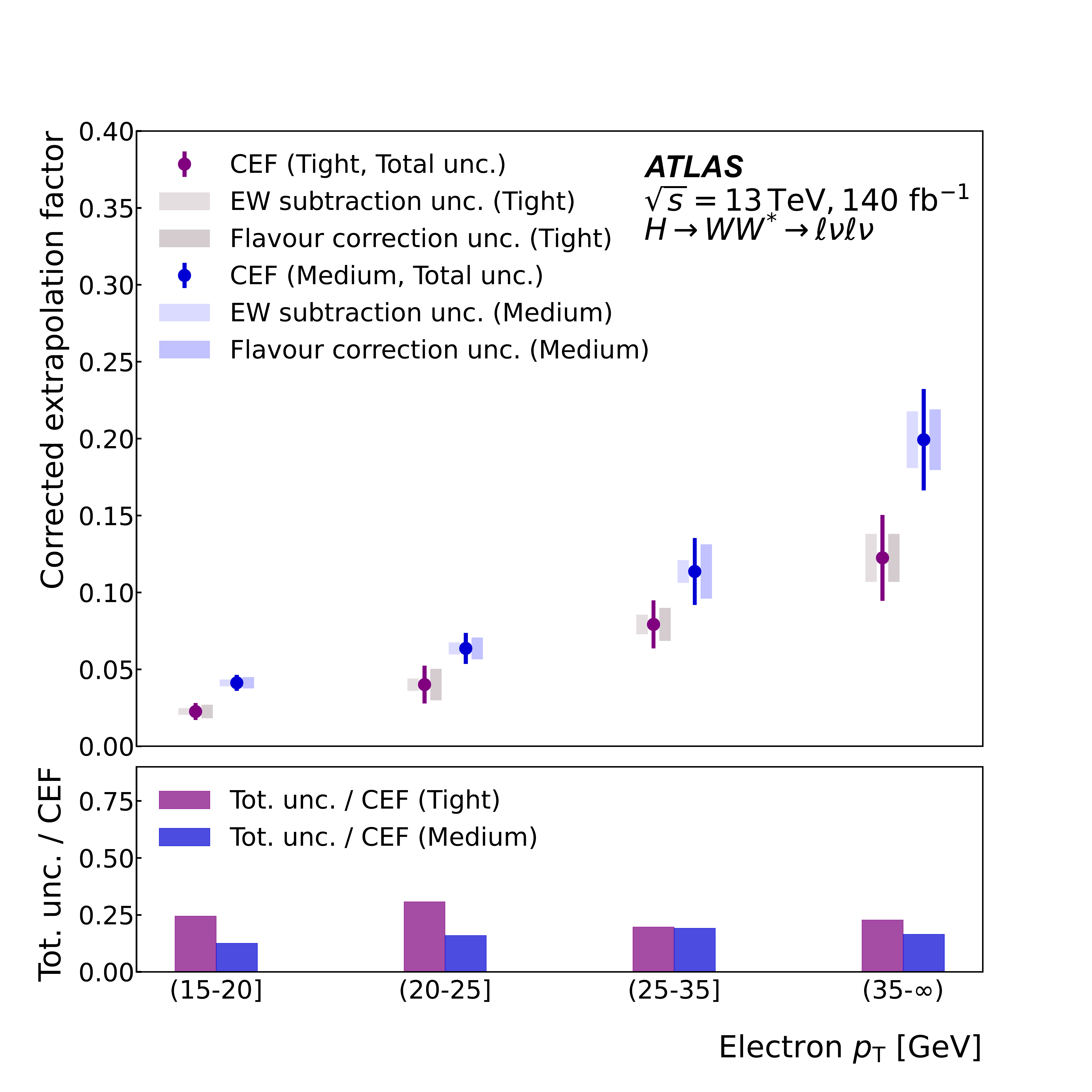}
\label{fig:FFs_electrons}
}
\hfill
\subfloat[]{
\includegraphics[width=0.48\textwidth]{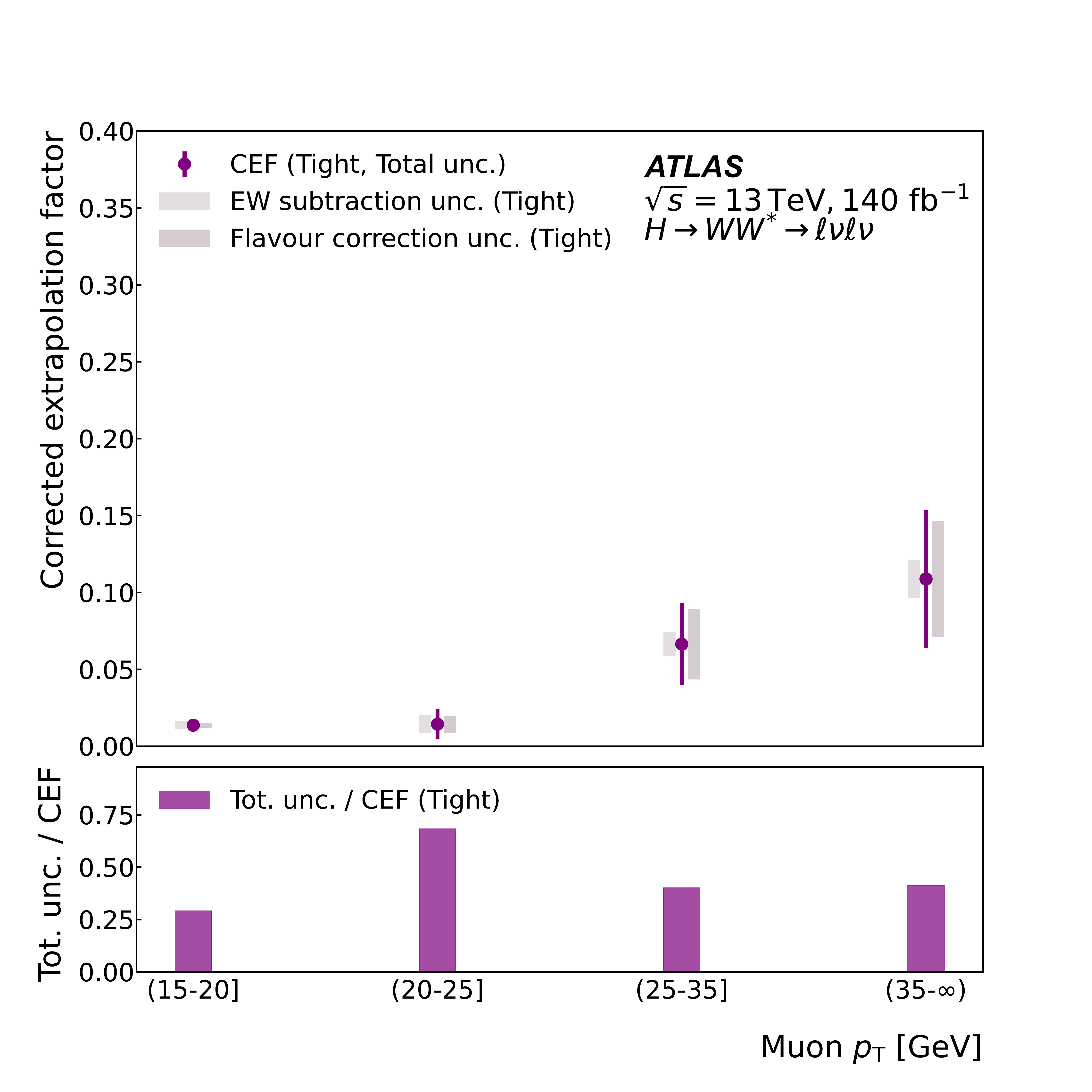}
\label{fig:FFs_muons}
} \\
\caption{Corrected extrapolation factors (CEF) for anti-identified (a) electrons and (b) muons. For electrons, the extrapolation factors are provided for those that satisfy both the Medium and Tight identification criteria. In the upper panels, the error bars indicate the total uncertainties in the corrected extrapolation factors and the shaded bars indicate systematic uncertainties, namely the uncertainty related to the subtraction of processes with exclusively prompt leptons (electroweak subtraction uncertainty) and the uncertainty related to the sample composition correction (flavour correction uncertainty). These sources are described in Section~\ref{sec:esystematics}. In the lower panels, the total relative uncertainties in the corrected extrapolation factors are shown.}
\label{fig:FFs}
\end{figure}

The corrected extrapolation factors are measured separately for anti-identified electrons and muons and differentially in the \pt of the anti-identified lepton, as shown in Figure~\ref{fig:FFs}. Systematic uncertainties are evaluated for the corrected extrapolation factors and are described in Section~\ref{sec:esystematics}.

The single-lepton triggers and some of the single-lepton parts (`legs') of the dilepton triggers apply selection criteria that are tighter than the anti-identification criteria considered by the analysis. As a result, if an anti-identified lepton is solely responsible for triggering an event, a bias occurs. For these events, a new method was developed, resulting in a more precise estimate of this contribution. This method employs an additional set of corrections that are applied to account for the bias. The corrections are derived using dijet events collected using single-lepton triggers that are looser than both the identification and anti-identification criteria considered by the analysis. From these events, two different extrapolation factors are calculated in the same manner as before: the estimation of the first additionally requires the anti-identified lepton to have fired one of the tight triggers above, whereas the estimation of the second places no such requirement. The trigger correction factor is then calculated as the ratio of the first extrapolation factor to the second. The corrected extrapolation factor is multiplied by this factor for events where the misidentified-lepton candidate solely fired a trigger that would induce a bias.

Like the extrapolation factors, the trigger correction factors are measured separately for anti-identified electrons and muons and differentially in the \pt of the anti-identified lepton. The corrected extrapolation factors are found to be consistent between the central and forward regions of the detector. This is not the case for the trigger correction factors, which are thus measured independently in central and forward regions of the detector defined by $|\eta|$ below and above 1.5 (1.05) for electrons (muons). The resulting trigger correction factors are shown in Figure~\ref{fig:TrigCFs} in Appendix~\ref{app:fakes}. Similarly to the corrected extrapolation factors, systematic uncertainties are evaluated for the trigger correction factors and are described in Section~\ref{sec:esystematics}.


\section{Systematic uncertainties}
\label{sec:systematics}

Uncertainties from both experimental and theory sources affect the results of the measurement. This section describes the estimation of their effects on the signal and background normalisations and, where applicable, on the shapes of the DNN discriminants in the SRs. Uncertainties from theory sources generally affect both the sample normalisations and the shapes; thus, their effects are calculated bin-by-bin in the DNN output score. Contrary to this, uncertainties from experimental sources mainly affect the sample normalisations, while their effect on the DNN SR shapes is often evaluated to be negligible and thus not considered.

For a given uncertainty in a region, the normalisation and shape components are subject to a pruning procedure, whereby a component is removed from that region if its impact is found to be negligible. Doing so improves the stability of the statistical analysis.

\subsection{Experimental uncertainties}
\label{sec:esystematics}

Uncertainties associated with the selected leptons originate from the trigger, reconstruction, identification, and isolation efficiencies~\cite{EGAM-2021-01, MUON-2018-03} and from the energy (for electrons) or momentum (for muons) scale and resolution~\cite{EGAM-2018-01, MUON-2022-01}. For jets, uncertainties arise from the jet energy scale and resolution~\cite{JETM-2018-05}, the performance of the jet-vertex taggers~\cite{PERF-2014-03, ATL-PHYS-PUB-2019-026}, and the \bjet identification~\cite{FTAG-2018-01}. Furthermore, uncertainties due to the trigger selection~\cite{TRIG-2018-05, TRIG-2018-01} as well as due to the soft term in the reconstruction of the \met~\cite{JETM-2020-03} are estimated. The uncertainty in the modelling of pile-up for simulated event samples is estimated by varying the reweighting to the pile-up profile in data, described in Section~\ref{sec:samples}, within its uncertainties. The uncertainty in the combined 2015--2018 integrated luminosity is 0.83\%~\cite{DAPR-2021-01}, obtained using the LUCID-2 detector~\cite{LUCID2} for the primary luminosity measurements. The integrated luminosity uncertainty is only considered for the Higgs boson signal and the background processes that are normalised to theory predictions.

For the data-driven estimate of the misidentified lepton background, see Section~\ref{sec:fakes}, three sources of uncertainty are considered for each corrected extrapolation factor: the statistical uncertainty in the extrapolation factor itself, an uncertainty related to the subtraction of processes with three prompt leptons from the \Zjets-enriched sample used to derive the extrapolation factors (`EW subtraction uncertainty'), and an uncertainty in the sample composition when deriving the correction factors (`flavour correction uncertainty'). The EW subtraction uncertainty is calculated using the theory uncertainties in the prompt lepton background, and the flavour correction uncertainty is given by the difference between the correction factors estimated by using \Vjets events simulated with \SHERPA[2.2.11] and those estimated by using \Vjets events simulated with $\POWHEG$+\PYTHIA[8]. This uncertainty is reflective of the generator dependence of the correction factors, which are estimated solely from simulation. The resulting uncertainties can be seen in Figure~\ref{fig:FFs}. For the estimation of the trigger correction factors, both statistical uncertainties and EW subtraction uncertainties are considered and similarly derived, as shown in Figure~\ref{fig:TrigCFs}.

The largest experimental uncertainties affecting the ggF measurement come from the lepton isolation, \bjet identification, and pile-up reweighting. The largest experimental uncertainty affecting the VBF measurement is the jet energy scale.

\subsection{Theory uncertainties}
\label{sec:tsystematics}

Sources of systematic uncertainty from theory are evaluated by reweighting simulated events or by comparing the nominal simulated samples with alternative samples. For all processes and unless stated otherwise, the uncertainty related to missing higher-order corrections is computed as the maximum variation of the envelope resulting from simultaneous variations of the renormalisation and factorisation scales, $\mu_R$ and $\mu_F$, by factors of 0.5 and 2, excluding the two variations with opposite extremes of $(0.5\times\mu_R, 2\times\mu_F)$ and $(2\times\mu_R, 0.5\times\mu_F)$. The uncertainty in the PDF is evaluated using the PDF error sets and following the recommendations of Ref.~\cite{Butterworth:2015oua}. It encapsulates the experimental and model-related uncertainties entering the PDF fit. The uncertainty in the value of the strong coupling constant is evaluated using the nominal PDF set with different values of the strong coupling constant and is combined with the PDF uncertainty following the recommendations of Ref.~\cite{Butterworth:2015oua}.

For signal processes, uncertainties from the matrix element matching and parton shower are estimated by comparing predictions from the nominal and alternative generators described in Section~\ref{sec:samples}. The approach described in Refs.~\cite{Stewart:2011cf, deFlorian:2016spz} is used to estimate the impact of missing higher-order corrections on the ggF and VBF production modes. The effects of these corrections on the overall predicted yield and on the distribution of events across STXS categories are considered, as described in Ref.~\cite{ATL-PHYS-PUB-2023-031}. The uncertainties in the Higgs boson branching ratios due to dependencies on SM parameter values (such as $b$- and $c$-quark masses) and missing higher-order effects are implemented using the correlation model described in Ref.~\cite{deFlorian:2016spz}, leading to an uncertainty of $\pm2.16\%$ in the \hwwlvlv branching fraction.

For the small contribution from the Higgs boson production not considered signal, a conservative normalisation uncertainty of $\pm50\%$ is assigned, resulting in a negligible effect on the measurements reported in this paper.

For the \VV/\Vys, \Vy, and \Zjets processes simulated using \SHERPA, variations of the matching scale and non-perturbative effects are considered; for parton shower uncertainties, variations of the resummation scale and the choice of parton shower recoil scheme are used. These variations are implemented in alternative samples, and comparisons with the nominal sample are made using particle-level events with selections similar to those applied to reconstructed events. Uncertainties in the choice of the implementation scheme for the approximate EW corrections to \VV/\Vys and \Zjets processes are also assigned~\cite{Kallweit:2014xda}.

Owing to an observed change in the modelling of the \ptZ spectrum going from \SHERPA[2.2.1] to 2.2.11/14, an additional `\pt spectrum' uncertainty is applied to the \SHERPA[2.2.11/14] \Zjets process, calculated as the difference between the 2.2.11/14 and 2.2.1 predictions, bin-by-bin in each SR and CR. The calculation closely matches the procedure followed by an ATLAS \VH, $H\to{b}\bar{b}/{c}\bar{c}$ measurement~\cite{HIGG-2020-20}, where a similar change in the modelling was observed.

The EW \WWjj process is assigned an additional normalisation uncertainty of 15\% due to NLO EW corrections, calculated using the leading-logarithm approximation~\cite{Denner:2000jv, Biedermann:2016yds, Denner:2019tmn}.

For top quark processes, uncertainties arising from the matrix element matching and parton shower are estimated by using the nominal and alternative samples described in Section~\ref{sec:samples} and in the same manner as for signal processes. Additional uncertainties relating to the choice of model for initial- and final-state radiation are computed by varying scale, resummation, and showering parameters. For \ttbar, uncertainties in the NNLO reweighting scheme are also considered. For \Wt, additional uncertainties obtained by comparing different diagram removal schemes~\cite{Frixione:2008yi} and different choices of the \hdamp scale~\cite{ATL-PHYS-PUB-2021-042} are applied.

The uncertainties described above are estimated for each SR separately, where the reduced (STXS or \STXSCP) fiducial categories are treated as different processes for the ggF and VBF signals, and cover the migration of events between SRs. When only a small number of events is available in the simulated event samples for certain phase spaces, the effect of the uncertainty is evaluated on an inclusive set of SRs and applied to the individual SRs within that set.

The largest theory uncertainties affecting the ggF measurement come from the choice of parton shower recoil scheme for \WW production, the \ptZ spectrum mismodelling for \Zjets production, and the choice of parton shower algorithm for ggF and \ttbar production. The largest theory uncertainties affecting the VBF measurement come from the choice of matrix element matching and parton shower algorithms for VBF production and the NLO EW corrections applied to the EW \WWjj process.


\section{Fit procedure}
\label{sec:statistics}

Results are obtained from a simultaneous profile likelihood fit~\cite{Cowan:2010js} to data using all SRs and CRs. In each SR for each DF or SF channel, the corresponding DNN output node is used as the final fit discriminant. The output values of the DNNs in the SRs described in Section~\ref{sec:channels} are binned to obtain an equal expected number of signal events in each bin, where signal refers to either ggF or VBF signal events depending on the SR. This procedure yields between two and six bins per SR. The chosen, relatively coarse, binning optimises the sensitivity while ensuring a sufficient number of events such that background templates are stable against fluctuations. Larger numbers of bins do not improve the expected sensitivity significantly in the presence of all systematic uncertainties. Background events tend to cluster in the low-score DNN bins.

The fit for the \STXSCP measurement can safely use the same $\geq$2-jets DF and SF DNNs in the SRs as the STXS measurement, as all training variables are CP-even. Due to the symmetric nature of the \DPhijjsgn distribution in the SM prediction, the simulated signal and background compositions are identical between the symmetric \DPhijjsgn SRs, $[-\pi, -\pi/2)$ and $[+\pi/2, +\pi)\,\si{rad}$ as well as $[-\pi/2, 0)$ and $[0, +\pi/2)\,\si{rad}$. To further ensure that these regions offer equal sensitivity to signal, the DNN binning is optimised and chosen to be the same in both regions. Maintaining this symmetry is essential for maximising the effectiveness of the analysis in detecting CP-violating effects.

Each CR is represented by a Poisson probability term to constrain the normalisation of its corresponding background process via a floating normalisation factor. Uncertainties enter the fit as nuisance parameters in the likelihood function. Theory uncertainties affecting the signal and the experimental uncertainties affecting both signal and background are correlated between SRs and CRs in all analysis channels. Theory uncertainties affecting the background modelling parameters and background normalisation factors are correlated across the SRs and CRs of a given jet multiplicity event class but are decorrelated between different event classes. This approach results in a conservative choice: the expected precision of all POIs in the STXS and \STXSCP measurements improves slightly when these uncertainties are correlated. This is caused by slight reductions of the post-fit nuisance parameter uncertainties, which result from interdependencies of the event yields between the event classes. Decorrelating makes the analysis less reliant on the modelling of these effects. The values of the POIs and of the background normalisation parameters in the observed fit change only slightly when the uncertainties are correlated instead of decorrelated, i.e.\ the changes are much smaller than the uncertainties of the parameters. Additionally, for cases where the underlying physics of the background is expected to differ substantially between DF and SF processes (e.g., \WZ/\ZZ/\Vy diboson production or triboson production), the theory uncertainties are also decorrelated between the DF and SF channels.

The cross-sections multiplied by the \hww branching ratio, $\XS\times\BRHWW$, are represented as POIs and inferred using only constraints from data. Two different configurations of this parameter set are measured from independent fits to the same regions. The first measures the 15~cross-sections defined by the STXS scheme as POIs, and the second measures the total ggF and VBF cross-sections as POIs and is referred to as the 2-POI fit. In both cases, the other Higgs boson production modes are fixed to their expected yields.

A third fit is used to measure a set of POIs that scale the expected signal yields for the 27~production categories considered by the \STXSCP measurement. These POIs are referred to as signal strengths, $\mu$.

An important distinction is made between the fit configurations measuring signal strengths versus those measuring cross-sections. The former considers both normalisation and shape components of the theory uncertainties affecting signal, while the latter removes the normalisation components from these same uncertainties (i.e., only considers the components affecting acceptance). When measuring a cross-section, the normalisation components are included in the uncertainty on the SM expectation that one compares to the measured cross-section. This avoids double counting of these uncertainties.

The individual contributions to the total uncertainty of each POI are determined via the breakdown method. In this method, the contributions of each source are evaluated as the difference in quadrature between the total uncertainty in the fit floating all sources of uncertainty and the uncertainty when the source is eliminated from the fit by fixing the corresponding nuisance parameters to their best-fit values.


\section{Results}
\label{sec:results}

In this section, the results for the STXS measurement (Section~\ref{sec:results_STXS}), the 2-POI fit using the STXS measurement (Section~\ref{sec:results_2POI}), and the \STXSCP measurement (Section~\ref{sec:results_DPhijj}) are presented.

The terms `pre-fit' and `post-fit' refer to the state of the statistical model before and after fitting to the observed data to find the best-fit values of the statistical model parameters. Prior to performing any fit, all parameters of the statistical model are initialised to values such that the statistical model corresponds to the SM expectation. Additionally, `expected results' correspond to those obtained using the pre-fit values of the statistical model parameters, with data substituted by the pre-fit yields from simulation.

\subsection{STXS measurement}
\label{sec:results_STXS}

\begin{figure}[htbp]
\centering
\subfloat[]{
\includegraphics[height=0.39\textheight]{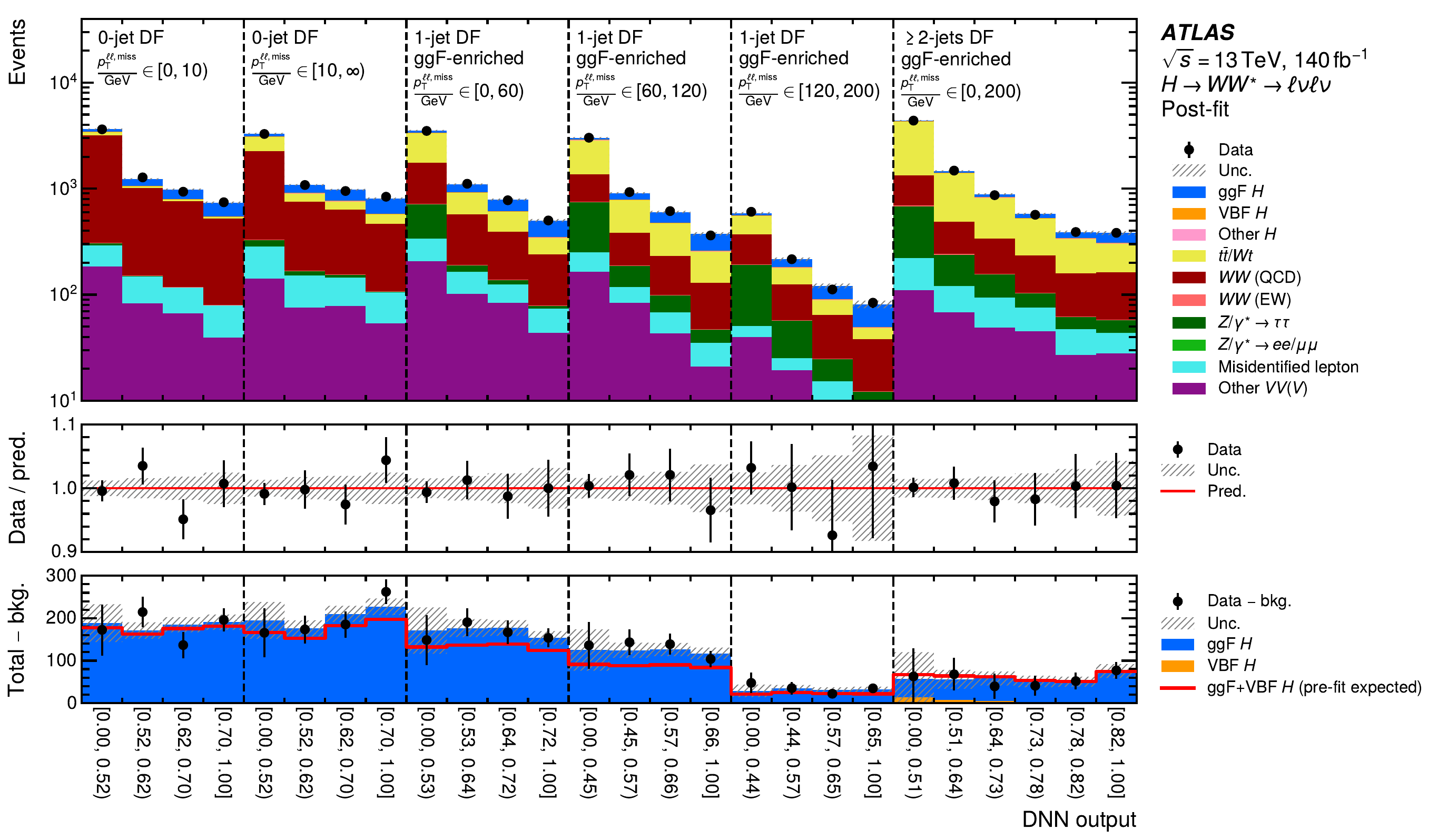}
\label{fig:STXS_SRs_ggF_a}
} \\
\subfloat[]{
\includegraphics[width=0.98\textwidth]{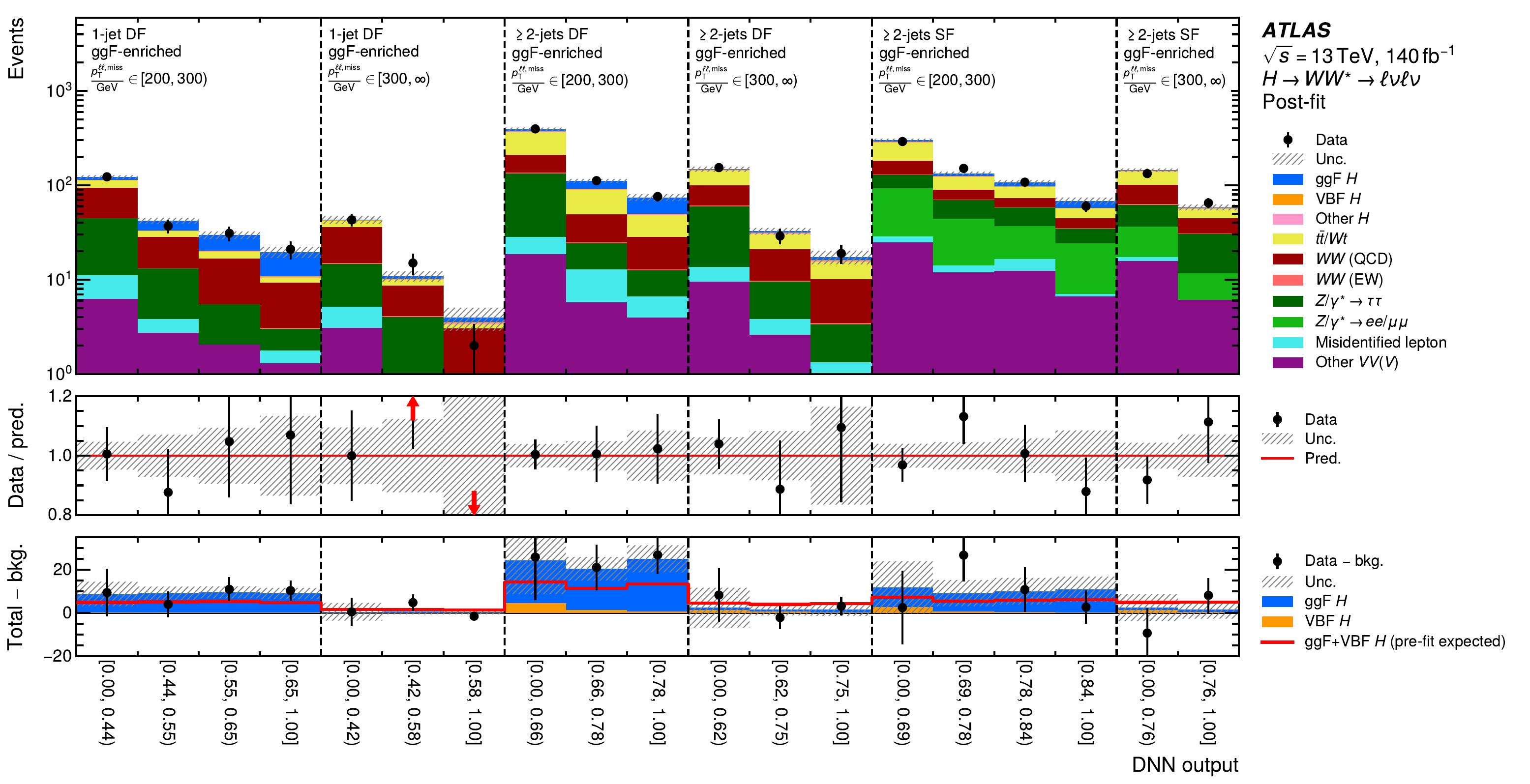}
\label{fig:STXS_SRs_ggF_b}
} \\
\caption{Post-fit DNN distributions in the ggF (a) low and (b) high \pTllmiss SRs entering the STXS measurement for all jet multiplicities; the post-fit results are obtained from the 15-POI fit. The vertical dashed lines delineate different SRs. The middle panels show the ratios of the data to the sum of the fitted signal and background, and the bottom panels show the differences between the data and the estimated background compared to the simulated post-fit signal distribution. In the middle panel in (b), the arrows indicate the positions of points outside the vertical axis range. In the bottom panels, the pre-fit expected prediction for the sum of ggF and VBF is overlaid. In all panels, the hatched band shows the total uncertainty.}
\label{fig:STXS_SRs_ggF}
\end{figure}

\begin{figure}[htbp]
\centering
\includegraphics[height=0.40\textheight]{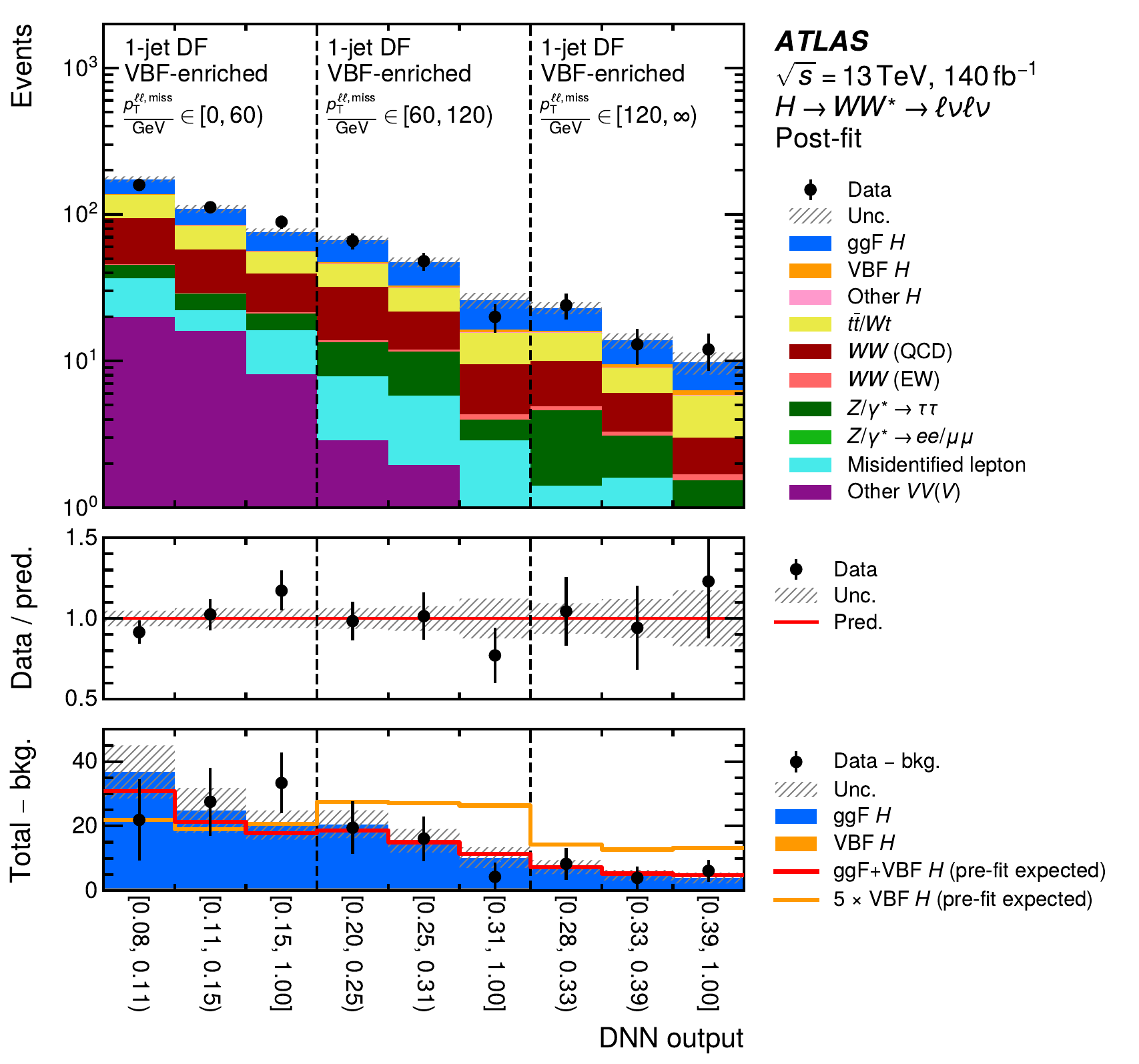}
\caption{Post-fit DNN distributions in the VBF 1-jet SRs entering the STXS measurement; the post-fit results are obtained from the 15-POI fit. The vertical dashed lines delineate different SRs. The DNN output ranges excluded in the SRs shown, which cover the interval between 0 and the first bin per SR, are instead covered by the corresponding ggF 1-jet SRs in Figure~\ref{fig:STXS_SRs_ggF}. The middle panel shows the ratios of the data to the sum of the fitted signal and background, and the bottom panel shows the differences between the data and the estimated background compared to the simulated post-fit signal distribution. In the bottom panel, the pre-fit expected prediction for the sum of ggF and VBF is overlaid; independently, the pre-fit expected prediction for VBF is overlaid and scaled up by a factor of five. In all panels, the hatched band shows the total uncertainty.}
\label{fig:STXS_SRs_1j_VBF}
\end{figure}

\begin{figure}[htbp]
\centering
\subfloat[]{
\includegraphics[width=0.98\textwidth]{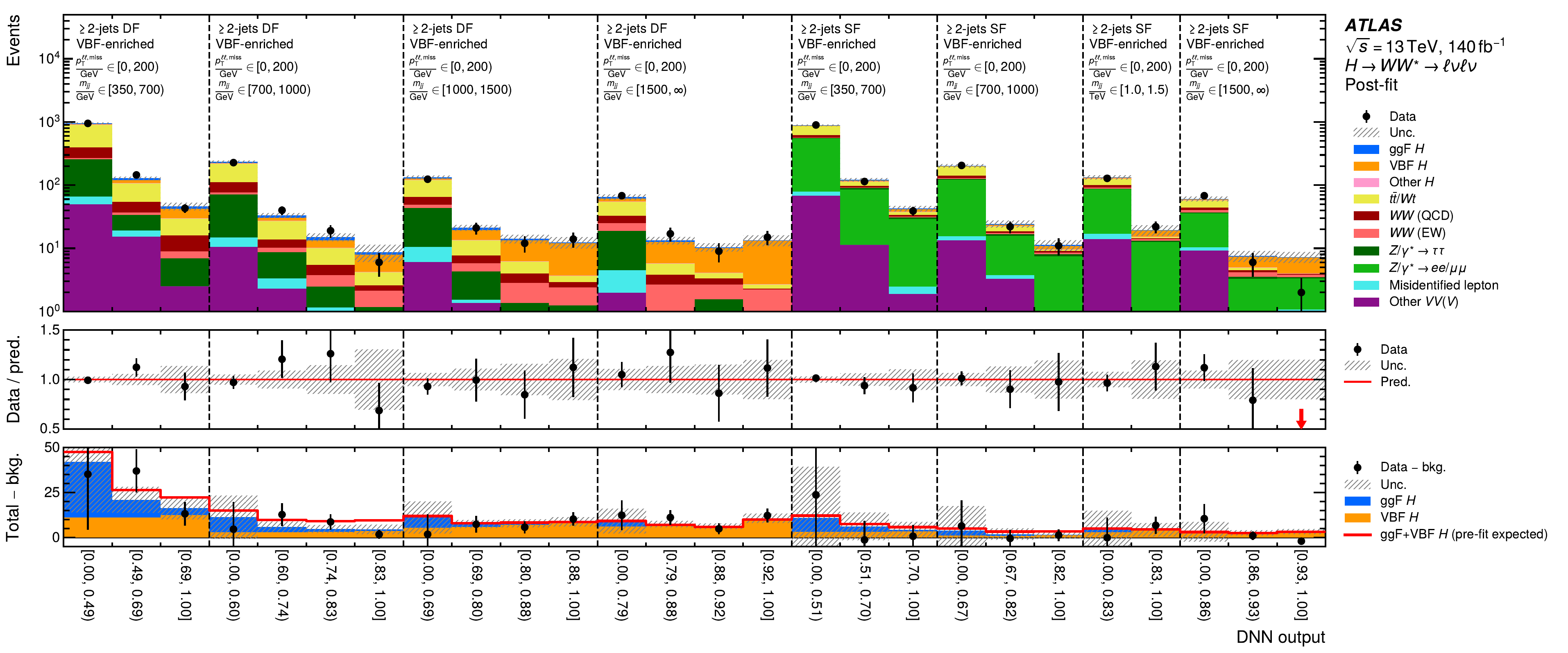}
} \\
\subfloat[]{
\includegraphics[width=0.98\textwidth]{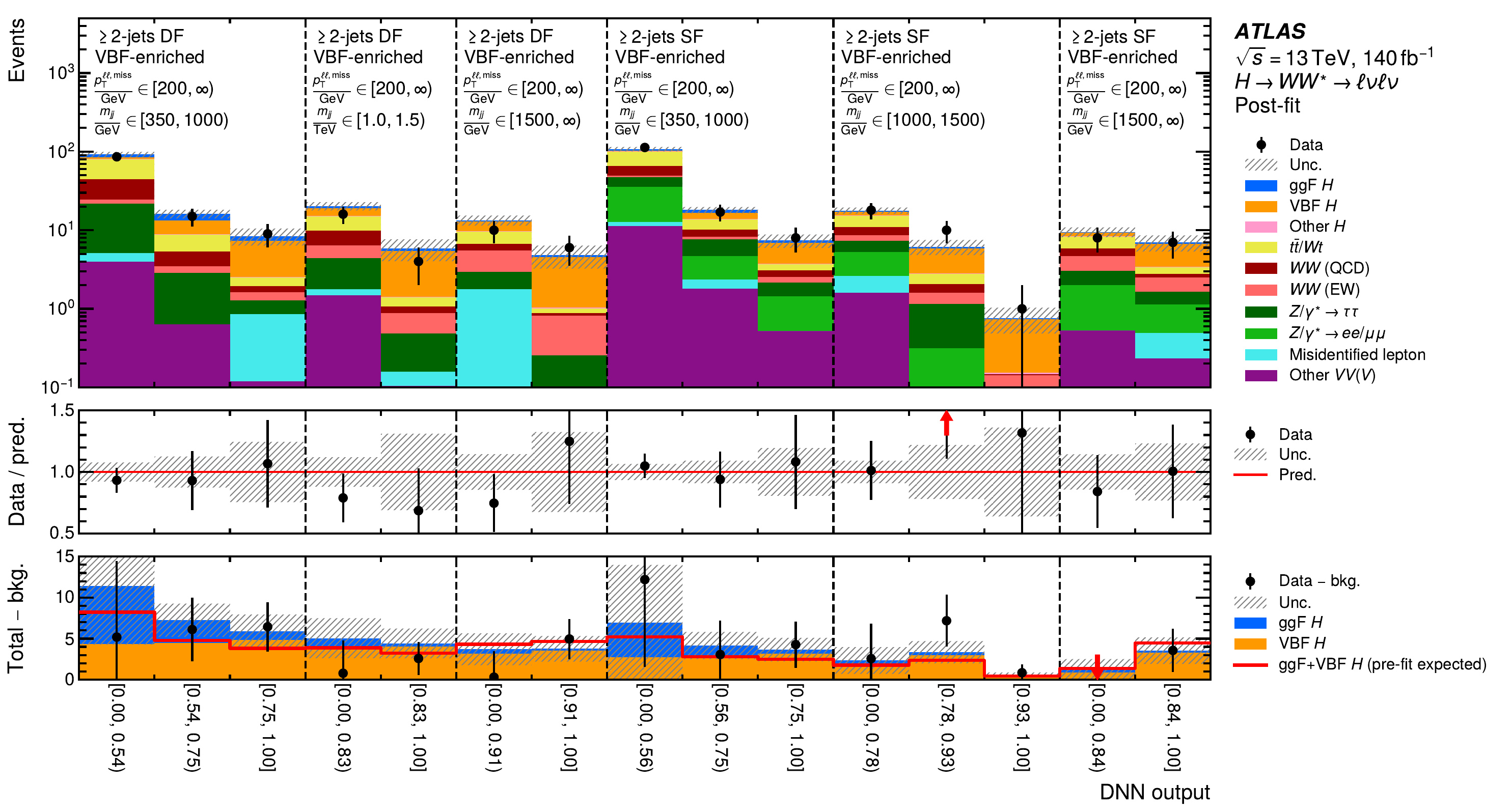}
} \\
\caption{Post-fit DNN distributions in the VBF $\geq$2-jets (a) low and (b) high \pTllmiss SRs entering the STXS measurement for both the DF and SF channels; the post-fit results are obtained from the 15-POI fit. The vertical dashed lines delineate different SRs. The middle panels show the ratios of the data to the sum of the fitted signal and background, and the bottom panels show the differences between the data and the estimated background compared to the simulated post-fit signal distribution. In the middle panels, the arrows indicate the positions of points outside the vertical axis range. In the bottom panels, the pre-fit expected prediction for the sum of ggF and VBF is overlaid. In all panels, the hatched band shows the total uncertainty.}
\label{fig:STXS_SRs_2j_VBF}
\end{figure}

Excellent post-fit background agreement with observed data in all CRs is shown in Figure~\ref{fig:STXS_CRs}. The corresponding post-fit background normalisation factors are consistent with unity, i.e., their pre-fit values, as shown in Figure~\ref{fig:normalization_factors} of Appendix~\ref{app:results}. Together, this gives a high confidence in the background modelling and the ability of the fit model to adjust the relevant background normalisations. Figures~\ref{fig:STXS_SRs_ggF} through \ref{fig:STXS_SRs_2j_VBF} show the post-fit yields in the SRs of the STXS measurement. The observed data yields agree, in both rate and shape, within uncertainties with the post-fit signal and background yields from the prediction in all SRs. Specifically, Figure~\ref{fig:STXS_SRs_ggF} shows the excellent agreement in all ggF SRs, including the ggF SF $\geq$2-jets SRs considered for the first time in this paper. Figure~\ref{fig:STXS_SRs_1j_VBF} shows the SRs that, for the first time in this channel, target the EW \qqH 1-jet STXS category; the observed SRs are in agreement with no VBF contribution and with the SM prediction. Lastly, Figure~\ref{fig:STXS_SRs_2j_VBF} demonstrates the excellent post-fit agreement of the data with the model for the VBF $\geq$2-jets SRs, including the VBF SF $\geq$2-jets SRs considered for the first time in this paper.

\begin{figure}[htbp]
\centering
\includegraphics[width=\textwidth]{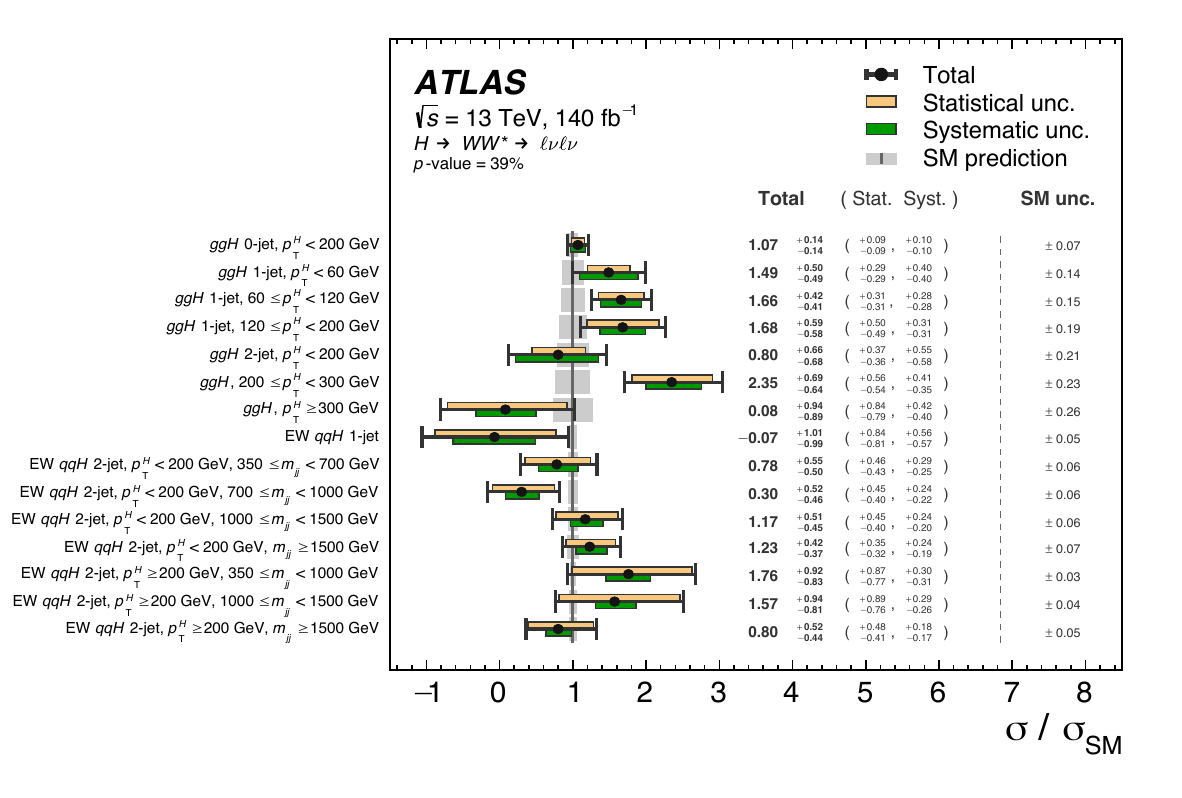}
\caption{Best-fit values and uncertainties for the \hww cross-sections measured in each of the STXS categories, normalised to the corresponding SM predictions. The error bars, upper boxes, and lower boxes show the total, statistical, and systematic uncertainties in the measurements, respectively. The shaded bands represent the theory uncertainty of the signal yield in the corresponding STXS category.}
\label{fig:STXS_POIs}
\end{figure}

Figure~\ref{fig:STXS_POIs} shows a summary of the cross-sections multiplied by the \hww branching ratios measured in each of the 15~STXS categories, normalised to the corresponding SM predictions. The results are compatible with the SM predictions with a $p$-value of 39\%. The correlations among the POIs are shown in Figure~\ref{fig:STXS_corr} in Appendix~\ref{app:results}. The strongest correlations are observed between the \ggH 1-jet and EW \qqH 1-jet cross-sections and between the \ggH $\geq$2-jets and EW \qqH $\geq$2-jets low \pTH low \mjj cross-sections, as expected from the SR composition shown in Figure~\ref{fig:STXS_comp}. No correlation is greater than 50\%.

Table~\ref{tab:STXS_POIs} in Appendix~\ref{app:results} provides the central values and uncertainties of each of the measured STXS cross-sections, together with the SM predictions. The measurements of the EW \qqH STXS cross-sections are dominated by statistical uncertainties. While the uncertainty is predominantly statistical for the \ggH category with $\pTH > \qty{300}{\GeV}$, experimental and theory uncertainties contribute significantly for the remaining \ggH categories. Systematic uncertainties exceed statistical uncertainties for the 0-jet category, the 1-jet category with $\pTH < \qty{60}{\GeV}$, and the $\geq$2-jets category with $\pTH < \qty{200}{\GeV}$. The largest uncertainties for the \ggH 0-jet category are muon efficiency calibration uncertainties and the theory uncertainties for the \WW background (dominated by the uncertainty originating from the choice of parton shower recoil scheme), which individually amount to 4\% relative to the POI value. The largest uncertainties for the \ggH 1-jet category with $\pTH < \qty{60}{\GeV}$ are the \WW and top theory uncertainties as well as the flavour tagging calibration uncertainties, which amount to about 10\% each relative to the POI value. The largest theory uncertainties for the \ggH $\geq2$-jets category with $\pTH < \qty{200}{\GeV}$ stem from the modelling of the top and \WW backgrounds. They amount to 51\% and 25\% relative to the POI value. The largest experimental uncertainties for that same category are the jet energy resolution and flavour tagging calibration uncertainties, which amount to 16\% and 13\% relative to the POI value, respectively.

Compared to the first full Run~2 publication measuring ggF and VBF production and eleven STXS cross-sections in the \hww decay channel~\cite{HIGG-2021-20}, this analysis refines the measurements of the STXS cross-sections. It measures more granular STXS cross-sections, particularly in the region of $\pTH > \qty{200}{\GeV}$, adds for the first time in this channel the EW \qqH STXS category with exactly one jet, and achieves higher precision on previously measured cross-sections. This increased performance can be attributed to the improvements in the lepton isolation algorithm that suppresses backgrounds due to misidentified leptons, the increased use of neural networks across all channels, and various other subdominant factors, as already mentioned in Section~\ref{sec:intro}. Owing to these improvements, the expected accuracy for the EW \qqH STXS categories with $\geq$2 jets has improved by approximately 10\% to 25\%. Improvements of the expected results for the \ggH STXS categories range from 10\% to 36\% and exceed 25\% for $\geq$2-jets and for \pTH above \qty{120}{\GeV}. The newly considered same-flavour channel in the $\geq$2-jets event class boosts the expected sensitivity for the STXS categories with large \pTH. The related gain is around 20\% for the EW \qqH STXS categories with \pTH above \qty{200}{\GeV} and 8\% for the \ggH STXS category with \pTH above \qty{300}{\GeV}.

\FloatBarrier

\subsection{Inclusive cross-section measurement}
\label{sec:results_2POI}

\begin{figure}[htbp]
\centering
\includegraphics[width=0.75\textwidth]{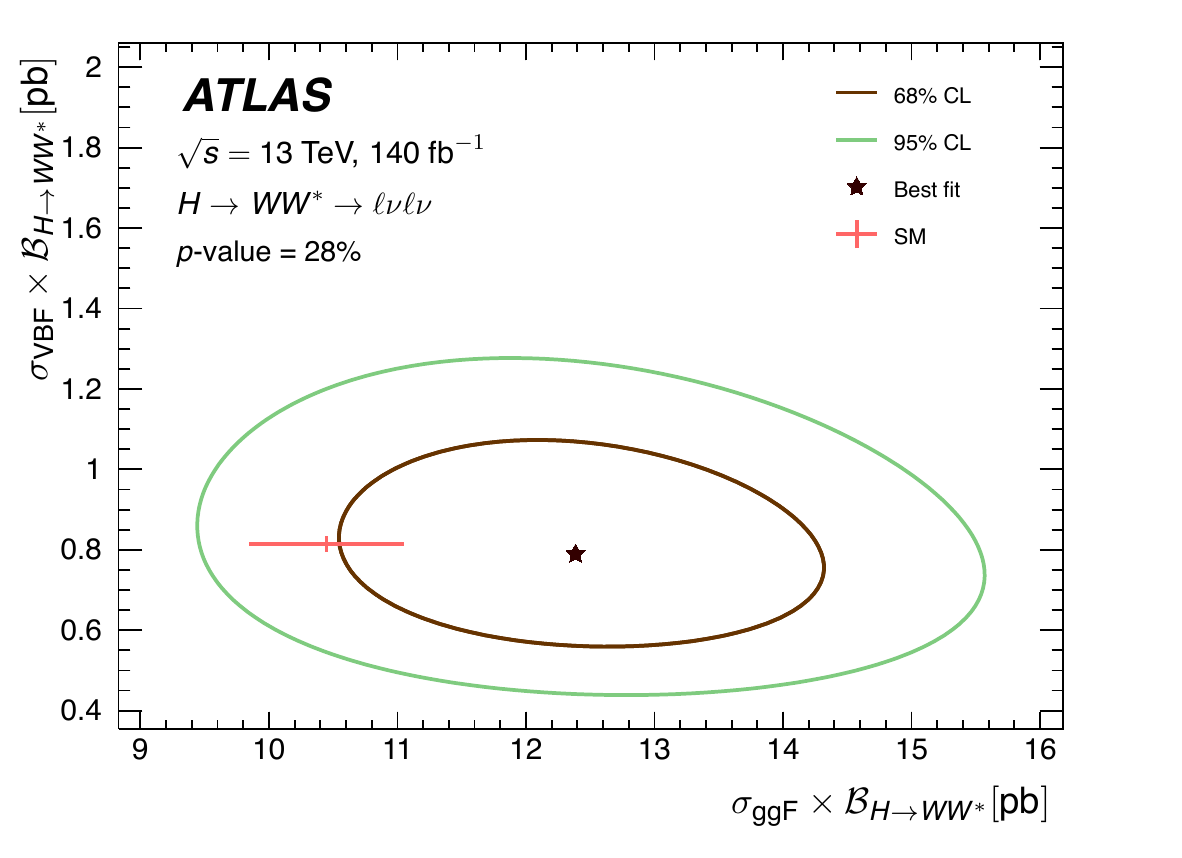}
\caption{Best-fit value and 68\% and 95\% CL two-dimensional contours of $\XSVBF\times\BRHWW$ versus $\XSggF\times\BRHWW$. The error bars on the SM prediction correspond to its 68\% CL error intervals, separately for $\XSVBF\times\BRHWW$ and $\XSggF\times\BRHWW$.}
\label{fig:2D_contours}
\end{figure}

The total ggF and VBF cross-sections multiplied by the \hww branching ratio, $\XSggF\times\BRHWW$ and $\XSVBF\times\BRHWW$, are measured using a fit with only two POIs. Here, all ggF POIs from the STXS fit are constrained to be identical and similarly for all VBF POIs. The expected precision of both measurements was improved by 11\% compared to those reported in the previous ATLAS Run~2 publication~\cite{HIGG-2021-20}, owing to several analysis improvements discussed earlier and briefly summarized in Section~\ref{sec:intro}. The observed results from this fit are:
\begin{equation*}
\begin{split}
\XSggF\times\BRHWW & = 12.4^{+1.3}_{-1.2}\,\si{pb} \\
& = 12.4 \pm 0.6\stat \pm 0.9\,(\textrm{exp. syst.}) \pm 0.3\,(\textrm{sig. theo.}) ^{+0.6}_{-0.5}\,(\textrm{bkg. theo.})\,\si{pb} \,, \\
\XSVBF\times\BRHWW & = 0.79^{+0.18}_{-0.16}\,\si{pb} \\
& = 0.79 \pm 0.11\stat ^{+0.07}_{-0.06}\,(\textrm{exp. syst.}) ^{+0.12}_{-0.09}\,(\textrm{sig. theo.}) \pm 0.05\,(\textrm{bkg. theo.})\,\si{pb} \,, \\
\end{split}
\end{equation*}
in agreement with their SM expectations of $10.4 \pm 0.6\,\si{pb}$ and $0.81 \pm 0.02\,\si{pb}$, respectively. Figure~\ref{fig:2D_contours} shows 68\% and 95\% confidence level (CL) two-dimensional contours of $\XSVBF\times\BRHWW$ versus $\XSggF\times\BRHWW$. The result is compatible with the SM prediction at about 68\% CL, which is driven by the slightly high observed, yet consistent with the SM, value of $\XSggF\times\BRHWW$. As expected, the observed correlation is very small. This is due to the fact that there is very little \ggH contamination in the VBF SRs, and only a tiny EW \qqH contribution in the ggF SRs. Only the challenging VBF 1-jet SRs are noteworthy exceptions, which, however, do not lead to an overall large correlation in this measurement. The expected compositions of the SRs are shown in Figure~\ref{fig:STXS_comp} and previously mentioned in Section~\ref{sec:results_STXS} in the context of the STXS results. Furthermore, the DNNs provide an excellent separation of ggF from VBF in the VBF $\geq$2-jets SRs, as shown in Figure~\ref{fig:STXS_SRs_2j_VBF}. These aspects were further improved compared to the corresponding compositions of the STXS SRs shown in the previous publication~\cite{HIGG-2022-17}.

Table~\ref{tab:2POI_breakdown} shows the relative impact of the main uncertainties on the measured values for $\XSggF\times\BRHWW$ and $\XSVBF\times\BRHWW$. The measurement of the ggF cross-section is dominated by systematic uncertainties while in the VBF case statistical and systematic uncertainties are comparable. The measurement of $\XSggF\times\BRHWW$ is limited approximately equally by theory and experimental uncertainties, with the single largest contribution arising from the systematic uncertainty associated with the efficiency of the muon isolation criteria. The largest systematic uncertainties in the $\XSVBF\times\BRHWW$ measurement are theory uncertainties in the signal process; in particular, the comparisons between different event generators for the parton shower and for the matrix-element matching are dominant.

\FloatBarrier

\begin{table}[htbp]
\centering
\caption{Breakdown of the main contributions to the total uncertainty in $\XSggF\times\BRHWW$ and $\XSVBF\times\BRHWW$, relative to the measured value. The individual sources of systematic uncertainties are grouped together. The indentation in the first column indicates this grouping. `Experimental uncertainties (excl. MC stat.)' corresponds to the group of experimental uncertainties excluding MC statistical uncertainties. The sum in quadrature of the individual components may differ from the total uncertainty due to correlations between the components.}
\label{tab:2POI_breakdown}
\begin{tabular}{l | S[table-format=3.2] S[table-format=3.2]}
\toprule
\Gape[3mm][3mm]{Source} & $\frac{\Delta(\XSggF\times\BRHWW)}{\XSggF\times\BRHWW}$ [\%] & $\frac{\Delta(\XSVBF\times\BRHWW)}{\XSVBF\times\BRHWW}$ [\%] \\
\midrule
\bfseries Total & \bfseries 10 & \bfseries 22 \\
\midrule
\bfseries Data statistical uncertainties & \bfseries 5.2 & \bfseries 14 \\
\midrule
\bfseries Total systematic uncertainties & \bfseries 8.7 & \bfseries 16 \\
\bfseries \hspace*{4mm} Experimental uncertainties & \bfseries 7.2 & \bfseries 7.9 \\
\bfseries \hspace*{8mm} MC statistical uncertainties & \bfseries 2.2 & \bfseries 4.5 \\
\bfseries \hspace*{8mm} Experimental uncertainties (excl. MC stat.) & \bfseries 7.0 & \bfseries 6.8 \\
\hspace*{12mm} Electrons & 1.9 & 1.3 \\
\hspace*{12mm} Muons & 4.1 & 0.7 \\
\hspace*{12mm} Jet energy scale & 1.6 & 5.6 \\
\hspace*{12mm} Jet energy resolution & 1.1 & 2.2 \\
\hspace*{12mm} Flavour tagging & 2.7 & 1.0 \\
\hspace*{12mm} Missing transverse momentum & 0.6 & 0.2 \\
\hspace*{12mm} Pile-up & 2.2 & 0.9 \\
\hspace*{12mm} Luminosity & 0.9 & 1.0 \\
\hspace*{12mm} Misidentified leptons & 2.5 & 0.9 \\
\bfseries \hspace*{4mm} Theory uncertainties & \bfseries 5.0 & \bfseries 15 \\
\hspace*{8mm} ggF $H$ & 2.2 & 4.3 \\
\hspace*{8mm} VBF $H$ & 1.1 & 13 \\
\hspace*{8mm} Other Higgs & 0.5 & 0.2 \\
\hspace*{8mm} Top quark & 2.2 & 1.8 \\
\hspace*{8mm} \WW & 2.8 & 2.2 \\
\hspace*{8mm} Other \VV & 1.2 & 0.9 \\
\hspace*{8mm} \Ztt & 1.9 & 3.3 \\
\hspace*{8mm} \Zeemm & 0.5 & 1.2 \\
\bottomrule
\end{tabular}
\end{table}

\FloatBarrier

\subsection{CP-sensitive measurement}
\label{sec:results_DPhijj}

\begin{figure}[htbp]
\centering
\subfloat[]{
\includegraphics[width=0.98\textwidth]{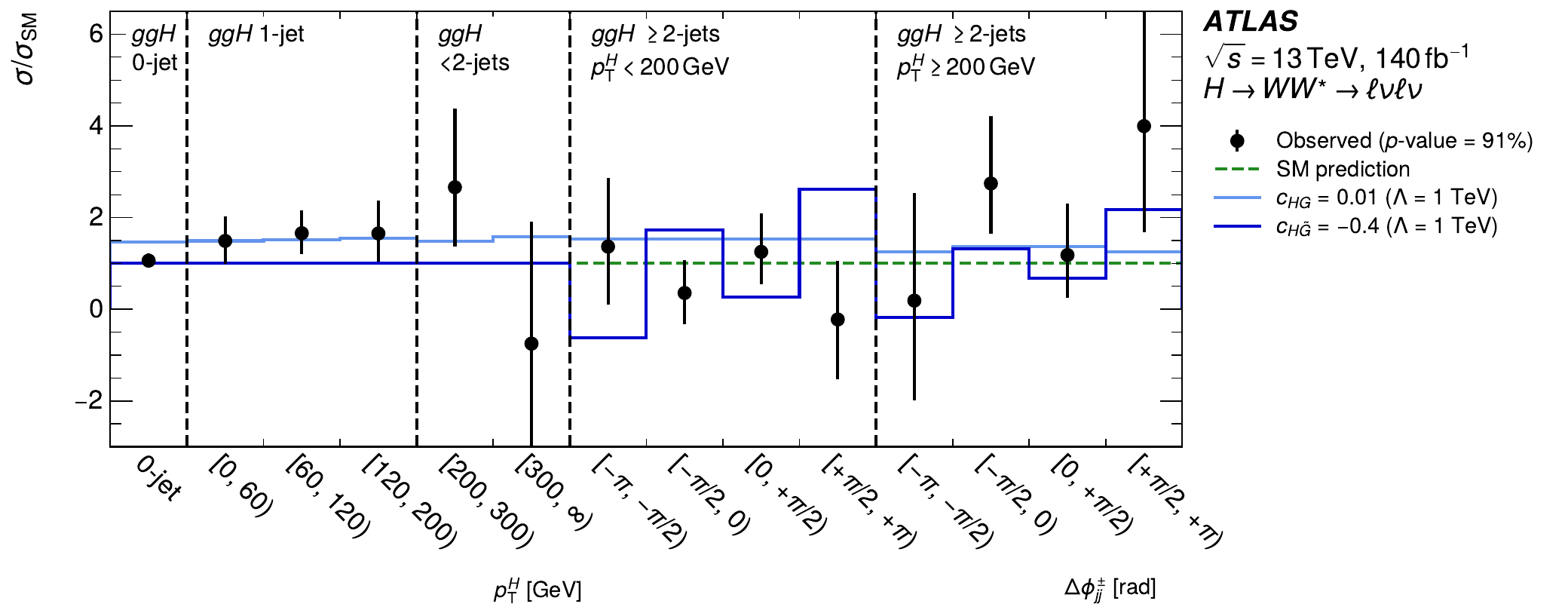}
\label{fig:DPhijj_POIs_ggF}
} \\
\subfloat[]{
\includegraphics[width=0.98\textwidth]{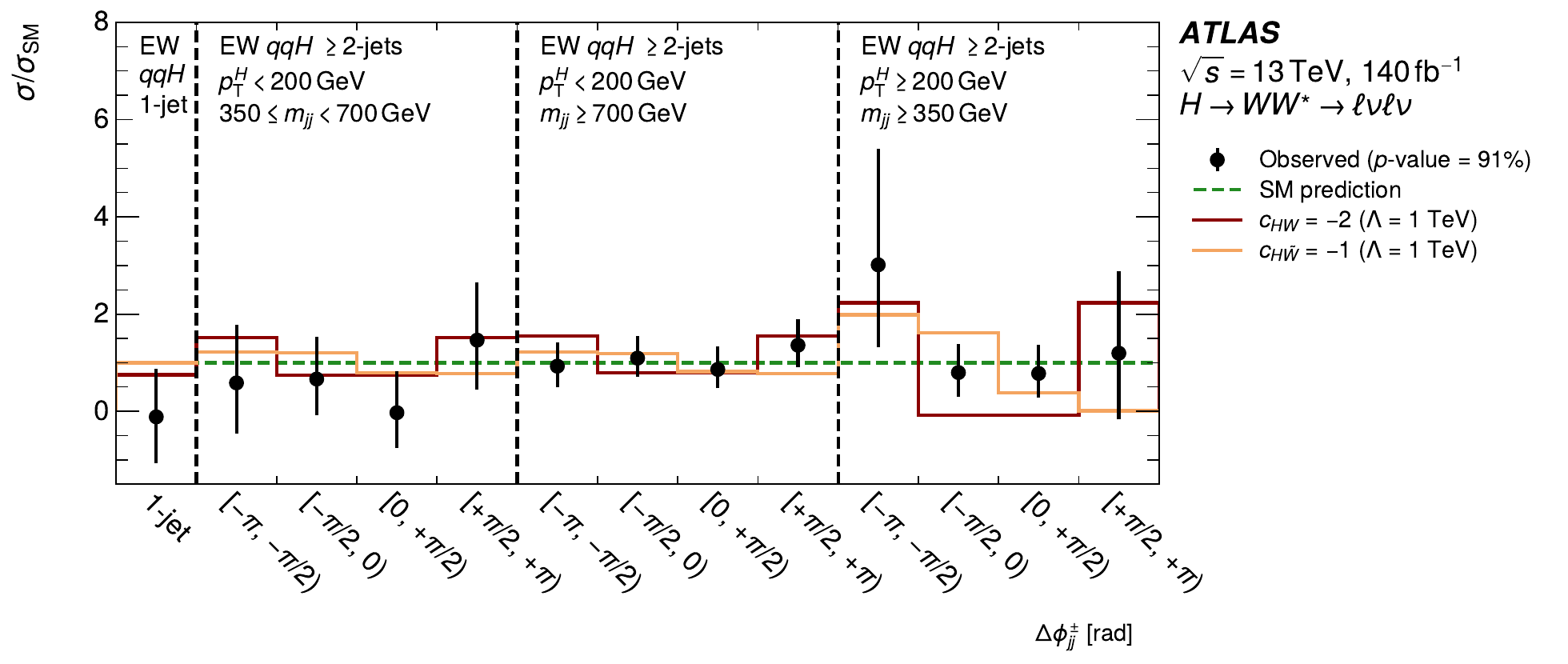}
\label{fig:DPhijj_POIs_VBF}
} \\
\caption{Best-fit values and uncertainties for the signal strengths measured in each of the \STXSCP categories, normalised to the corresponding SM predictions, for (a) \ggH and (b) EW \qqH production. The error bars show the total uncertainty in each signal strength. The $p$-value quoted in the legends collectively considers the data in (a) and (b). Also shown is the modification to the SM prediction induced by the relevant CP-even/odd \dimSix operators with a BSM physics scale $\Lambda = \qty{1}{\TeV}$. Only inference terms which are linear in the corresponding Wilson coefficients are considered. Each Wilson coefficient value corresponds to the 95\% CL limit furthest from zero for \cHW, \cHWtil, and \cHG, as obtained by the statistical interpretation of data. The value of \cHGtil is chosen to be the secondary local minimum identified in its likelihood scan, which is particularly instructive for this Wilson coefficient. See Section~\ref{sec:interpretation} for the definitions of the Wilson coefficients.}
\label{fig:DPhijj_POIs}
\end{figure}

In this section, the results of the \STXSCP measurement are shown and discussed. Figure~\ref{fig:DPhijj_POIs} shows a summary of the 27~measured signal strengths and uncertainties in each of the \STXSCP categories. The results are compatible with the SM predictions with a $p$-value of 91\%. While the impact of experimental and theory systematic uncertainties on the POIs remains of a similar magnitude to that of the nominal STXS measurement, the \STXSCP measurement is subject to increased statistical uncertainties due to the finer classification, as discussed in Section~\ref{sec:dphijj_cat}. The correlations among the POIs are shown in Figure~\ref{fig:DPhijj_corr} in Appendix~\ref{app:results}. The estimated background contribution aligns with the STXS measurement as the fit model uses nearly the same background estimation strategy, as described at the end of Section~\ref{sec:control_regions}. Additionally, the post-fit values of the background normalisation factors, shown in Figure~\ref{fig:normalization_factors} of Appendix~\ref{app:results}, are consistent with unity and consistent with the corresponding results from the STXS measurement discussed in the previous section.

Also shown in Figure~\ref{fig:DPhijj_POIs} are the expected BSM effects arising from CP-conserving (\cHG, \cHW) and CP-violating (\cHGtil, \cHWtil) anomalous couplings, as formulated within the SMEFT framework (see their definitions in Section~\ref{sec:interpretation}). This measurement demonstrates significant potential for constraining these effects, with the results from such an interpretation provided in Section~\ref{sec:cp_interp}. The compatibility of the observed data with a CP-conserving scenario was tested by checking the consistency of the data with totally symmetric \DPhijjsgn distributions (i.e., correlating the signal between the $[-\pi, -\pi/2)$ and $[+\pi/2, +\pi)\,\si{rad}$ bins and between the $[-\pi/2, 0)$ and $[0, +\pi/2)\,\si{rad}$ bins) in all SRs, resulting in a $p$-value of 75\%.

\FloatBarrier


\section{Effective field theory interpretations of results}
\label{sec:interpretation}

Effective field theory interpretations in CP-conserving and CP-violating scenarios are performed in the context of SMEFT~\cite{Brivio:2017vri, ATL-PHYS-PUB-2019-042} for the measurements presented in Sections~\ref{sec:results_STXS} and \ref{sec:results_DPhijj}, respectively. For both interpretations, the signal strength POIs are parameterised in terms of the modifications induced by selected \dimSix operators and their corresponding Wilson coefficients.

For the modification of the SM scattering amplitudes, only terms that are linear in a Wilson coefficient are considered by this measurement, corresponding to the interference between the SM and a \dimSix operator. Terms that are quadratic in a Wilson coefficient, corresponding to the interference between two \dimSix operators, are not considered. It is expected that quadratic effects will have a mild impact on the interpretation of the results presented in this paper; such studies were performed and reported in Refs.~\cite{HIGG-2020-25} and \cite{HIGG-2022-17} on the same set of Wilson coefficients. Furthermore, constraints obtained using only a linear parameterisation are generally looser and therefore more conservative than those that include quadratic terms, as demonstrated by these previous results.

The following sections describe the formulation and derivation of the SMEFT parameterisation (Section~\ref{sec:parameterisation}), the results of the SMEFT interpretation of the STXS measurement in a CP-conserving scenario (Section~\ref{sec:smeft_interp}), and the results of the SMEFT interpretation of the \STXSCP measurement in a CP-violating scenario (Section~\ref{sec:cp_interp}).

\subsection{SMEFT parameterisation}
\label{sec:parameterisation}

By introducing \dimSix operators, the expected Higgs boson production cross-sections ($\XS_i \,\forall\, i=1,\ldots,N$ where $N$ is the number of particle-level categories), the Higgs boson partial width (\GHtoWW), and the Higgs boson total width (\GHtot) are scaled by the linear functions of the Wilson coefficients:
\begin{equation*}
\begin{split}
& \XS_i \to \XS_i \times \left(1 + \sum_{j=1}^M A^{\XS_i}_j \times c_j\right) \,\forall\, i=1,\ldots,N \,, \\
& \GHtoWW \to \GHtoWW \times \left(1 + \sum_{j=1}^M A^{\GHtoWW}_j \times c_j\right) \,, \\
& \GHtot \to \GHtot \times \left(1+\sum_{j=1}^M A^{\GHtot}_j \times c_j\right) \,,
\end{split}
\end{equation*}
where $M$ is the number of Wilson coefficients considered and $c_j$ is the Wilson coefficient for \dimSix operator~$j$. Additionally, $A^{\XS_i}_j$, $A^{\GHtoWW}_j$, and $A^{\GHtot}_j$ are the factors obtained from the matrix-element calculation for $\XS_i$, \GHtoWW, and \GHtot, respectively, using simulated event samples for a single \dimSix operator~$j$ at $c_j = 1$. Therefore, the full parameterisation becomes:
\begin{equation}
\mu_i \to \mu_i \times \frac{\left(1 + \sum_{j=1}^M (A^{\XS_i}_j + A^{\GHtoWW}_j) \times c_j \right)}{1 + \sum_{j=1}^M A^{\GHtot}_j \times c_j} \,\forall\, i=1,\ldots,N \,,
\label{eqn:smeft_param}
\end{equation}
where $\mu_i$ is signal strength for category~$i$ and is fixed to the SM prediction of 1. In doing so, the POIs of the fit model become the Wilson coefficients, $c_j \,\forall\, j=1,\ldots,M$.

For each \dimSix operator, the corresponding modifications to the Higgs boson production cross-sections, partial width, and total width are calculated following the same procedure as the one used in the ATLAS interpretation of combined Higgs boson measurements~\cite{HIGG-2022-17}; the modifications induced by these same operators on background processes are neglected. Dedicated \ggH and EW \qqH samples (including \VH contributions for the latter) were simulated using \MGNLO~\cite{Alwall:2014hca}, with the effects of \dimSix operators introduced via the \SMEFTatNLO~\cite{Degrande:2020evl} and \SMEFTsim~\cite{Brivio:2017btx, Brivio:2020onw} UFO models~\cite{Degrande:2011ua}. The operators are given in the Warsaw basis~\cite{Grzadkowski:2010es}. For loop-induced \ggH production, \SMEFTatNLO was used to account for top quark loop amplitudes at high \pt; in contrast, \SMEFTsim treats $\H\to\gg$ as an effective vertex. For EW \qqH production and the decay, \SMEFTsim was used. Both UFO models were implemented using top-flavour symmetry and the \mW-input scheme~\cite{Aguilar-Saavedra:2018ksv}, and a BSM physics scale $\Lambda = \qty{1}{\TeV}$ was used. The BSM predictions were implemented by reweighting the generated events~\cite{Mattelaer:2016gcx}, and the events were showered using \PYTHIA[8]. For each operator $j$, the impact is derived as the relative change in the predicted rate for each STXS category as shown in Eq.~\ref{eqn:smeft_param}, taking into account the operator's modification of the production cross-section, partial width, and total width as a function of $c_j$. Only operators that have a combined impact on the production cross-section and partial width of greater than 0.001 for at least one STXS category are considered. The EFT operator corresponding to \cHW (see Table~\ref{tab:smeft_defs}) has a non-negligible impact on the signal acceptance which originates from the \mll selections applied by this analysis, as was demonstrated in Ref.~\cite{HIGG-2022-17}. This is accounted for by a correction factor applied to the \GHtoWW partial width parameterisation coefficient.

\paragraph{Fully-linearised parameterisation}

As long as the EFT effects on the POIs are small, the parameterisation in Eq.~\ref{eqn:smeft_param} may be Taylor-expanded to leading-order in the Wilson coefficients, resulting in the `fully-linearised' parameterisation:
\begin{equation*}
\mu_i \to \mu_i \times \left(1 + \sum_{j=1}^M P_{ij} \times c_j \right) \,\forall\, i=1,\ldots,N \,,
\label{eqn:smeft_param_lin}
\end{equation*}
where $P_{ij}$ is the (fully-linearised) impact of \dimSix operator $j$ at $c_j = 1$ in STXS category $i$. The elements of the $N \times M$ parameterisation matrix $\mathbf{P}$ are given by:
\begin{equation}
[\mathbf{P}]_{ij} \equiv P_{ij} = A^{\XS_i}_j + A^{\GHtoWW}_j - A^{\GHtot}_j \,,
\label{eqn:smeft_param_mat}
\end{equation}
where $[\mathbf{P}]_{ij}$ corresponds to the element in $i$-th row and $j$-th column of $\mathbf{P}$.

\paragraph{Principal component analysis}

It is expected that the various BSM couplings may have similar impacts on the signal strengths in STXS categories, resulting in highly correlated Wilson coefficients when they are fitted to the data simultaneously. This may also lead to poor convergence of the fit. Therefore, a rotation of the Warsaw basis is performed to identify the directions, i.e.\ the set of mutually orthogonal linear combinations of Wilson coefficients for which the analysis has the greatest sensitivity, a procedure known as principal component analysis (PCA).

The covariance matrix of the Wilson coefficients, $\mathbf{C}_\textrm{SMEFT}$, is obtained by transforming the expected covariance matrix of the STXS signal strengths, $\mathbf{C}_\textrm{STXS}$:
\begin{equation*}
\mathbf{C}_\textrm{SMEFT}^{-1} = \mathbf{P}^\top \mathbf{C}_\textrm{STXS}^{-1} \mathbf{P} \,,
\end{equation*}
where $\mathbf{P}$ is the fully-linearised parameterisation matrix defined in Eq.~\ref{eqn:smeft_param_mat}. An eigenvector decomposition of $\mathbf{C}_\textrm{SMEFT}^{-1}$ is performed, yielding a matrix of eigenvectors $\mathbf{E}$ where each row corresponds to an eigenvector and each column corresponds to the component of each eigenvector along a (physical) direction in the Warsaw basis. The parameterisation matrix in this rotated basis $\mathbf{P}^\prime$ is given by:
\begin{equation*}
\mathbf{P}^\prime = \mathbf{P} \mathbf{E}^{-1} \,.
\end{equation*}
Only the directions with eigenvalues greater than 0.01 are kept, corresponding to an expected uncertainty in the associated Wilson coefficient for each eigenvector of less than $1 / \sqrt{0.01} = 10$ at the 68\% CL. After obtaining the most sensitive directions, the parameterisation is implemented as in Eq.~\ref{eqn:smeft_param} but with the Wilson coefficients in the Warsaw basis replaced by the rotated basis.

\subsection{SMEFT interpretation in a CP-conserving scenario}
\label{sec:smeft_interp}

\begin{figure}[t]
\centering
\includegraphics[width=\textwidth]{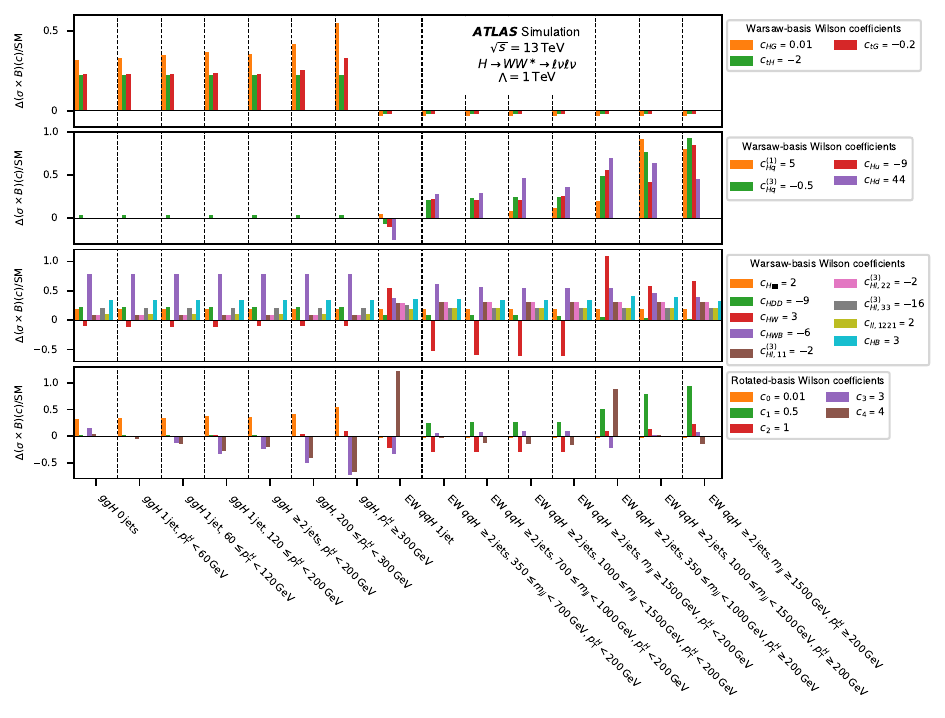}
\caption{Fractional variation of each of the STXS POIs measured relative to their respective SM expectation for the values of the Wilson coefficients shown in the legend. The operators are grouped according to those affecting (top row) primarily \ggH production, (second row) primarily EW \qqH production, and (third row) all categories equally. These three rows present the fractional variations for operators in the Warsaw basis; the bottom row presents the fractional variations for operators in the rotated basis, following the PCA described in Section~\ref{sec:parameterisation}. The fractional variations are independently evaluated using Eq.~\ref{eqn:smeft_param}, with each Wilson coefficient set to its approximate 95\% CL interval bound as reported by this measurement.}
\label{fig:smeft_impacts}
\end{figure}

Following the prescription for selecting the Warsaw-basis operators relevant to the STXS measurement as described in the previous section, there are sixteen \dimSix operators considered by the SMEFT interpretation. They are shown in Table~\ref{tab:smeft_defs}, and their expected impact on the STXS signal strengths is visualised graphically in Figure~\ref{fig:smeft_impacts}. When fitting each Wilson coefficient independently (i.e., with all others fixed to their SM expectation of 0), consistency with the SM expectation is observed for each of them, as shown in Figure~\ref{fig:smeft_1poi_fits} of Appendix~\ref{app:EFT}.

\begin{figure}[b]
\centering
\includegraphics[width=\textwidth]{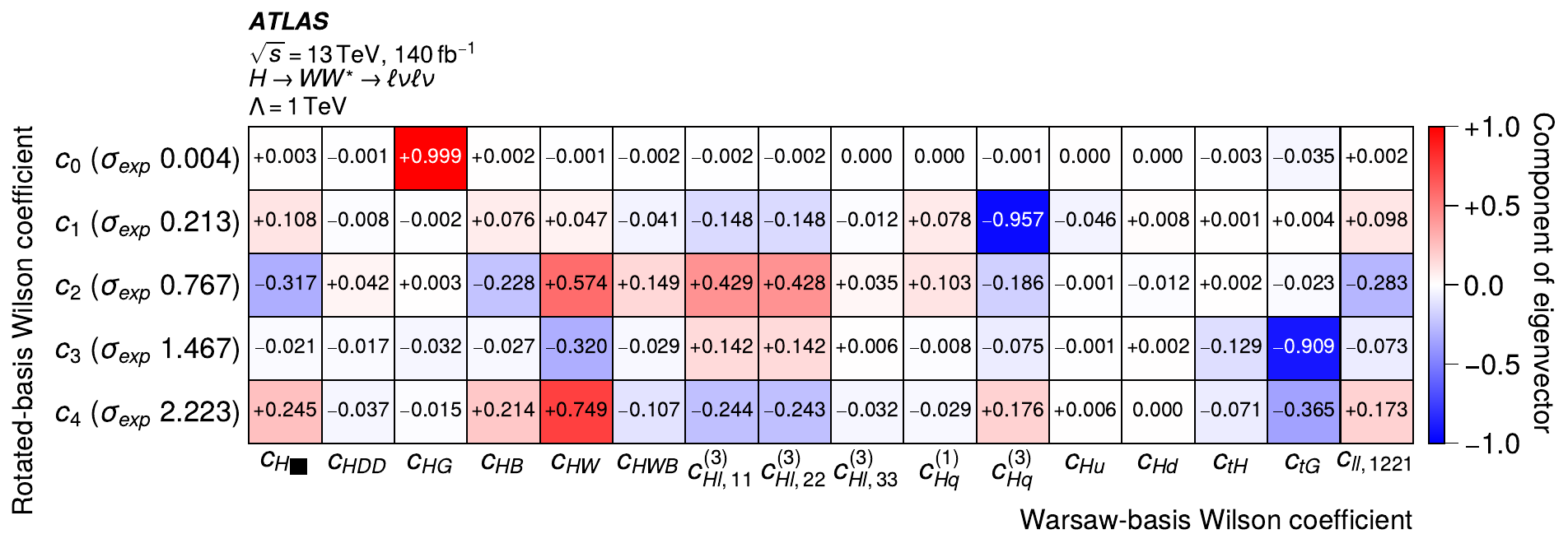}
\caption{The composition of the (orthonormal) eigenvectors defining the rotated basis in terms of operators in the Warsaw basis, labelled in terms of their corresponding Wilson coefficients. Also shown is the expected 68\% CL error for each Wilson coefficient in the rotated basis, $\sigma_\textrm{exp}$, calculated as $1/\sqrt{\textrm{EV}}$ where $\textrm{EV}$ is the eigenvalue for the corresponding eigenvector.}
\label{fig:smeft_eigenvectors}
\end{figure}

Subsequently and also following the prescription described in the previous section, PCA is performed using the expected covariance matrix of STXS POIs, yielding a reduced set of five operators in a rotated basis. The eigenvectors defining this basis are shown in Figure~\ref{fig:smeft_eigenvectors} with the corresponding Wilson coefficients labelled as $c_0$, $c_1$, $c_2$, $c_3$, and $c_4$. The operators corresponding to $c_0$, $c_1$, $c_3$, and $c_4$ primarily point along the operators corresponding to \cHG, \cHqthree, \ctG, and \cHW, respectively, while $c_2$ exhibits a mixture of several physical operators. The impacts for these operators are also visualised in Figure~\ref{fig:smeft_impacts}.

\begin{figure}[htbp]
\centering
\includegraphics[width=0.75\textwidth]{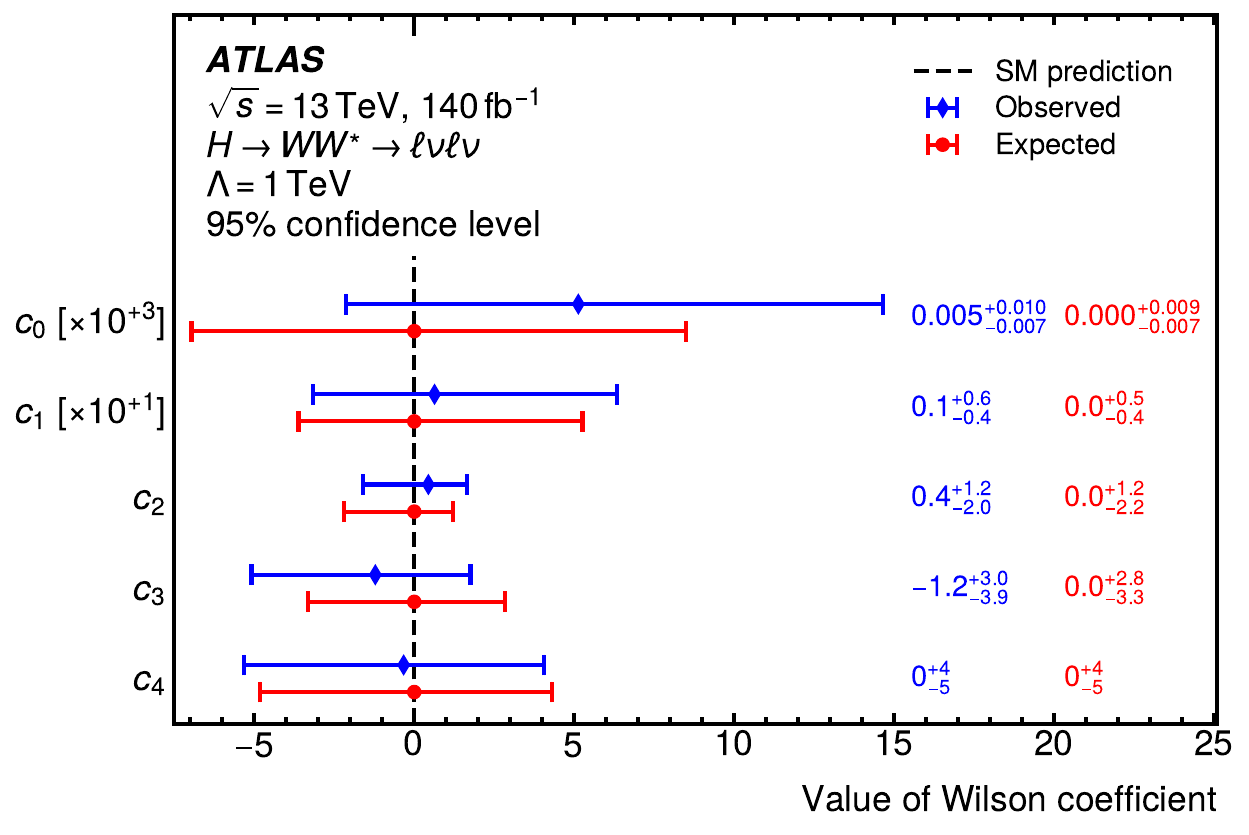}
\caption{Best-fit values and uncertainties for the Wilson coefficients of the \dimSix operators considered by the SMEFT interpretation of the STXS measurement in a CP-conserving scenario. The \dimSix operators are given in the rotated basis, and all Wilson coefficients are fitted simultaneously. Both the observed (diamond) and expected (circle) results are shown. The best-fit values and uncertainties for the observed (expected) results are shown outside (inside) of the parentheses on each line. The error bars and quoted errors correspond to the total 95\% CL interval on each Wilson coefficient.}
\label{fig:smeft_pca_fit}
\end{figure}

Equation~\ref{eqn:smeft_param} is re-parameterised in terms of $c_0$, $c_1$, $c_2$, $c_3$, and $c_4$ using the eigenvectors calculated above, and a final fit is performed where these Wilson coefficients are simultaneously fitted. The results of this fit are shown in Figure~\ref{fig:smeft_pca_fit}. Each Wilson coefficient is consistent with the SM expectation. The observed values of the correlations between different measured Wilson coefficient are generally small, $\leq15\%$ in absolute value, but non-zero, as shown in Figure~\ref{fig:EFT_correlations} of Appendix~\ref{app:EFT}. The smallness is due to the fact that eigenvectors are fitted, while the non-zero size of the observed correlations in the fit to data is because the directions of the eigenvectors were derived using the expected STXS covariance matrix; the correlations between the expected values of different Wilson coefficients are all less than 1\%.

\begin{table}[htbp]
\centering
\caption{Definitions for relevant \dimSix operators impacting Higgs boson production and decay. The operators are given in the Warsaw basis, and the operator and coefficient notation follows the convention of Ref.~\cite{Brivio:2017btx}. The horizontal lines delineate groups of operators affecting (a) primarily \ggH production, (b) primarily EW \qqH production, and (c) a mixture of both production modes and \hww decays, in that order. Also shown are the effects of each operator or example diagrams that are affected via the modification of SM vertices (open dots) or insertion of BSM vertices (filled dots).}
\label{tab:smeft_defs}
\resizebox{0.78\textwidth}{!}{
\begin{tabular}{c c c}
\toprule
Wilson coefficient & Operator & Effect or affected diagram \\
\midrule
\cHG & \opHG & \diagHG \\
\ctG & \deftG & \diagtG \\
\ctH & \deftH & \diagtH \\
\midrule
\cHqone & \opHqOne & \diagHqone \\
\cHqthree & \opHqThree & \diagHqthree \\
\cHu & \opHu & \diagHu \\
\cHd & \opHd & \diagHd \\
\midrule
\cHbox & \opHbox & Rescales all Higgs boson couplings \\
\cHDD & \opHDD & \diagHDD \\
\cHB & \opHB & \diagHB \\
\cHW & \opHW & \diagHW \\
\cHWB & \opHWB & \diagHWB \\
\cHlthreeone & \defHlThreeOne & \multirow{3}{*}{\diagHlthreei $\forall\,i = 1, 2, 3$} \\
\cHlthreetwo & \defHlThreeTwo \\
\cHlthreethree & \defHlThreeThree \\
\cllonetwoone & \defllOneTwoOne & Rescales Fermi constant \\
\bottomrule
\end{tabular}
}
\end{table}

\FloatBarrier

\FloatBarrier

\subsection{SMEFT interpretation in a CP-violating scenario}
\label{sec:cp_interp}

The SMEFT interpretation of the \STXSCP measurement follows a similar approach to that of the STXS measurement, with the key difference being that only the shape of the \DPhijjsgn distribution is used to investigate potential CP violation. The interpretation only considers four EFT operators: the two CP-odd operators to which the measurement has sensitivity and their CP-even counterparts. Two of the operators, \opHG (with Wilson coefficient \cHG) and \defHGtil (with Wilson coefficient \cHGtil), induce CP-even and CP-odd modifications to \ggH production, respectively. The other two operators, \opHW (with Wilson coefficient \cHW) and \defHWtil (with Wilson coefficient \cHWtil) induce CP-even and CP-odd modifications to EW \qqH production, respectively. The corresponding operators for the Wilson coefficients \cHWB and \cHB, along with their CP-odd counterparts, are not included as their influence on the shape of the \DPhijjsgn distribution is negligible. The Wilson coefficients for all other EFT operators are also set to 0 in the interpretation.

CP violation will manifest as a modification of the shape of the \DPhijjsgn spectrum~\cite{ATL-PHYS-PUB-2024-019}. To ensure that only shape information is used to constrain CP-violating effects from \dimSix operators, and therefore provide a genuine test of CP violation, each $\geq$2-jets \STXSCP category is scaled by a `signal normalisation factor' (signal NF), \sigNF, that coherently scales the normalisation of the same category across each of its \DPhijjsgn bins and is treated as a nuisance parameter. The parameterisation of the $\geq$2-jets categories becomes:
\begin{equation*}
\mu_{ij} \to \mu_{ij} \times \sigNF_i \times \frac{\left(1 + \sum_k (A^{\XS_{ij}}_k + A^{\GHtoWW}_k) \times c_k \right)}{1 + \sum_k A^{\GHtot}_k \times c_k} \,,
\label{eqn:smeft_param_sigNF}
\end{equation*}
where $i$ corresponds to each of the $\geq$2-jets categories:
\begin{itemize}
\item \ggH $\geq$2-jets, $\pTH < \qty{200}{\GeV}$;
\item \ggH $\geq$2-jets, $\pTH \geq \qty{200}{\GeV}$;
\item EW \qqH $\geq$2-jets, $\pTH < \qty{200}{\GeV}$, $350 \leq \mjj < \qty{700}{\GeV}$;
\item EW \qqH $\geq$2-jets, $\pTH < \qty{200}{\GeV}$, $\mjj \geq \qty{700}{\GeV}$;
\item EW \qqH $\geq$2-jets, $\pTH \geq \qty{200}{\GeV}$, $\mjj \geq \qty{350}{\GeV}$,
\end{itemize}
$j$ corresponds to the index of the \DPhijjsgn bin, $j \in [1, 2, 3, 4]$, and $k$ corresponds to the \dimSix operator, $k \in [HG, H\tilde{G}, HW, H\tilde{W}]$. Following the parameterisation, each $\mu_{ij}$ is fixed to 1 and each $\sigNF_i$ is floated. All of the 0- and 1-jet categories are parameterised according to Eq.~\ref{eqn:smeft_param} and help to constrain the effects of the CP-even operators.

\begin{figure}[htbp]
\centering
\subfloat[]{
\includegraphics[width=0.65\textwidth]{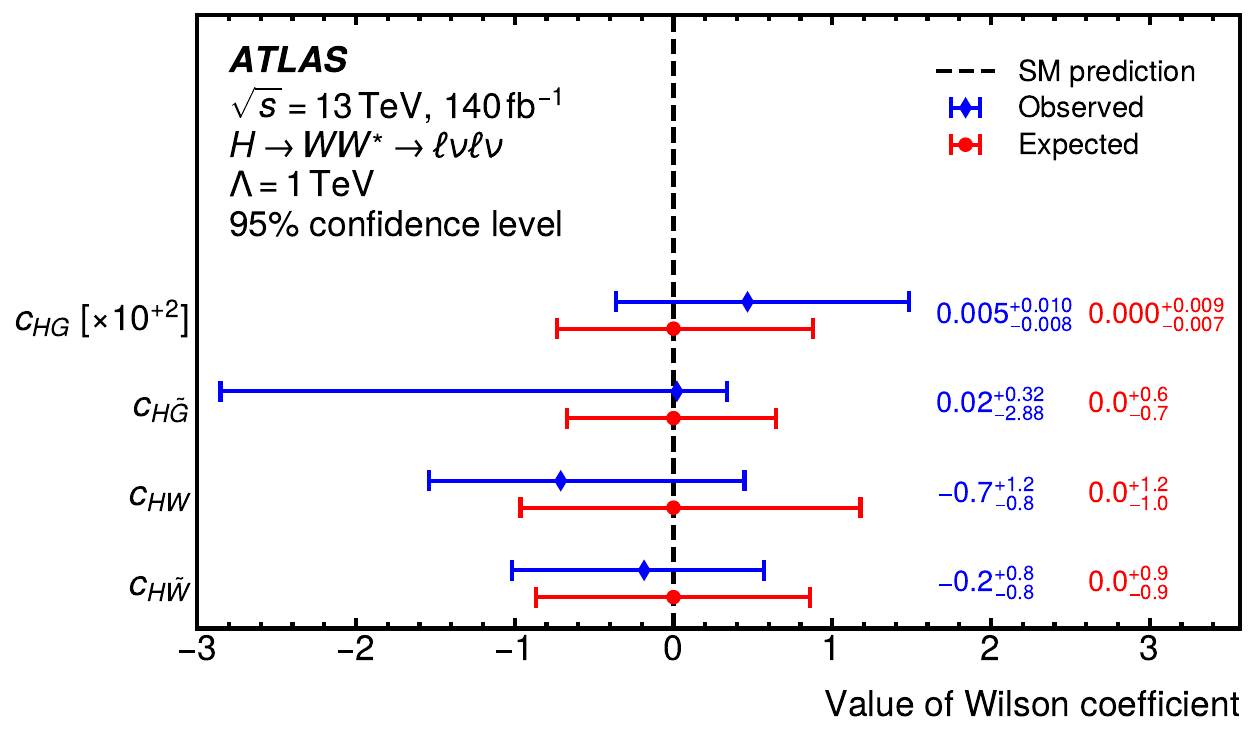}
\label{fig:cp_fit_pois}
}
\subfloat[]{
\raisebox{9mm}{\includegraphics[width=0.34\textwidth]{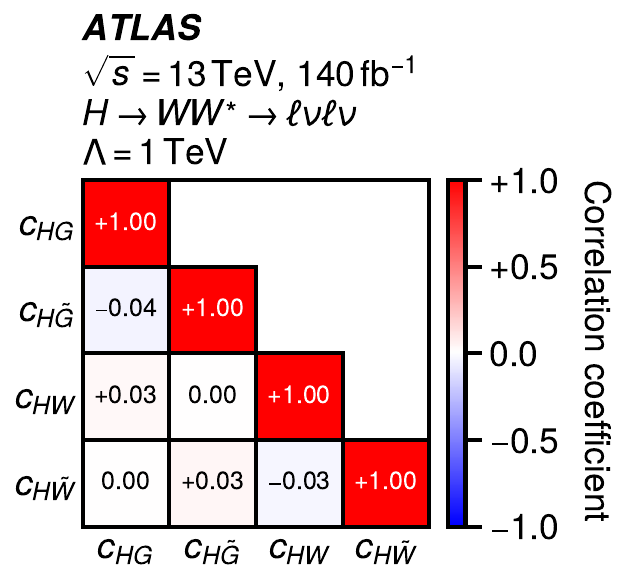}}
\label{fig:cp_fit_corr}
} \\
\caption{(a) Best-fit values and uncertainties and (b) correlations for the Wilson coefficients of the \dimSix operators considered by the SMEFT interpretation of the \STXSCP measurement in a CP-violating scenario. The \dimSix operators are given in the Warsaw basis, and all Wilson coefficients are fitted simultaneously. In (a), both the observed (diamond) and expected (circle) results are shown. The best-fit values and uncertainties for the observed (expected) results are given by the left (right) numbers on each line. The error bars and quoted errors correspond to the total 95\% CL interval in each Wilson coefficient.}
\label{fig:cp_fit}
\end{figure}

All four Wilson coefficients are obtained from a simultaneous fit to data, the results of which are shown in Figure~\ref{fig:cp_fit_pois}. Each Wilson coefficient is consistent with the SM expectation within its 95\% CL interval. An additional visualization of these observed results is shown in Figure~\ref{fig:Lambda_limits_CPmixed} in Appendix~\ref{app:EFT}, where instead of assuming an energy scale $\Lambda = \qty{1}{\TeV}$, three different fixed values for the Wilson coefficients are assumed, and a lower limit on the energy scale is extracted. This results in lower limits of approximately $\qty{100}{\TeV}$ for the operator corresponding to \cHG and approximately $\qty{10}{\TeV}$ for each of the operators corresponding to \cHGtil, \cHW, and \cHWtil, using the most aggressive scenario for each Wilson coefficient. Despite a non-negligible cross-contamination of \ggH and EW \qqH in their respective ggF and VBF SRs, the correlations between different observed values of the Wilson coefficient are all less than 5\% in absolute value, as can be seen in Figure~\ref{fig:cp_fit_corr}. This may stem from the distinct impact between \cHWtil and \cHGtil on the \DPhijjsgn distribution, as shown in Figure~\ref{fig:DPhijj_POIs}.

\begin{figure}[htbp]
\centering
\subfloat[]{
\includegraphics[width=0.48\textwidth]{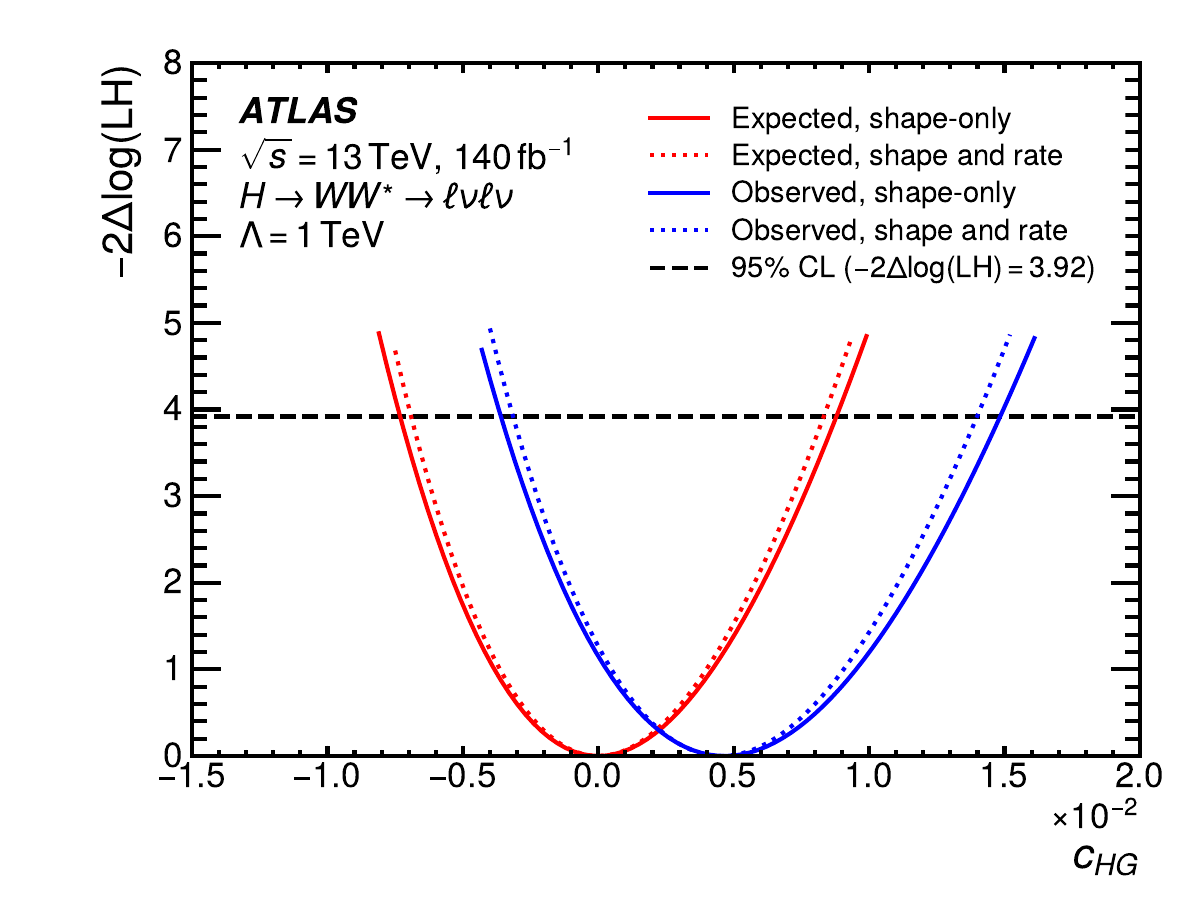}
\label{fig:CP_LLH_cHG}
}
\hfill
\subfloat[]{
\includegraphics[width=0.48\textwidth]{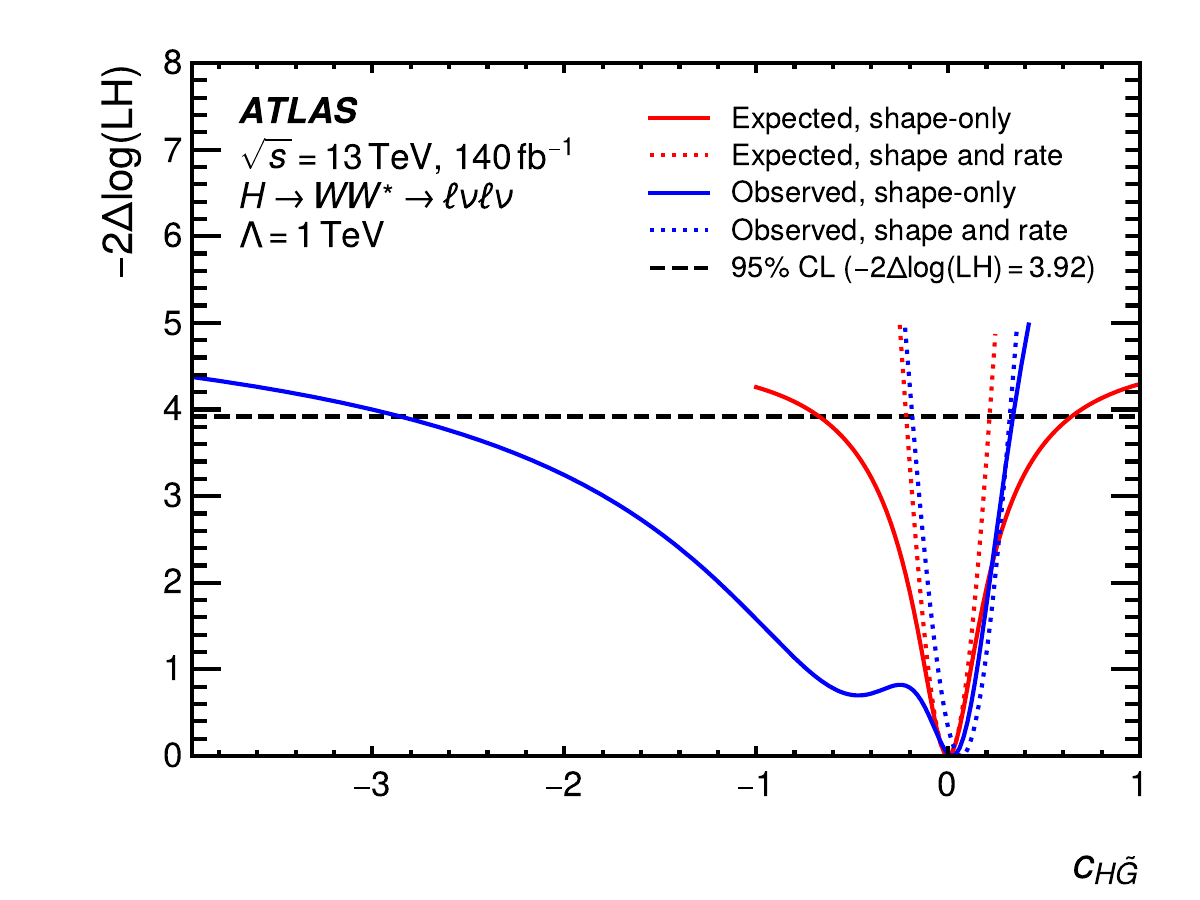}
\label{fig:CP_LLH_cHGtil}
} \\
\subfloat[]{
\includegraphics[width=0.48\textwidth]{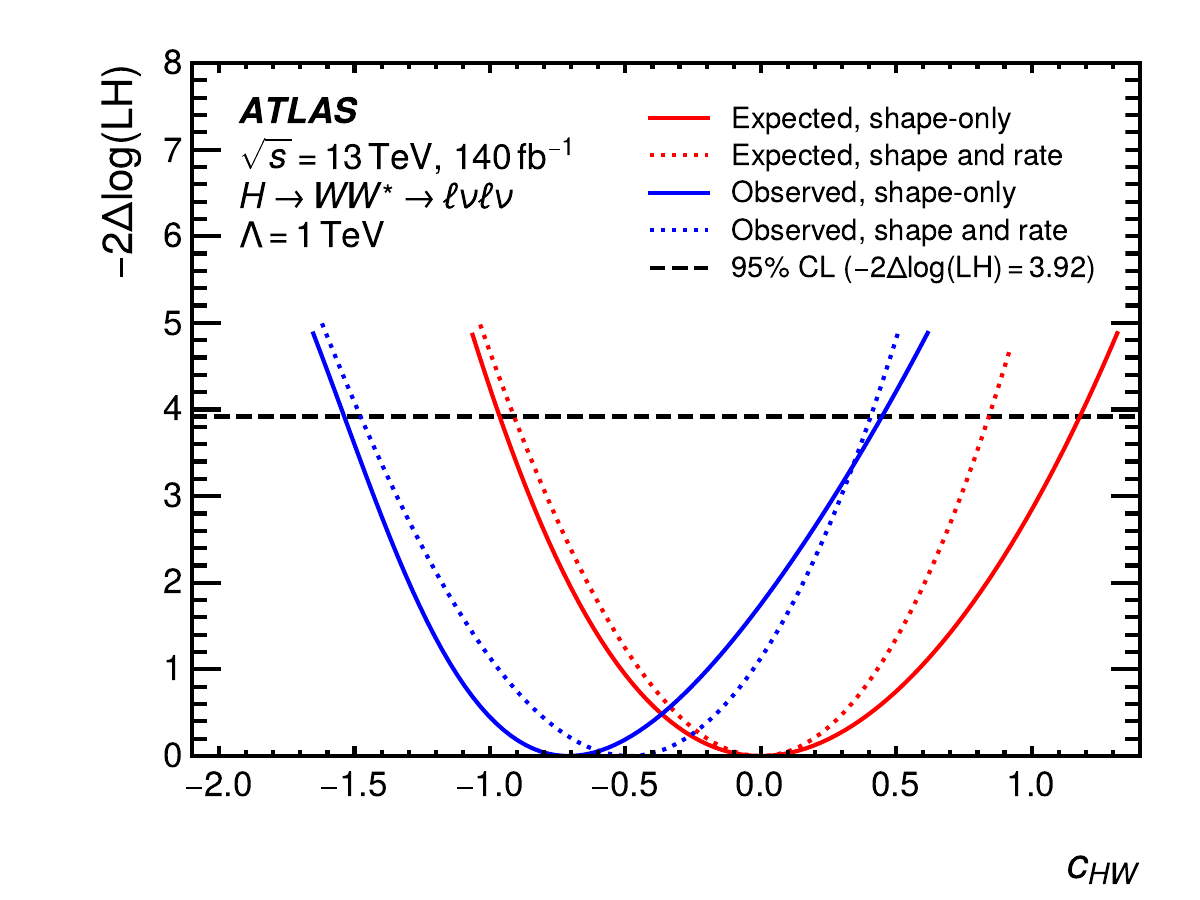}
\label{fig:CP_LLH_cHW}
}
\hfill
\subfloat[]{
\includegraphics[width=0.48\textwidth]{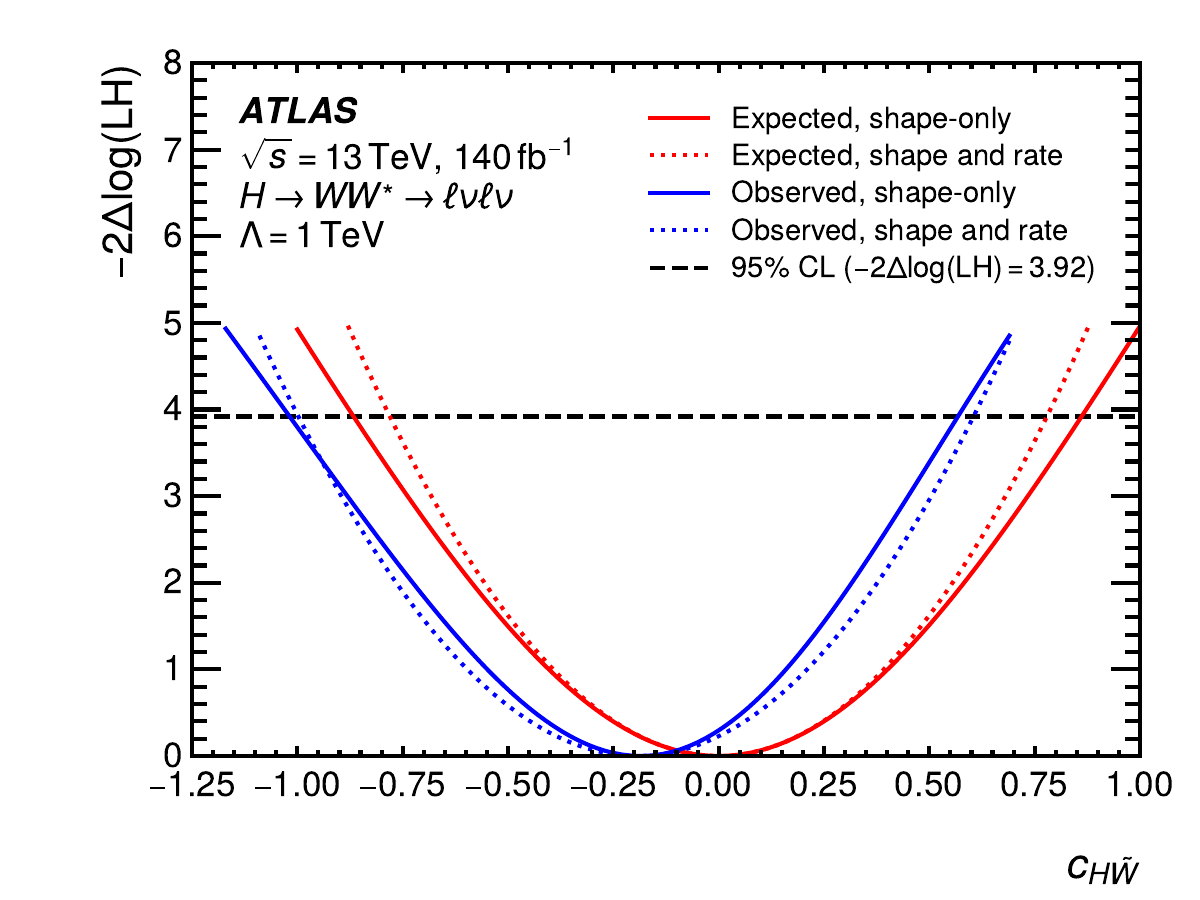}
\label{fig:CP_LLH_cHWtil}
} \\
\caption{Negative log-likelihood scans for the Wilson coefficients of the \dimSix operators considered by the SMEFT interpretation of the \STXSCP measurement in a CP-violating scenario. Shown are the expected and observed results for the fits utilising shape-only information (i.e., with signal NFs floated) and shape and rate information (i.e., with signal NFs fixed); the shape-only fits are the primary results of this measurement, as they constitute a genuine test of CP violation. The 95\% CL is indicated by the horizontal dashed line.}
\label{fig:CP_LLH}
\end{figure}

The asymmetry between the lower and upper uncertainties for the observed value of \cHGtil is best understood by studying the negative log-likelihood scan for \cHGtil, which is shown alongside likelihood scans for the other three Wilson coefficients in Figure~\ref{fig:CP_LLH}. Floating the signal NFs (which is the default in this analysis in order to remove the effects of rate changes from the measurement of the Wilson coefficients) significantly broadens the likelihood curve for \cHGtil, while such a significant effect is not observed for any of the other three Wilson coefficients. The operator associated with \cHGtil induces large absolute yield variations in \DPhijjsgn bins; for example, at $\cHGtil = 0.7$, the ggF signal yield varies by $+375\%$ in a given \DPhijjsgn bin and $-375\%$ in the symmetric \DPhijjsgn bin. To mitigate these large and asymmetric yield variations, the corresponding signal NFs are driven to small values, which in turn requires the fit to tune the nuisance parameters for background theory systematics (and primarily those attributed to \WW and \ttbar) in order to maintain adequate agreement between the model and data. Because these background theory systematic errors are large in the ggF-enriched SRs, the fit has significant freedom to adjust these nuisance parameters before the agreement degrades, resulting in the broadening in the likelihood scan. This behaviour is not possible when the signal NFs or the background theory systematic errors are fixed. Additionally, there is a second local minimum at $\cHGtil \approx -0.4$, further contributing to the asymmetry in the observed \cHGtil limits. As shown in Figure~\ref{fig:DPhijj_POIs_ggF}, the measured \ggH $\geq$2-jets \STXSCP POIs exhibit opposite trends below and above $\pTH = \qty{200}{\GeV}$. The global minimum is at $\cHGtil \approx 0.0$, which aligns with the POIs measured in the $\pTH < \qty{200}{\GeV}$ categories. Figure~\ref{fig:DPhijj_POIs_ggF} also illustrates the EFT effect at the second minimum of \cHGtil, which can modify the SM prediction for the ggF signal to better align with data in the $\pTH \geq \qty{200}{\GeV}$ regions.

Altogether, this interpretation demonstrates the potential to simultaneously constrain CP-even and CP-odd anomalous $Hgg$ and $HVV$ couplings in \hww decays using an STXS-based approach incorporating the \DPhijjsgn variable. The precise measurements of the \ggH 0- and 1-jet categories lead to the same stringent constraint on \cHG as the one imposed by the STXS measurement when other anomalous couplings are neglected. In contrast, incorporating \DPhijjsgn binning significantly enhances the observed 95\% confidence limits on \cHW, which goes from $[-2.3, 2.2]$ in a single-parameter fit using the STXS scheme (shown in Figure~\ref{fig:smeft_1poi_fits} of Appendix~\ref{app:EFT}) to $[-1.5, 0.4]$ in the four-parameter fit using the \STXSCP scheme. Compared to the results on the VBF differential cross-sections
in the \hww channel published by ATLAS~\cite{HIGG-2020-25}, the additional splitting in \pTH enhances the sensitivity to \cHWtil, improving its constraints by over 50\%. Additionally, the constraint for \cHW (\cHWtil) is similar to (slightly looser than) that obtained in the \Htautau channel~\cite{HIGG-2022-07}. Both measurements are statistically limited, complementary, and consistent with the SM.

\clearpage

\section{Conclusions}
\label{sec:conclusion}

This paper presents the measurement of Higgs boson production by gluon--gluon fusion and vector-boson fusion in the \hwwlvlv decay channel. The measurement is based on a data sample of \pp collisions with an integrated luminosity of \qty{140}{\ifb} recorded with the ATLAS detector at the LHC in 2015--2018 at a centre-of-mass energy of \qty{13}{\TeV}. The ggF and VBF cross-sections multiplied by the \hww branching ratio are simultaneously measured to be $12.4^{+1.3}_{-1.2}\,\si{pb}$ and $0.79^{+0.18}_{-0.16}\,\si{pb}$, in agreement with the SM predictions of $10.4 \pm 0.6\,\si{pb}$ and $0.81 \pm 0.02\,\si{pb}$, respectively. Higgs boson production in the \hwwlvlv decay channel is further characterised through STXS measurements in a total of 15~categories. The measurements of the inclusive ggF and VBF cross-sections are improved over a previous measurement with the same data sample by 11\%. The improvements amount to up to 36\% for the STXS categories, where the gains are largest for categories with large \pTH. The STXS measurement along with an STXS-like \dphijjsgn measurement are interpreted in the context of an effective field theory, with and without CP-violating operators. All measurements and their interpretations are compatible with the SM predictions.


%
\section*{Acknowledgements}
%

%
%

%
%

We thank CERN for the very successful operation of the LHC and its injectors, as well as the support staff at
CERN and at our institutions worldwide without whom ATLAS could not be operated efficiently.

The crucial computing support from all WLCG partners is acknowledged gratefully, in particular from CERN, the ATLAS Tier-1 facilities at TRIUMF/SFU (Canada), NDGF (Denmark, Norway, Sweden), CC-IN2P3 (France), KIT/GridKA (Germany), INFN-CNAF (Italy), NL-T1 (Netherlands), PIC (Spain), RAL (UK) and BNL (USA), the Tier-2 facilities worldwide and large non-WLCG resource providers. Major contributors of computing resources are listed in Ref.~\cite{ATL-SOFT-PUB-2025-001}.

We gratefully acknowledge the support of ANPCyT, Argentina; YerPhI, Armenia; ARC, Australia; BMWFW and FWF, Austria; ANAS, Azerbaijan; CNPq and FAPESP, Brazil; NSERC, NRC and CFI, Canada; CERN; ANID, Chile; CAS, MOST and NSFC, China; Minciencias, Colombia; MEYS CR, Czech Republic; DNRF and DNSRC, Denmark; IN2P3-CNRS and CEA-DRF/IRFU, France; SRNSFG, Georgia; BMFTR, HGF and MPG, Germany; GSRI, Greece; RGC and Hong Kong SAR, China; ICHEP and Academy of Sciences and Humanities, Israel; INFN, Italy; MEXT and JSPS, Japan; CNRST, Morocco; NWO, Netherlands; RCN, Norway; MNiSW, Poland; FCT, Portugal; MNE/IFA, Romania; MSTDI, Serbia; MSSR, Slovakia; ARIS and MVZI, Slovenia; DSI/NRF, South Africa; MICIU/AEI, Spain; SRC and Wallenberg Foundation, Sweden; SERI, SNSF and Cantons of Bern and Geneva, Switzerland; NSTC, Taipei; TENMAK, T\"urkiye; STFC/UKRI, United Kingdom; DOE and NSF, United States of America.

Individual groups and members have received support from BCKDF, CANARIE, CRC and DRAC, Canada; CERN-CZ, FORTE and PRIMUS, Czech Republic; COST, ERC, ERDF, Horizon 2020, ICSC-NextGenerationEU and Marie Sk{\l}odowska-Curie Actions, European Union; Investissements d'Avenir Labex, Investissements d'Avenir Idex and ANR, France; DFG and AvH Foundation, Germany; Herakleitos, Thales and Aristeia programmes co-financed by EU-ESF and the Greek NSRF, Greece; BSF-NSF and MINERVA, Israel; NCN and NAWA, Poland; La Caixa Banking Foundation, CERCA Programme Generalitat de Catalunya and PROMETEO and GenT Programmes Generalitat Valenciana, Spain; G\"{o}ran Gustafssons Stiftelse, Sweden; The Royal Society and Leverhulme Trust, United Kingdom.

In addition, individual members wish to acknowledge support from Armenia: Yerevan Physics Institute (FAPERJ); CERN: European Organization for Nuclear Research (CERN DOCT); Chile: Agencia Nacional de Investigaci\'on y Desarrollo (FONDECYT 1230812, FONDECYT 1240864, Fondecyt 3240661); China: Chinese Ministry of Science and Technology (MOST-2023YFA1605700, MOST-2023YFA1609300), National Natural Science Foundation of China (NSFC - 12175119, NSFC 12275265); Czech Republic: Czech Science Foundation (GACR - 24-11373S), Ministry of Education Youth and Sports (ERC-CZ-LL2327, FORTE CZ.02.01.01/00/22\_008/0004632), PRIMUS Research Programme (PRIMUS/21/SCI/017); EU: H2020 European Research Council (ERC - 101002463); European Union: European Research Council (BARD No. 101116429, ERC - 948254, ERC 101089007), European Regional Development Fund (SMASH COFUND 101081355, SLO ERDF), Horizon 2020 Framework Programme (MUCCA - CHIST-ERA-19-XAI-00), European Union, Future Artificial Intelligence Research (FAIR-NextGenerationEU PE00000013), Horizon 2020 (EuroHPC - EHPC-DEV-2024D11-051), Italian Center for High Performance Computing, Big Data and Quantum Computing (ICSC, NextGenerationEU); France: Agence Nationale de la Recherche (ANR-21-CE31-0022, ANR-22-EDIR-0002); Germany: Baden-Württemberg Stiftung (BW Stiftung-Postdoc Eliteprogramme), Deutsche Forschungsgemeinschaft (DFG - 469666862, DFG - CR 312/5-2); China: Research Grants Council (GRF); Italy: Istituto Nazionale di Fisica Nucleare (ICSC, NextGenerationEU), Ministero dell'Università e della Ricerca (NextGenEU 153D23001490006 M4C2.1.1, NextGenEU I53D23000820006 M4C2.1.1, NextGenEU I53D23001490006 M4C2.1.1); Japan: Japan Society for the Promotion of Science (JSPS KAKENHI JP22H01227, JSPS KAKENHI JP22H04944, JSPS KAKENHI JP22KK0227, JSPS KAKENHI JP23KK0245, JSPS KAKENHI JP24K23939); Norway: Research Council of Norway (RCN-314472); Poland: Ministry of Science and Higher Education (IDUB AGH, POB8, D4 no 9722), Polish National Science Centre (NCN 2021/42/E/ST2/00350, NCN OPUS 2023/51/B/ST2/02507, NCN OPUS nr 2022/47/B/ST2/03059, NCN UMO-2019/34/E/ST2/00393, UMO-2022/47/O/ST2/00148, UMO-2023/49/B/ST2/04085, UMO-2023/51/B/ST2/00920, UMO-2024/53/N/ST2/00869); Portugal: Foundation for Science and Technology (FCT); Spain: Ministry of Science and Innovation (MCIN \& NextGenEU PCI2022-135018-2, MICIN \& FEDER PID2021-125273NB, RYC2019-028510-I, RYC2020-030254-I, RYC2021-031273-I, RYC2022-038164-I); Sweden: Carl Trygger Foundation (Carl Trygger Foundation CTS 22:2312), Swedish Research Council (Swedish Research Council 2023-04654, VR 2021-03651, VR 2022-03845, VR 2022-04683, VR 2023-03403, VR 2024-05451), Knut and Alice Wallenberg Foundation (KAW 2018.0458, KAW 2022.0358, KAW 2023.0366); Switzerland: Swiss National Science Foundation (SNSF - PCEFP2\_194658); United Kingdom: Leverhulme Trust (Leverhulme Trust RPG-2020-004), Royal Society (NIF-R1-231091); United States of America: U.S. Department of Energy (ECA DE-AC02-76SF00515), Neubauer Family Foundation.

%
%


%
\clearpage
\appendix
\part*{Appendix}
\addcontentsline{toc}{part}{Appendix}

\section{Additional material on backgrounds with misidentified leptons}
\label{app:fakes}

\begin{figure}[htbp]
\centering
\subfloat[]{
\includegraphics[width=0.37\textwidth]{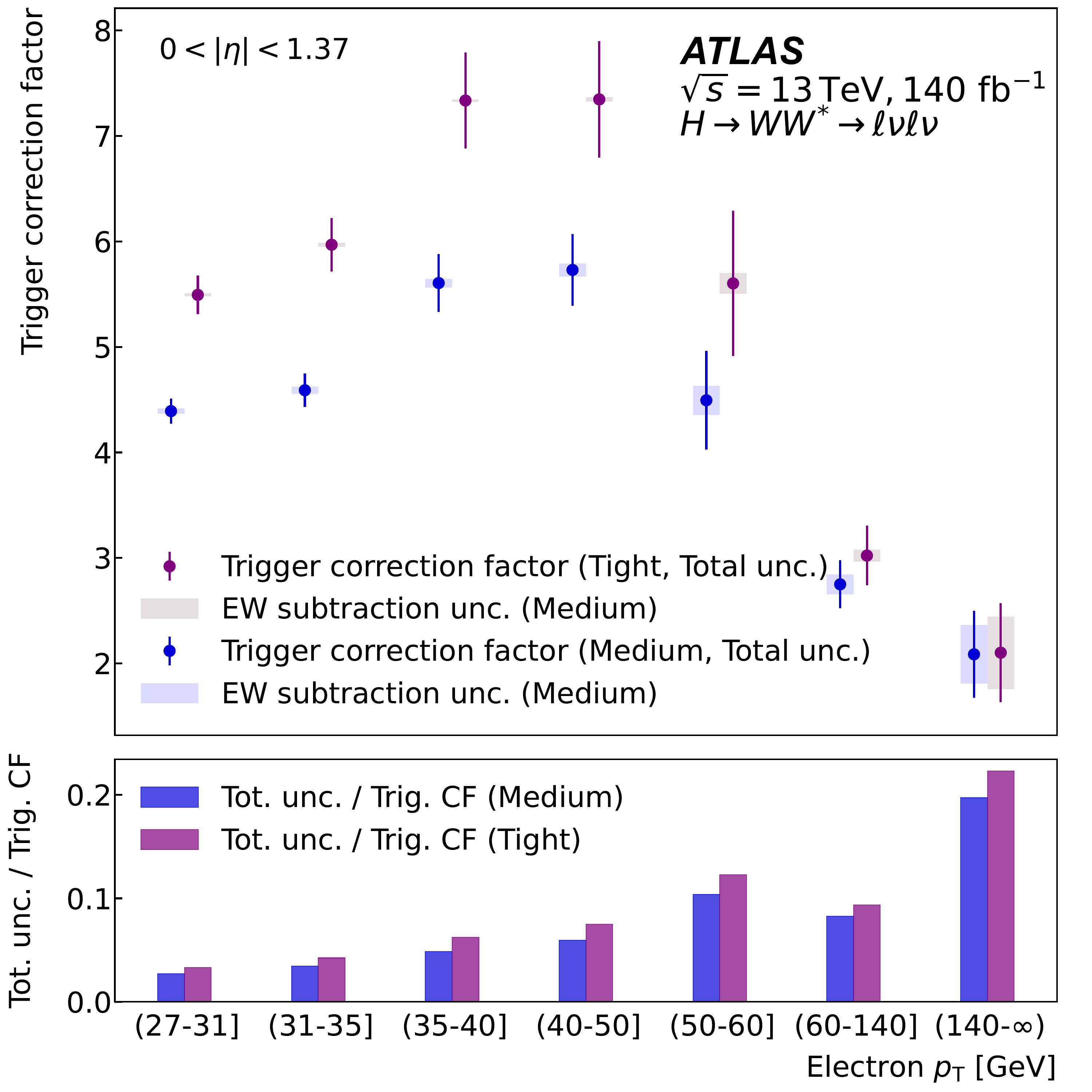}
\label{fig:TrigCFs_electrons_central}
}
\subfloat[]{
\includegraphics[width=0.37\textwidth]{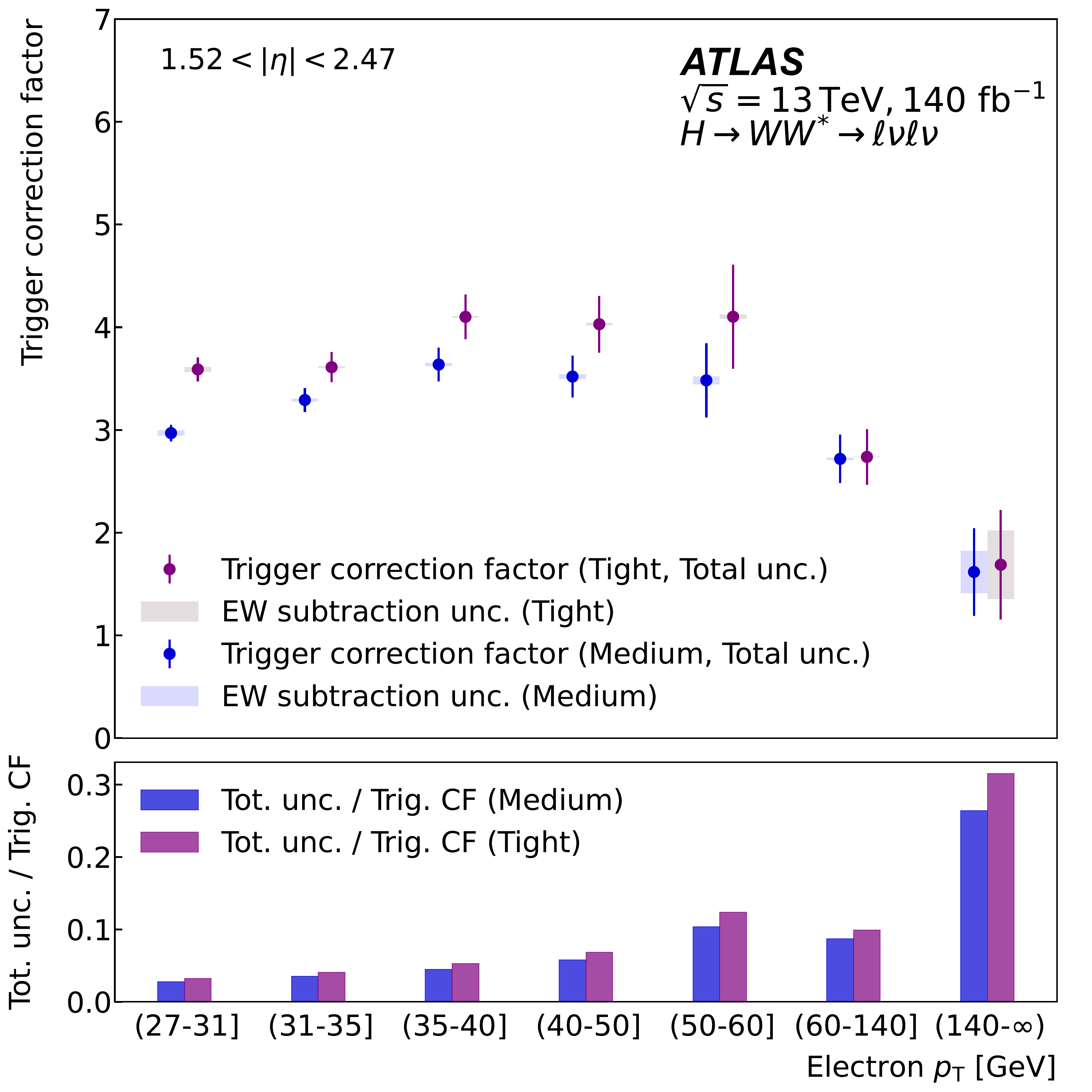}
\label{fig:TrigCFs_electrons_forward}
} \\
\subfloat[]{
\includegraphics[width=0.37\textwidth]{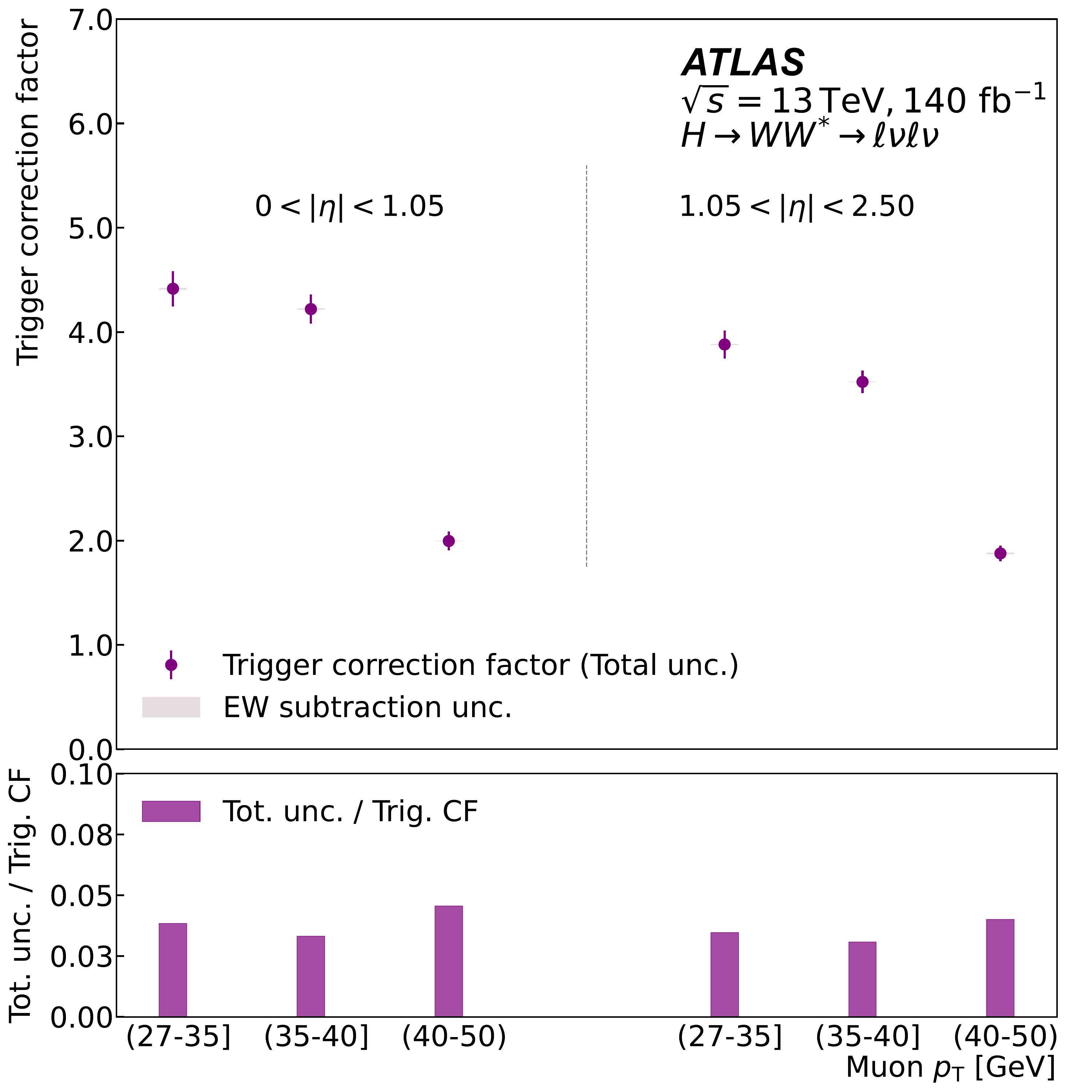}
\label{fig:TrigCFs_muons}
} \\
\caption{Trigger correction factors for anti-identified electrons in the (a) central and (b) forward regions of the detector and (c) for anti-identified muons. For electrons, the correction factors are provided for those that satisfy both the Medium and Tight identification criteria. As no trigger correction factor is necessary for a muon \pt above $\qty{50}{\GeV}$, the entries in (c) stop there. In the upper panels, the error bars indicate the total uncertainties in the trigger correction factors and the shaded bars indicate systematic uncertainties arising from the sources described in Section~\ref{sec:esystematics}. In the lower panels, the total relative uncertainties in the trigger correction factors are shown.}
\label{fig:TrigCFs}
\end{figure}

Single-lepton triggers and some dilepton triggers use stricter selection criteria than the anti-identification criteria used by this measurement for the data-driven estimation of the misidentified lepton background. Consequently, when only an anti-identified lepton triggers an event, it introduces bias. To account for this, an additional correction is applied to the extrapolation factor. Details of this additional correction are discussed in Section~\ref{sec:fakes}, while the resulting trigger correction factors are shown in Figure~\ref{fig:TrigCFs}.

\FloatBarrier

\section{Additional material on the STXS and CP-sensitive measurements}
\label{app:results}

\begin{figure}[htbp]
\centering
\includegraphics[width=\textwidth]{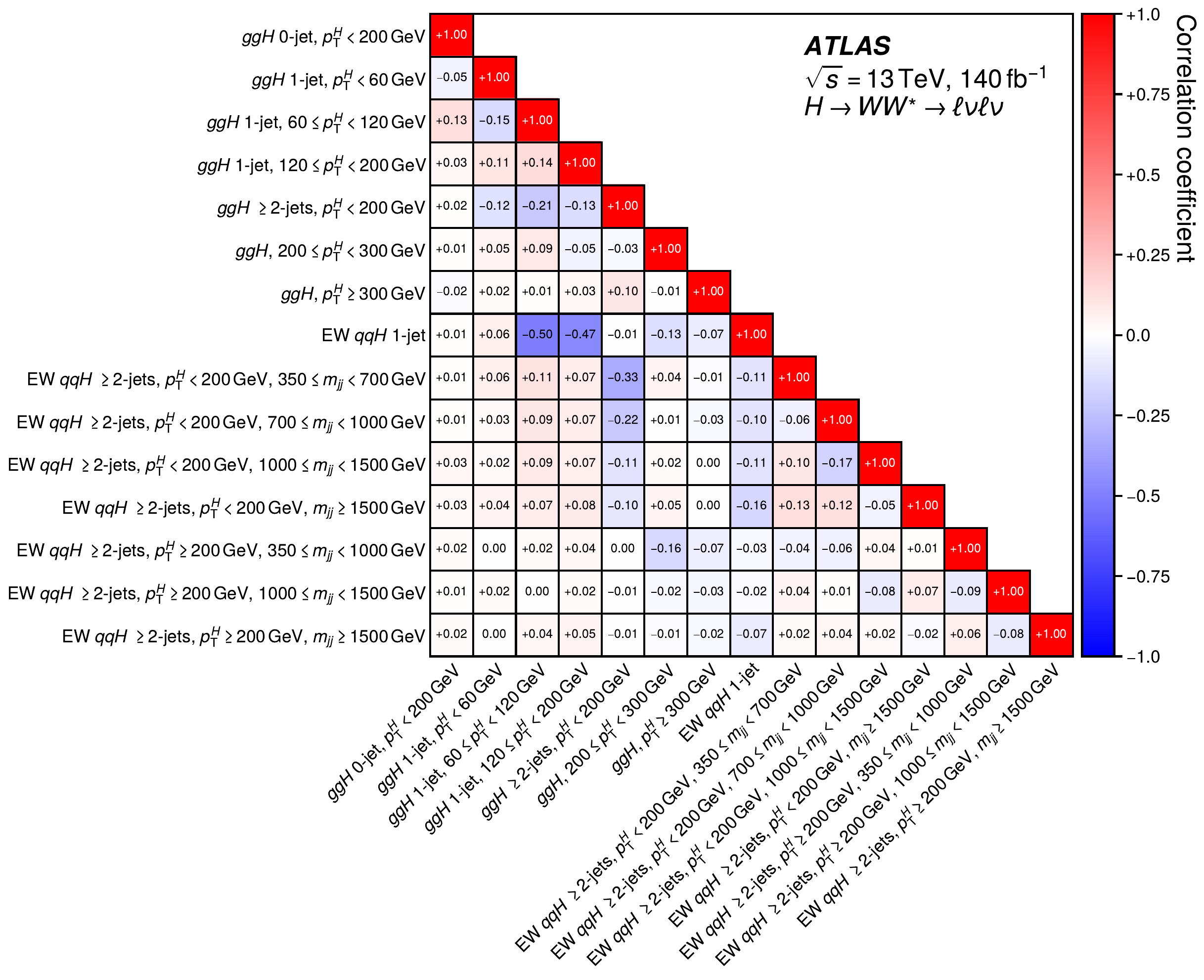}
\caption{Observed correlations between the production cross-sections multiplied by the \HWW branching ratio measured in data for each of the STXS categories, as discussed in Section~\ref{sec:results_STXS}.}
\label{fig:STXS_corr}
\end{figure}

\begin{figure}[htbp]
\centering
\includegraphics[width=\textwidth]{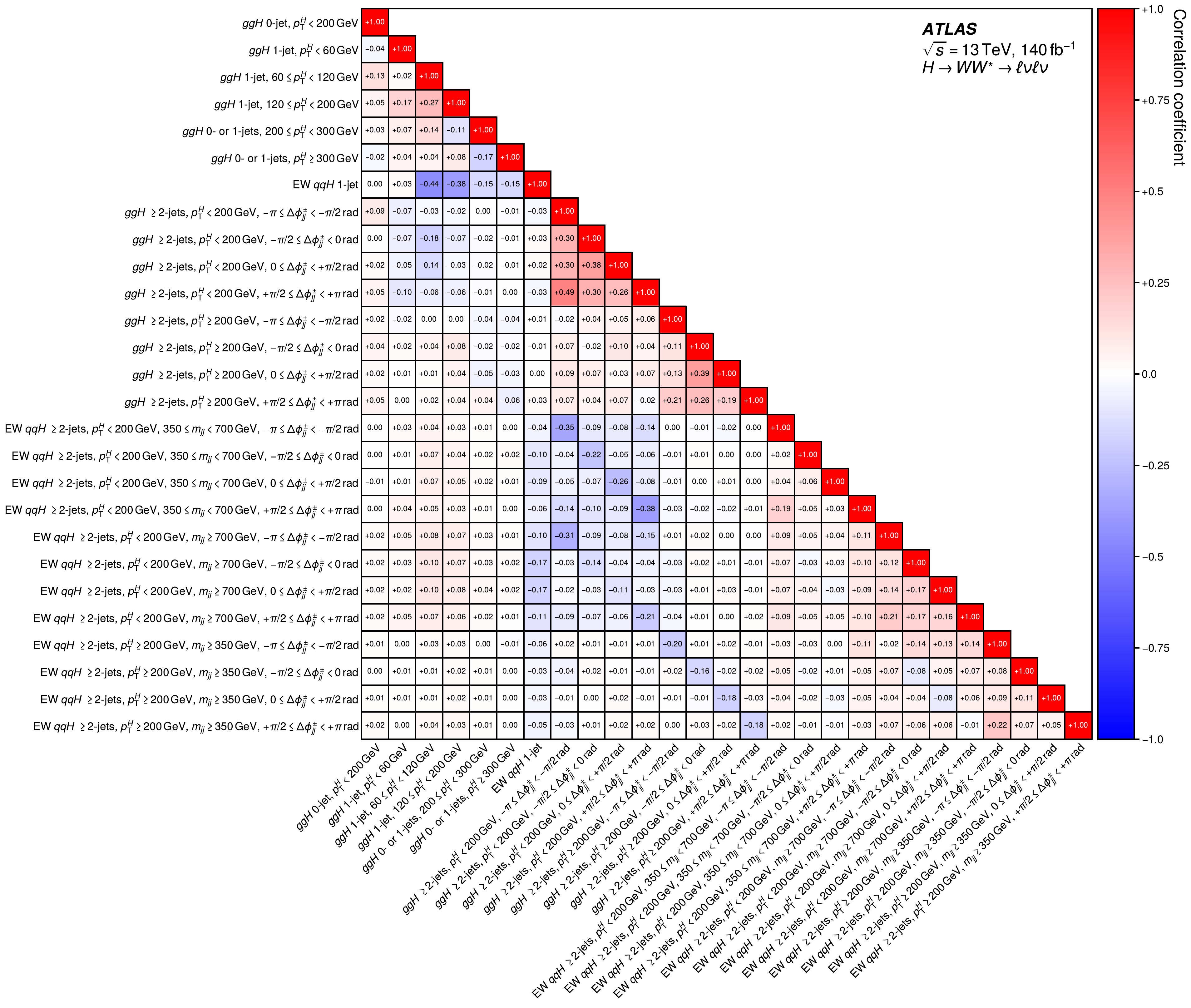}
\caption{Observed correlations between the production cross-sections multiplied by the \HWW branching ratio measured in data for each of the \STXSCP categories as discussed in Section~\ref{sec:results_DPhijj}.}
\label{fig:DPhijj_corr}
\end{figure}

This section provides additional information about the results of the STXS measurement, whose main results are shown and discussed in Section~\ref{sec:results_STXS}, and on the results of the \STXSCP measurement, whose main results are shown and discussed in Section~\ref{sec:results_DPhijj}. The observed central values and uncertainties of each of the measured STXS cross-sections, discussed in Section~\ref{sec:results_STXS} and compared with their SM prediction in Figures~\ref{fig:STXS_POIs}, are provided in Table~\ref{tab:STXS_POIs}. The observed correlation matrix for the measured STXS cross-sections is shown in Figure~\ref{fig:STXS_corr}, while the observed correlation matrix for the measured \STXSCP categories is shown in Figure~\ref{fig:DPhijj_corr}.

\begin{figure}[htbp]
\centering
\includegraphics[width=0.80\textwidth]{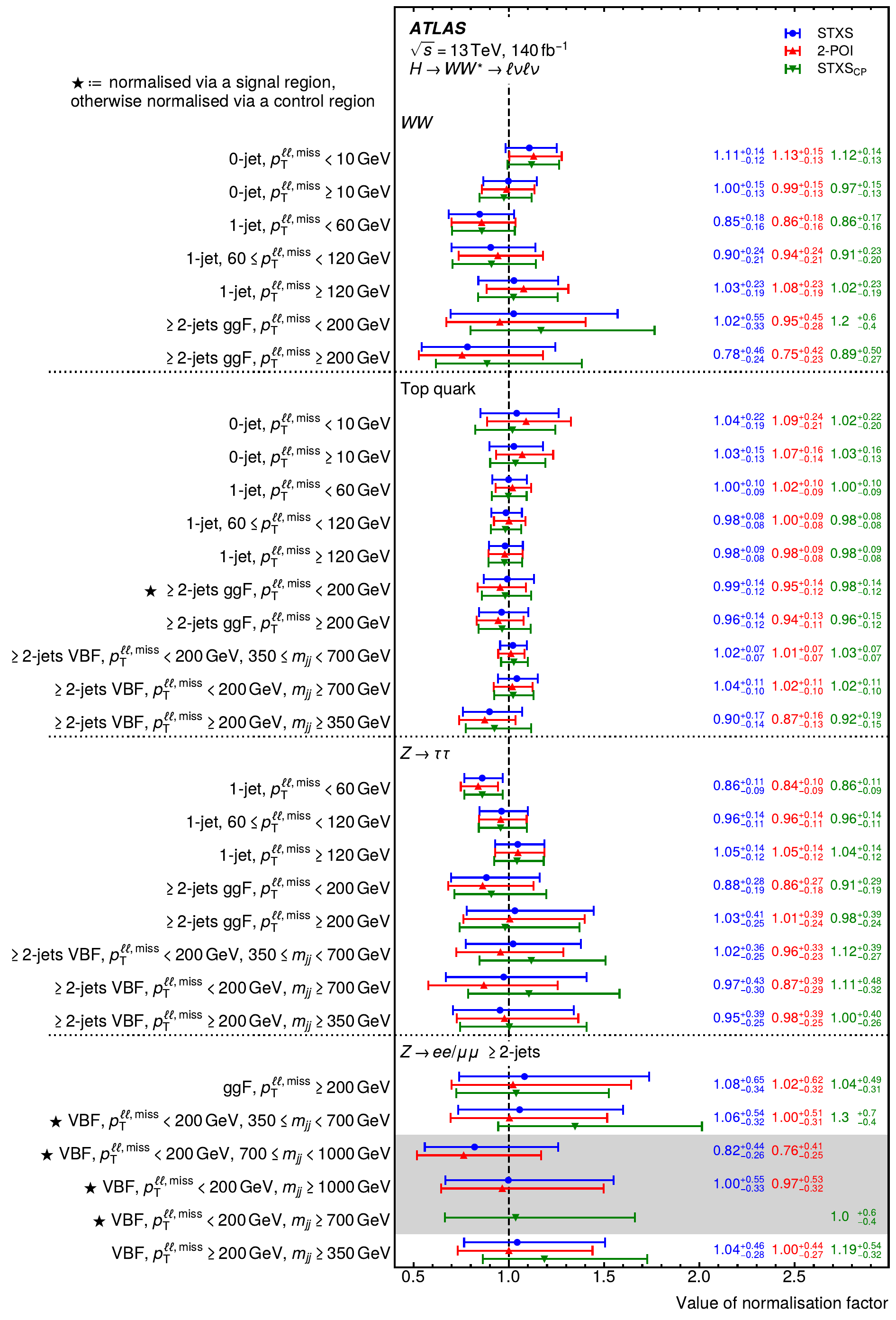}
\caption{Best-fit values and uncertainties for the background normalisation factors (NFs) from the STXS, 2-POI, and \STXSCP measurements. The process, jet multiplicity, and \ptllmiss/\mjj range to which each NF is applied are indicated. Each NF has a corresponding CR unless prefixed by a star ($\bigstar$), in which case the NF is fitted using only the corresponding SR(s). The shaded area denotes the NFs that use different \mjj splits between the STXS/2-POI and \STXSCP measurements.}
\label{fig:normalization_factors}
\end{figure}

The observed background normalisation factors and their uncertainties obtained from the STXS, 2-POI, and \STXSCP fits to data are shown in Figure~\ref{fig:normalization_factors} and are in good agreement with each other and with unity. The uncertainties in the NFs are generally large due to correlations with the uncertainties in the backgrounds they normalise, and differences between the best-fit values between the STXS/2-POI and \STXSCP measurements are due to differences in how the SRs are defined and the post-fit values of the nuisance parameters.

\begin{table}[htbp]
\centering
\caption{Best-fit values and uncertainties for the production cross-sections multiplied by the \HWW branching ratio in each of the STXS categories and the corresponding SM predictions, as discussed in Section~\ref{sec:results_STXS}.}
\label{tab:STXS_POIs}
\resizebox{\textwidth}{!}{
\begin{tabular}{l | r c c c c c | S[table-format=5.2,table-number-alignment=right]@{$\,\pm\,$}S[table-format=3.2,table-number-alignment=left]}
\toprule
\multirow{2}{*}{STXS category ($\XS_i\times\BRHWW$)} & \multicolumn{1}{c}{Value} & \multicolumn{5}{c|}{Uncertainty [fb]} & \multicolumn{2}{c}{SM prediction} \\
& \multicolumn{1}{c}{[fb]} & Total & Stat. & Exp. syst. & Sig. theo. & Bkg. theo. & \multicolumn{2}{c}{[fb]} \\
\midrule
\makecell[l]{\ggH 0-jet  \\ \hspace*{4mm}{\scriptsize $\pTH < \qty{200}{\GeV}$}} & $6300$ & $\pm800$ & $_{-500}^{+600}$ & $\pm500$ & $\pm100$ & $_{-350}^{+360}$ & 5900 & 400 \\
\midrule
\makecell[l]{\ggH 1-jet low \pTH \\ \hspace*{4mm}{\scriptsize $\pTH < \qty{60}{\GeV}$}} & $2100$ & $\pm700$ & $\pm400$ & $_{-350}^{+370}$ & $_{-80}^{+110}$ & $\pm400$ & 1400 & 190 \\
\midrule
\makecell[l]{\ggH 1-jet medium \pTH \\ \hspace*{4mm}{\scriptsize $60 \leq \pTH < \qty{120}{\GeV}$}} & $1600$ & $\pm400$ & $\pm300$ & $_{-180}^{+190}$ & $_{-40}^{+50}$ & $_{-200}^{+190}$ & 970 & 150 \\
\midrule
\makecell[l]{\ggH 1-jet high \pTH \\ \hspace*{4mm}{\scriptsize $120 \leq \pTH < \qty{200}{\GeV}$}} & $270$ & $\pm90$ & $\pm80$ & $_{-33}^{+34}$ & $\pm7$ & $\pm34$ & 160 & 30 \\
\midrule
\makecell[l]{\ggH $\geq$2-jets  \\ \hspace*{4mm}{\scriptsize $\pTH < \qty{200}{\GeV}$}} & $800$ & $\pm700$ & $\pm400$ & $_{-290}^{+260}$ & $_{-150}^{+140}$ & $\pm500$ & 1010 & 210 \\
\midrule
\makecell[l]{\ggH very high \pTH \\ \hspace*{4mm}{\scriptsize $200 \leq \pTH < \qty{300}{\GeV}$}} & $230$ & $_{-60}^{+70}$ & $\pm50$ & $_{-18}^{+20}$ & $_{-15}^{+24}$ & $_{-24}^{+26}$ & 96 & 22 \\
\midrule
\makecell[l]{\ggH highest \pTH \\ \hspace*{4mm}{\scriptsize $\pTH \geq \qty{300}{\GeV}$}} & $2$ & $_{-23}^{+25}$ & $_{-21}^{+22}$ & $\pm6$ & $_{-3.1}^{+3.3}$ & $_{-8}^{+9}$ & 26 & 7 \\
\midrule
\makecell[l]{EW \qqH 1-jet  \\ \hspace*{4mm}{\scriptsize ----}} & $-20$ & $_{-270}^{+280}$ & $_{-220}^{+230}$ & $\pm120$ & $_{-60}^{+40}$ & $\pm90$ & 279 & 14 \\
\makecell[l]{EW \qqH $\geq$2-jets low \pTH low \mjj \\ \hspace*{4mm}{\scriptsize $\pTH < \qty{200}{\GeV}$, $350 \leq \mjj < \qty{700}{\GeV}$}} & $90$ & $\pm60$ & $\pm50$ & $_{-17}^{+18}$ & $_{-10}^{+17}$ & $_{-19}^{+20}$ & 109 & 7 \\
\makecell[l]{EW \qqH $\geq$2-jets low \pTH medium \mjj \\ \hspace*{4mm}{\scriptsize $\pTH < \qty{200}{\GeV}$, $700 \leq \mjj < \qty{1000}{\GeV}$}} & $17$ & $_{-26}^{+29}$ & $_{-23}^{+25}$ & $\pm9$ & $_{-5}^{+6}$ & $_{-8}^{+9}$ & 56.0 & 3.5 \\
\makecell[l]{EW \qqH $\geq$2-jets low \pTH high \mjj \\ \hspace*{4mm}{\scriptsize $\pTH < \qty{200}{\GeV}$, $1000 \leq \mjj < \qty{1500}{\GeV}$}} & $59$ & $_{-23}^{+26}$ & $_{-20}^{+23}$ & $_{-6}^{+7}$ & $_{-4}^{+7}$ & $_{-7}^{+8}$ & 50.6 & 2.9 \\
\makecell[l]{EW \qqH $\geq$2-jets low \pTH highest \mjj \\ \hspace*{4mm}{\scriptsize $\pTH < \qty{200}{\GeV}$, $\mjj \geq \qty{1500}{\GeV}$}} & $62$ & $_{-19}^{+21}$ & $_{-16}^{+18}$ & $_{-6}^{+7}$ & $_{-6}^{+9}$ & $_{-4}^{+5}$ & 50 & 4 \\
\makecell[l]{EW \qqH $\geq$2-jets high \pTH medium low \mjj \\ \hspace*{4mm}{\scriptsize $\pTH \geq \qty{200}{\GeV}$, $350 \leq \mjj < \qty{1000}{\GeV}$}} & $25$ & $_{-12}^{+13}$ & $_{-11}^{+12}$ & $_{-3.5}^{+3.2}$ & $_{-1.1}^{+1.2}$ & $_{-2.2}^{+2.3}$ & 14.3 & 0.5 \\
\makecell[l]{EW \qqH $\geq$2-jets high \pTH high \mjj \\ \hspace*{4mm}{\scriptsize $\pTH \geq \qty{200}{\GeV}$, $1000 \leq \mjj < \qty{1500}{\GeV}$}} & $12$ & $_{-6}^{+7}$ & $_{-6}^{+7}$ & $_{-1.4}^{+1.7}$ & $_{-0.6}^{+1.0}$ & $_{-1.2}^{+1.1}$ & 7.55 & 0.32 \\
\makecell[l]{EW \qqH $\geq$2-jets high \pTH highest \mjj \\ \hspace*{4mm}{\scriptsize $\pTH \geq \qty{200}{\GeV}$, $\mjj \geq \qty{1500}{\GeV}$}} & $8$ & $_{-4}^{+5}$ & $_{-4}^{+5}$ & $\pm1.1$ & $_{-0.5}^{+1.1}$ & $_{-1.2}^{+1.1}$ & 9.9 & 0.5 \\
\bottomrule
\end{tabular}
}
\end{table}

\FloatBarrier

\section{Additional material on the SMEFT interpretations}
\label{app:EFT}

\begin{figure}[htbp]
\centering
\includegraphics[width=0.40\textwidth]{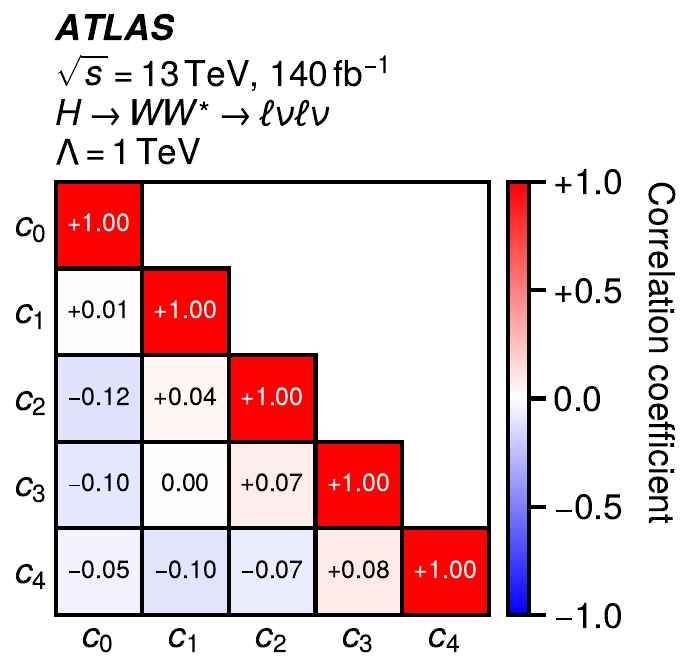}
\caption{Observed correlations for the Wilson coefficients of the \dimSix operators considered by the SMEFT interpretation of the STXS measurement in a CP-conserving scenario. The \dimSix operators are given in the rotated basis. All Wilson coefficients are fitted simultaneously.}
\label{fig:EFT_correlations}
\end{figure}

The correlations for the observed values of the Wilson coefficients in SMEFT interpretation of the STXS measurement are generally small, as shown in Figure~\ref{fig:EFT_correlations}, owing to the method of simultaneously fitting eigenvectors described in Section~\ref{sec:smeft_interp}.

\begin{figure}[htbp]
\centering
\includegraphics[width=0.75\textwidth]{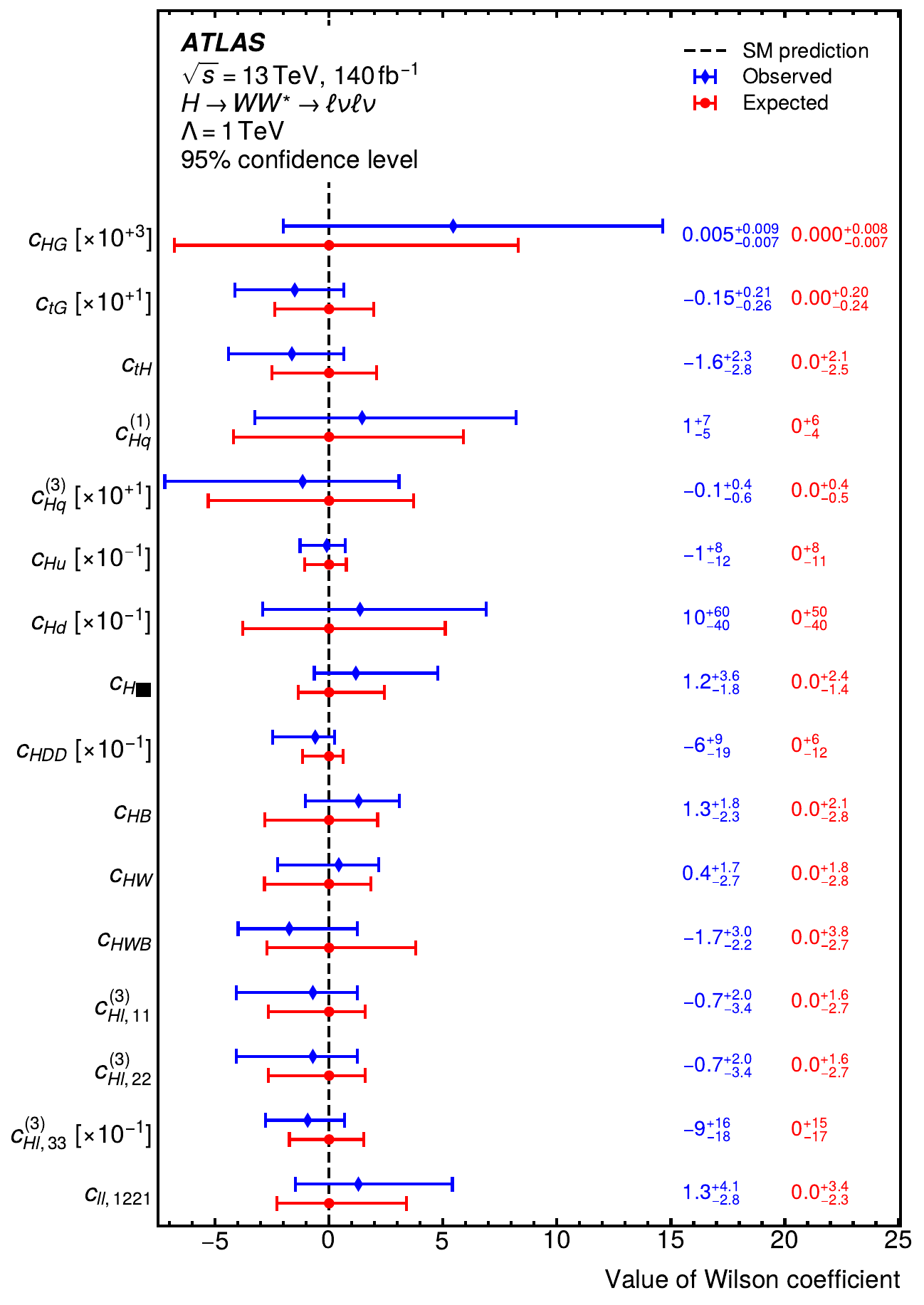}
\caption{Best-fit values and uncertainties for the Wilson coefficients of the \dimSix operators considered by the SMEFT interpretation of the STXS measurement in a CP-conserving scenario. The \dimSix operators are given in the Warsaw basis, and each Wilson coefficient is fitted independently with the others fixed to 0. Both the observed (diamond) and expected (circle) results are shown. The best-fit values and uncertainties for the observed (expected) results are given by the left (right) numbers on each line. The error bars and quoted errors correspond to the total 95\% CL uncertainty in each Wilson coefficient.}
\label{fig:smeft_1poi_fits}
\end{figure}

A SMEFT interpretation of the STXS measurement is also performed using the Warsaw (i.e., not rotated) basis for the 16~operators shown in Table~\ref{tab:smeft_defs}. Constraints on the corresponding Wilson coefficients are obtained by fitting each Wilson coefficient one at a time while fixing all others to zero, i.e.\ their SM value. The expected and observed results of this procedure are summarised in Figure~\ref{fig:smeft_1poi_fits}. All measured Wilson coefficients are in agreement with the SM within their 95\% CL intervals. While such an approach has limited validity regarding its physics interpretation, these results demonstrate the sensitivity of the analysis to each measured Wilson coefficient.

\begin{figure}[htbp]
\centering
\includegraphics[width=0.48\textwidth]{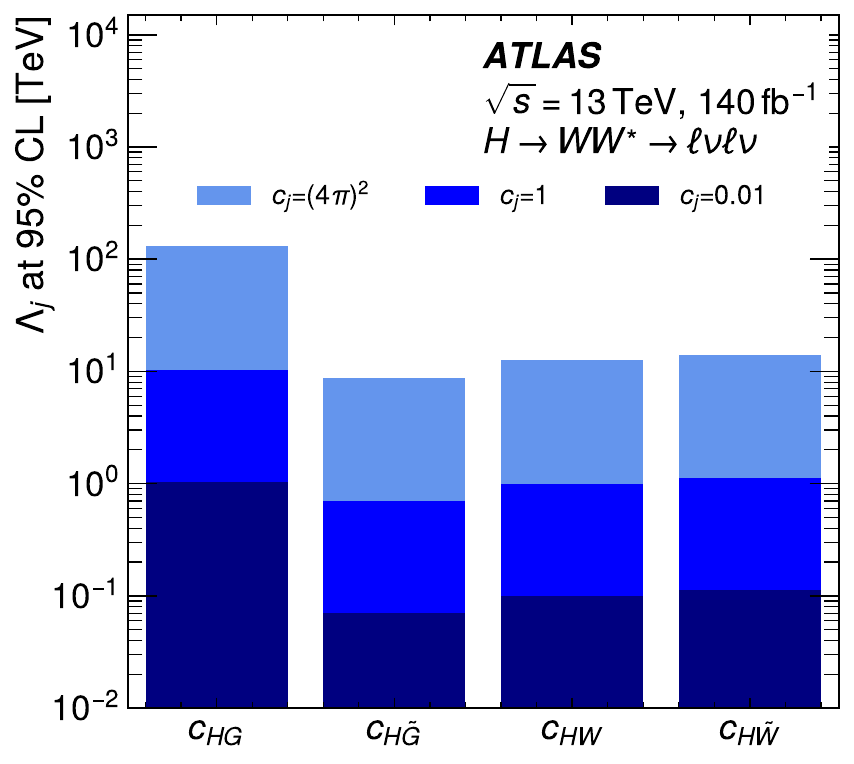}
\caption{Summary of the 95\% confidence level (CL) observed lower limits on the scale of new physics, $\Lambda_j$, for the indicated values of the Wilson coefficients, $c_j$. These limits are derived using the results of the SMEFT interpretation of the \STXSCP measurement in a CP-violating scenario, shown in Figure~\ref{fig:cp_fit_pois}. Asymmetric uncertainties are symmetrised using the root mean square.}
\label{fig:Lambda_limits_CPmixed}
\end{figure}

\begin{figure}[htbp]
\centering
\includegraphics[width=0.94\textwidth]{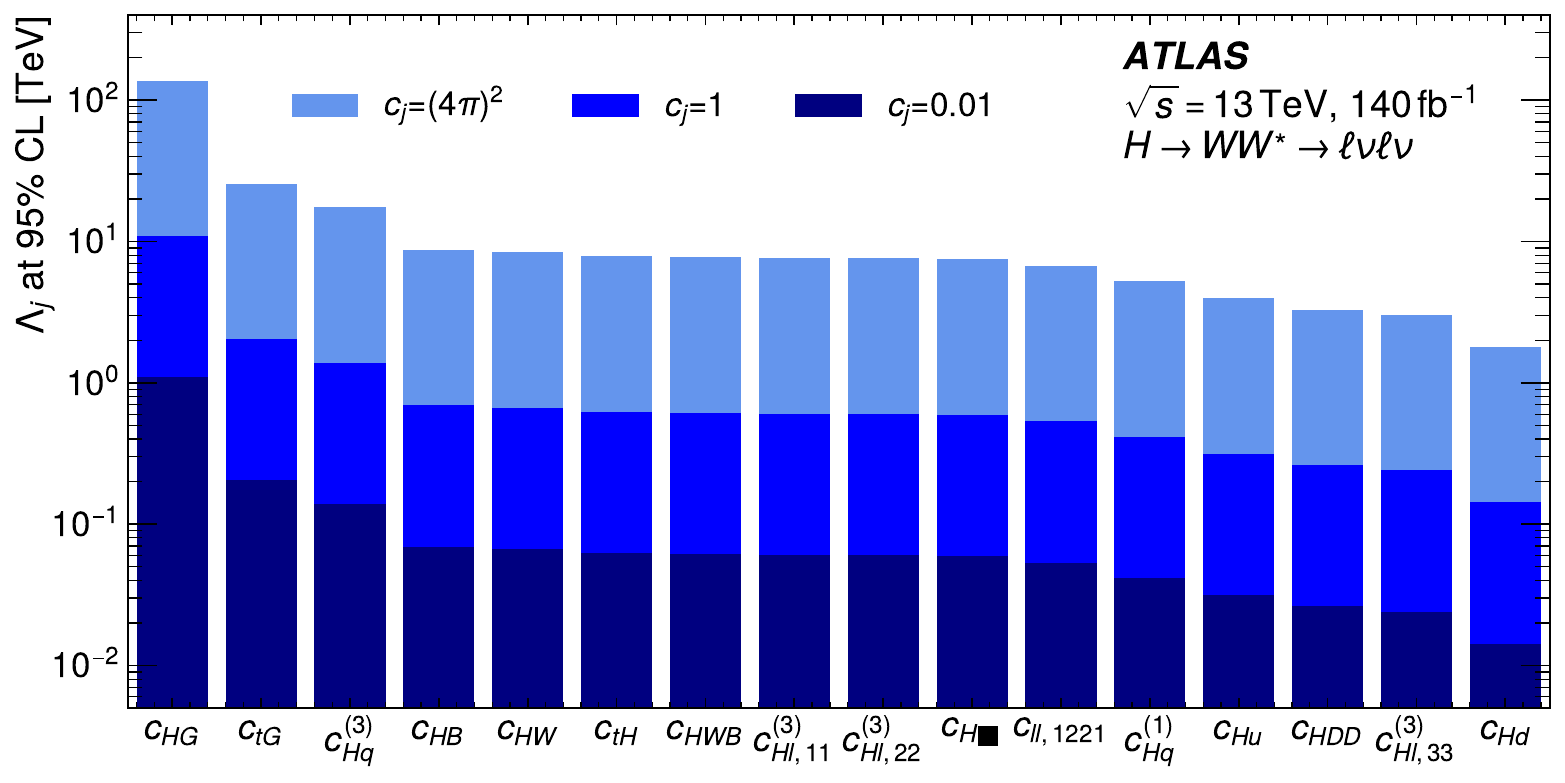}
\caption{Summary of the 95\% confidence level (CL) observed lower limits on the scale of new physics, $\Lambda_j$, for the indicated values of the Wilson coefficients, $c_j$. These limits are derived using the results of the SMEFT interpretation of STXS measurement in a CP-conserving scenario, shown in Figure~\ref{fig:smeft_1poi_fits}. Asymmetric uncertainties are symmetrised using the root mean square.}
\label{fig:Lambda_limits_STXS}
\end{figure}

Limits on Wilson coefficients, where a fixed value of the energy scale $\Lambda = \qty{1}{\TeV}$ is assumed, can also be interpreted in another manner. One can assume a certain value of the Wilson coefficient and thus express results as a lower limit on the possible energy scale where the corresponding operators would enter. For the 95\% CL interval on the Wilson coefficient observed in data, one can assume three different fixed values of the Wilson coefficient instead. These three values are $(4\pi)^2$, 1, and 0.01 and result in three different 95\% CL observed limits on the scale of new physics $\Lambda_j$. The results of this procedure are shown in Figure~\ref{fig:Lambda_limits_CPmixed} for the SMEFT interpretation of the \STXSCP measurement in a CP-violating scenario described in Section~\ref{sec:cp_interp}. Figure~\ref{fig:Lambda_limits_STXS} shows these same results for the SMEFT of the STXS measurement in a CP-conserving scenario where only one coefficient is measured at a time, as described in the previous paragraph.

\FloatBarrier
\printbibliography
\clearpage
 
\begin{flushleft}
\hypersetup{urlcolor=black}
{\Large The ATLAS Collaboration}

\bigskip

\AtlasOrcid[0000-0002-6665-4934]{G.~Aad}$^\textrm{\scriptsize 104}$,
\AtlasOrcid[0000-0001-7616-1554]{E.~Aakvaag}$^\textrm{\scriptsize 17}$,
\AtlasOrcid[0000-0002-5888-2734]{B.~Abbott}$^\textrm{\scriptsize 123}$,
\AtlasOrcid[0000-0002-0287-5869]{S.~Abdelhameed}$^\textrm{\scriptsize 119a}$,
\AtlasOrcid[0000-0002-1002-1652]{K.~Abeling}$^\textrm{\scriptsize 55}$,
\AtlasOrcid[0000-0001-5763-2760]{N.J.~Abicht}$^\textrm{\scriptsize 49}$,
\AtlasOrcid[0000-0002-8496-9294]{S.H.~Abidi}$^\textrm{\scriptsize 30}$,
\AtlasOrcid[0009-0003-6578-220X]{M.~Aboelela}$^\textrm{\scriptsize 45}$,
\AtlasOrcid[0000-0002-9987-2292]{A.~Aboulhorma}$^\textrm{\scriptsize 36e}$,
\AtlasOrcid[0000-0001-5329-6640]{H.~Abramowicz}$^\textrm{\scriptsize 157}$
% ...
\bigskip
\\

% ...
$^{17}$Department for Physics and Technology, University of Bergen, Bergen; Norway.\\
% ...
$^{30}$Physics Department, Brookhaven National Laboratory, Upton NY; United States of America.\\
% ...
$^{36}$$^{(a)}$Facult\'e des Sciences Ain Chock, Universit\'e Hassan II de Casablanca;$^{(b)}$Facult\'{e} des Sciences, Universit\'{e} Ibn-Tofail, K\'{e}nitra;$^{(c)}$Facult\'e des Sciences Semlalia, Universit\'e Cadi Ayyad, LPHEA-Marrakech;$^{(d)}$LPMR, Facult\'e des Sciences, Universit\'e Mohamed Premier, Oujda;$^{(e)}$Facult\'e des sciences, Universit\'e Mohammed V, Rabat;$^{(f)}$Institute of Applied Physics, Mohammed VI Polytechnic University, Ben Guerir; Morocco.\\
% ...
$^{45}$Physics Department, Southern Methodist University, Dallas TX; United States of America.\\
% ...
$^{49}$Fakult\"{a}t Physik , Technische Universit{\"a}t Dortmund, Dortmund; Germany.\\
% ...
$^{55}$II. Physikalisches Institut, Georg-August-Universit\"{a}t G\"ottingen, G\"ottingen; Germany.\\
% ...
$^{104}$CPPM, Aix-Marseille Universit\'e, CNRS/IN2P3, Marseille; France.\\
% ...
$^{119}$$^{(a)}$New York University Abu Dhabi, Abu Dhabi;$^{(b)}$United Arab Emirates University, Al Ain; United Arab Emirates.\\
% ...
$^{123}$Homer L. Dodge Department of Physics and Astronomy, University of Oklahoma, Norman OK; United States of America.\\
% ...
$^{157}$Raymond and Beverly Sackler School of Physics and Astronomy, Tel Aviv University, Tel Aviv; Israel.\\
% ...
$^{a}$ Also at Affiliated with an institute formerly covered by a cooperation agreement with CERN.\\
$^{b}$ Also at An-Najah National University, Nablus; Palestine.\\
$^{c}$ Also at Borough of Manhattan Community College, City University of New York, New York NY; United States of America.\\
$^{d}$ Also at Center for Interdisciplinary Research and Innovation (CIRI-AUTH), Thessaloniki; Greece.\\
$^{e}$ Also at Centre of Physics of the Universities of Minho and Porto (CF-UM-UP); Portugal.\\
$^{f}$ Also at CERN, Geneva; Switzerland.\\
% ...

\end{flushleft}


%

%
%

%

%
%
%
%

\end{document}